%% file: main.tex
\documentclass[aps,pra,twocolumn,floatfix,amsmath,amssymb,eqsecnum,nofootinbib,superscriptaddress]{revtex4-1}
\usepackage{graphicx}
\usepackage{dcolumn}
\usepackage{bm}
\usepackage{epstopdf}
\usepackage{framed}
\usepackage[toc,page]{appendix}

\usepackage[utf8]{inputenc}

\newcommand{\beq}{\begin{equation}}
\newcommand{\eeq}{\end{equation}}
\newcommand{\ba}{\begin{array}{ccc}}
\newcommand{\ea}{\end{array}}

\def\bea{\begin{eqnarray}}
\def\eea{\end{eqnarray}}

\def\Tr{ {\rm Tr} }
\def\<{\langle}
\def\>{\rangle}

\def\bZ{\mathbb{Z}}
\def\bR{\mathbb{R}}

\def\cH{\mathcal{H}}
\usepackage{amsmath}
\usepackage{amssymb}
\usepackage{amsthm}
\newtheorem{thm}{Theorem}
\newtheorem{lemma}{Lemma}

\theoremstyle{definition}
\newtheorem{defin}{Definition}
\newtheorem{q}{Question}
\usepackage{color}
\usepackage{tikz-cd}
\usepackage{enumerate}
\usepackage[caption=false]{subfig}

\newcommand{\ket}[1]{|#1\rangle}

\newcommand{\U}{\mathrm{U}}

\usepackage{tikz}
\usetikzlibrary{decorations.markings}
\begin{document}

\title{Gauging spatial symmetries and the classification of topological crystalline phases}
\author{Ryan Thorngren}
\affiliation{Department of Mathematics, University of California, Berkeley, CA 94720, USA}
\affiliation{Kavli Institute for Theoretical Physics, University of California, Santa Barbara, CA 93106, USA}
\author{Dominic V. Else}
\affiliation{Department of Physics, University of California, Santa Barbara, CA 93106, USA}
\begin{abstract}
We put the theory of interacting topological crystalline phases on a systematic footing. These are topological phases protected by space-group symmetries. Our central tool is an elucidation of what it means to ``gauge'' such symmetries. We introduce the notion of a \emph{crystalline topological liquid}, and argue that most (and perhaps all) phases of interest are likely to satisfy this criterion. We prove a \emph{Crystalline Equivalence Principle}, which states that in Euclidean space, crystalline topological liquids with symmetry group $G$ are in one-to-one correspondence with topological phases protected by the same symmetry $G$, but acting \emph{internally},
where if an element of $G$ is orientation-reversing, it is realized as an anti-unitary symmetry in the internal symmetry group.
As an example, we explicitly compute, using group cohomology, a partial classification of bosonic symmetry-protected topological (SPT) phases protected by crystalline symmetries in (3+1)-D for 227 of the 230 space groups. For the 65 space groups not containing orientation-reversing elements (Sohncke groups), there are no cobordism invariants which may contribute phases beyond group cohomology, and so we conjecture our classification is complete.

\end{abstract}
\maketitle

Symmetry is an important feature of many physical systems. Many phases of matter can be characterized in part by the way the symmetry is implemented. For example, liquids and solids are distinguished by whether or not they spontaneously break spatial symmetries. In fact, it was once thought that \emph{all} known phases could be distinguished by their symmetries and that all continuous phase transitions were spontaneous symmetry breaking transitions. The discovery of \emph{topological order}\cite{Wen1990} showed that, at zero temperature, there are quantum phases of matter that can be distinguished by patterns of long-range entanglement without the need to invoke symmetry. However, even for topological phases symmetry is important. Any symmetry that is not spontaneously broken in a topological phase must have some action on the topological structure of the phase, and different such patterns can distinguish different phases. Even a phase of matter that is trivial without symmetry can become non-trivial when considering how symmetry is implemented. Topological phases distinguished by symmetry are known as \emph{symmetry-enriched topological (SET)}\cite{Maciejko2010,Essin2013,Lu2013,Mesaros2013,Hung2013,BBCW,Cheng2015a} or \emph{symmetry-protected topological (SPT)}\cite{Gu2009,Pollmann2010,Pollmann2012,Fidkowski2010,Chen2010,Chen2011,Schuch2011a,Fidkowski2011,CGLW,Chen2011b,Vishwanath2013,Wang2014, K,Gu2014,Else2014,Burnell2014,
Wang2015,Cheng2015} depending on whether they are nontrivial or trivial without symmetry, respectively.

For internal symmetries, which do not move points in space around, very general and powerful ways of understanding SPT and SET phases have been formulated in terms of mathematical notions such as group cohomology\cite{CGLW}, category theory\cite{BBCW}, and cobordisms\cite{K,KTTW}. On the other hand, such techniques have not, so far, been extended to the case of \emph{space group} symmetries. We refer to these topological phases enriched by space-group symmetries as \emph{topological crystalline phases}. This is a significant omission because any system which arranges itself into a regular crystal lattice is invariant under one of 230 space groups in three dimensions. Fermionic phases of matter protected by space-group symmetries are called \emph{topological crystalline insulators} or \emph{topological crystalline superconductors} depending on whether charge is conserved \cite{Fu2011,Hsieh2012,Dziawa2012,Tanaka2012,Xu2012,Zhang2013}. Progress towards a general classification in free-fermion systems has been made \cite{Fang2012,Fang2013,Chiu2013,Morimoto2013,Slager2013,Shiozaki2014,Chiu2016} and some understanding of the effect of interactions been achieved \cite{Isobe2015,Hsieh2014a,Qi2015,Yoshida2015a,SS}. Meanwhile, intrinsically strongly interacting phases protected by spatial symmetries have also been found \cite{Wen2002,Essin2013,Hsieh2014,Cho2015,Yoshida2015,BBCW,Lapa2016,You2014,Hermele2016,Cheng2015c,Jiang2016}. In particular Ref.~\cite{SHFH} gave an approach for deriving the general classification of interacting SPT phases protected by a group of spatial symmetries that leave a given point invariant. However, for SETs and/or general space groups, there is so far no systematic theory analogous to the one that exists for internal symmetries, except in one dimension \cite{Chen2011d}.
Our goal in this paper is to fill this gap.

We will adopt two complementary and related viewpoints to the classification. The first viewpoint is in terms of topological quantum field theories (TQFTs), which are believed to describe the low-energy physics of topological phases. We state and motivate a proposal for how to implement a spatial symmetry in a TQFT.

Our second, more concrete, viewpoint is based on the idea of understanding the SPT or SET order of a system by studying its response to a gauge field. For example, SPTs in (2+1)-D protected by an internal $U(1)$ symmetry can be identified by the topological response to a $U(1)$ gauge field. All such possible responses are described by the Chern-Simons action
\begin{equation}
    S = \frac{k}{4\pi} \int A \wedge dA.
\end{equation}
The coefficient $k$ has a physical interpretation as the quantized Hall conductance. Because it is quantized, the only way to get between systems with different values of $k$ is if $\U(1)$ symmetry is broken or the gap closes. Further, since this is the only term that may appear, we learn that the different $\U(1)$ SPTs in 2+1D are labelled by this integer. We call this procedure of coupling a $G$-symmetric system to a background $G$ gauge field ``gauging" the $G$ symmetry, though strictly speaking we do not consider making the gauge field dynamical. Stricter terminology would call the dynamical gauge theory the result of gauging and our procedure the first step, called equivariantization, a mouthful, or pregauging. Many of the general approaches to SPT and SET phases can be formulated in terms of gauging\cite{DW,Levin2012,Hung2013,BBCW}.

We want to apply similar approaches to the study of systems with spatial symmetry. So we will ask the question
\begin{q}
What does it mean to gauge a spatial symmetry?
\end{q}
We will give what we believe to be the definitive answer to this question, motivated by the intuition of ``gauge fluxes'' which for spatial symmetries are crystallographic defects such as dislocations and disclinations. There seems to be a natural generalization of this to symmetries which act on spacetime as well, such as time reversal symmetry or time translation. We will mention briefly this generalization and how the classification extends to these spacetime symmetries, where it agrees with known group cohomology classifications of time reversal-invariant and Floquet SPTs, respectively.

Using the two viewpoints mentioned above, we will elucidate the general theory of crystalline topological phases. Our results are based on a key physical assumption, namely that the phases of matter under consideration are \emph{crystalline topological liquid}, which roughly means that, although crystalline, they preserve a certain degree of ``fluidity'' in the low-energy limit. The idea is motivated by the notion of ``topological liquids'' which have an IR limit that is described by a topological quantum field theory (TQFT), i.e.\ the long-range physics is only sensitive to the topology of the background manifold. This is in contrast to ``fracton'' topological phases\cite{Haah,Yoshida,Vijay2016,Williamson2016} where no such topological IR limit exists\footnote{Although see Section \ref{sec_generalizations}.}. Crystalline topological liquids are a generalization of topological liquids to systems with crystal symmetries.

The main result of this paper is the following.
\begin{framed}
\textbf{Crystalline Equivalence Principle}: The classification of crystalline topological liquids with spatial symmetry group $G$ is the \emph{same} as the classification of topological phases with \emph{internal} symmetry $G$.
\end{framed}
Compare Ref. \onlinecite{DN}, where a similar principle was conjectured for symmetry groups containing \emph{time translation symmetry}.
This result holds for systems living on a \emph{contractible} space, ie. Euclidean space in $d$ dimensions. On other manifolds, for example Euclidean space with some holes, some new things happen. We note for this correspondence, orientation-reversing symmetries in the space group must correspond to anti-unitary symmetries in the internal group.

\begin{table}
\begin{tabular}{|l|l|l|}
\hline
Number & Name & Classification \\ \hline
1 & $\mathrm{p1}$ & 0\\
2 & $\mathrm{p2}$ & $\mathbb{Z}_{2}^{\times 4}$\\
3 & $\mathrm{pm}$ & $\mathbb{Z}_{2}^{\times 2}$\\
4 & $\mathrm{pg}$ & 0\\
5 & $\mathrm{cm}$ & $\mathbb{Z}_{2}$\\
6 & $\mathrm{p2mm}$ & $\mathbb{Z}_{2}^{\times 8}$\\
7 & $\mathrm{p2mg}$ & $\mathbb{Z}_{2}^{\times 3}$\\
8 & $\mathrm{p2gg}$ & $\mathbb{Z}_{2}^{\times 2}$\\
9 & $\mathrm{c2mm}$ & $\mathbb{Z}_{2}^{\times 5}$\\
10 & $\mathrm{p4}$ & $\mathbb{Z}_{2}\times\mathbb{Z}_{4}^{\times 2}$\\
11 & $\mathrm{p4mm}$ & $\mathbb{Z}_{2}^{\times 6}$\\
12 & $\mathrm{p4gm}$ & $\mathbb{Z}_{2}^{\times 2}\times\mathbb{Z}_{4}$\\
13 & $\mathrm{p3}$ & $\mathbb{Z}_{3}^{\times 3}$\\
14 & $\mathrm{p3m1}$ & $\mathbb{Z}_{2}$\\
15 & $\mathrm{p31m}$ & $\mathbb{Z}_{2}\times\mathbb{Z}_{3}$\\
16 & $\mathrm{p6}$ & $\mathbb{Z}_{2}^{\times 2}\times\mathbb{Z}_{3}^{\times 2}$\\
17 & $\mathrm{p6mm}$ & $\mathbb{Z}_{2}^{\times 4}$\\
\hline
\end{tabular}
\caption{\label{bosonic2d}The classification of bosonic SPT phases in (2+1)-D protected by space group symmetries, for each of the 17 2-D space groups (sometimes known as ``wallpaper groups'').}
\end{table}

We emphasize that the Crystalline Equivalence Principle is expected to hold for both bosonic and fermionic\footnote{There are some caveats for fermionic systems: systems with $R^2 = +1$, where $R$ is a reflection, are in correspondence with systems with $T^2 = (-1)^F$, where $T$ is time-reversal, and vice versa.} systems, and for both SPT and SET phases. As an example of results that one can deduce from this general principle, we find that bosonic SPT phases protected by orientation-preserving unitary spatial symmetry $G$ are classified by the group cohomology $\mathcal{H}^{d+1}(G, \mathrm{U}(1))$, since that is the classification of internal SPTs with symmetry $G$ (See Appendix \ref{topphases} for more details on the definition of $\mathcal{H}$.) This agrees with a recent classification of a class of tensor networks with spatial symmetries\cite{Jiang2016}.
In (3+1)-D, for space groups containing orientation-reversing transformations, this classification is expected to be incomplete, just as it is for internal symmetry groups containing anti-unitary symmetries\cite{K}. Applying the principle to fermionic systems, one obtains a partial classification of fermionic SPT's protected by space-group symmetries from ``group supercohomology'' \cite{Gu2014} and a complete classification of fermionic crystalline SPT's from cobordism theory \cite{KTTW}, with some caveats. We attempt this in section \ref{fermions} for crystalline topological superconductors and insulators.

Our results allow for the classification to be explicitly computed in many cases. For example, Table \ref{bosonic2d} shows the classification of bosonic SPT phases protected by space-group symmetry in (2+1)-D as obtained from group cohomology. For more details of how Table \ref{bosonic2d} was computed, and the (3+1)-D version of the table, see Appendix \ref{sec_bosonic_classif}.

The outline of our paper is as follows. In Section \ref{sec_toplimit}, we introduce the notion of a crystalline topological liquid. Then, in Section \ref{sec_core_concepts} we introduce the key ideas involved in gauging a spatial symmetry. Specifically, in section \ref{sec_spatial_gauge_fields} we discuss our definition of crystalline gauge field. Then in \ref{sec_tcl} we argue that crystalline topological liquids naturally couple to such crystalline gauge fields. In \ref{sec_crystalline_equivalence} we use the gauging picture to derive the Crystalline Equivalence Principle, which applies to the physically relevant case of phases of matter in contractible space $\mathbb{R}^d$. In \ref{sec_beyond} we discuss extensions to non-contractible spaces and a general classification result for crystalline gauge fields.

In Section \ref{sec_bootstrap} we give a construction of many crystalline topological liquids from ordinary topological liquids by considering systems which carry both a spatial $G$ symmetry and an internal $G$ symmetry. 

In Section \ref{sec_classification} we describe our approach towards classifying crystalline topological liquids using topological response. In \ref{sec_flux_fusion}, this is defined in terms of fusion and braiding of symmetry fluxes. In \ref{sec_topological_actions} it is described in terms of effective topological actions. Finally, in \ref{sec_examples}, we give many examples of crystalline gauge backgrounds and compute the resulting partition functions in exactly solvable models. This section is particularly important, as it elucidates where some of the familiar features of ordinary SPT phases appear in crystalline SPT phases.

In Section \ref{sec_space_dep_tqft}, we describe how our methods can be placed into a general context of position-dependent topological limit, and discuss implications of emergent Lorentz invariance or lack thereof.

In Section \ref{noncontractibleproperties}, we derive several general structural results about the classification of crystalline SPTs (invertible crystalline topological liquids).

We give generalizations of our methods in Section \ref{sec_generalizations}, including Floquet phases in \ref{sec_floquet} and phases beyond ordinary equivariant cohomology in \ref{fermions} including fermions. In \ref{sigma} we discuss how our methods apply to topological terms of sigma models.

In Section \ref{sec_beyond_tcl}, we describe some ways in which our crystalline topological liquid assumption can fail and include some comments about fracton phases. 

In Section \ref{open} we discuss questions for future work.


We hope this paper will inspire the discovery of many curious quantum crystals.

\section{The topological limit of a crystalline topological phase}
\label{sec_toplimit}

\begin{figure*}
    \subfloat[][Smooth state]{\includegraphics[scale=0.5]{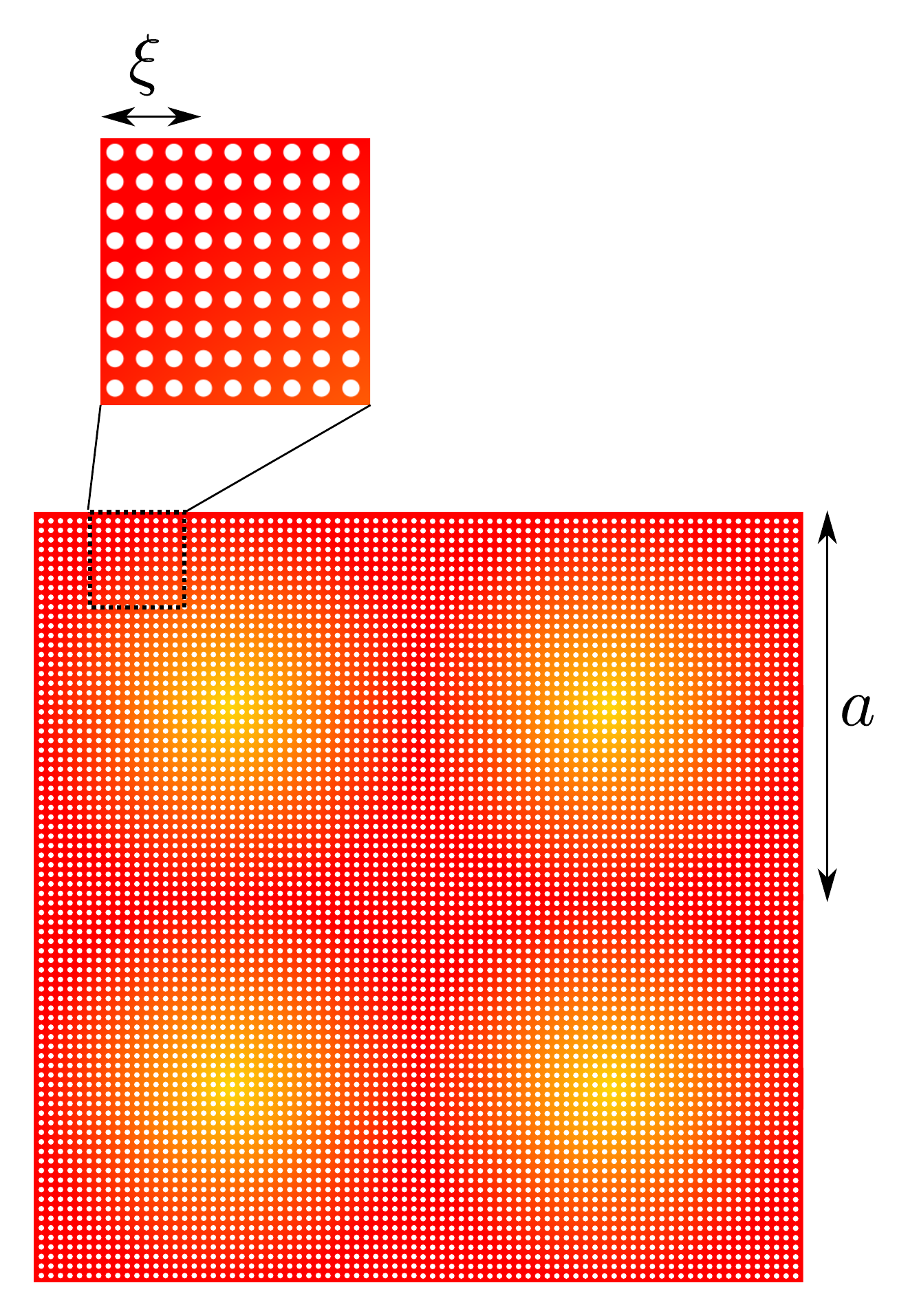}}
    \subfloat[Topological limit]{\includegraphics[scale=0.5]{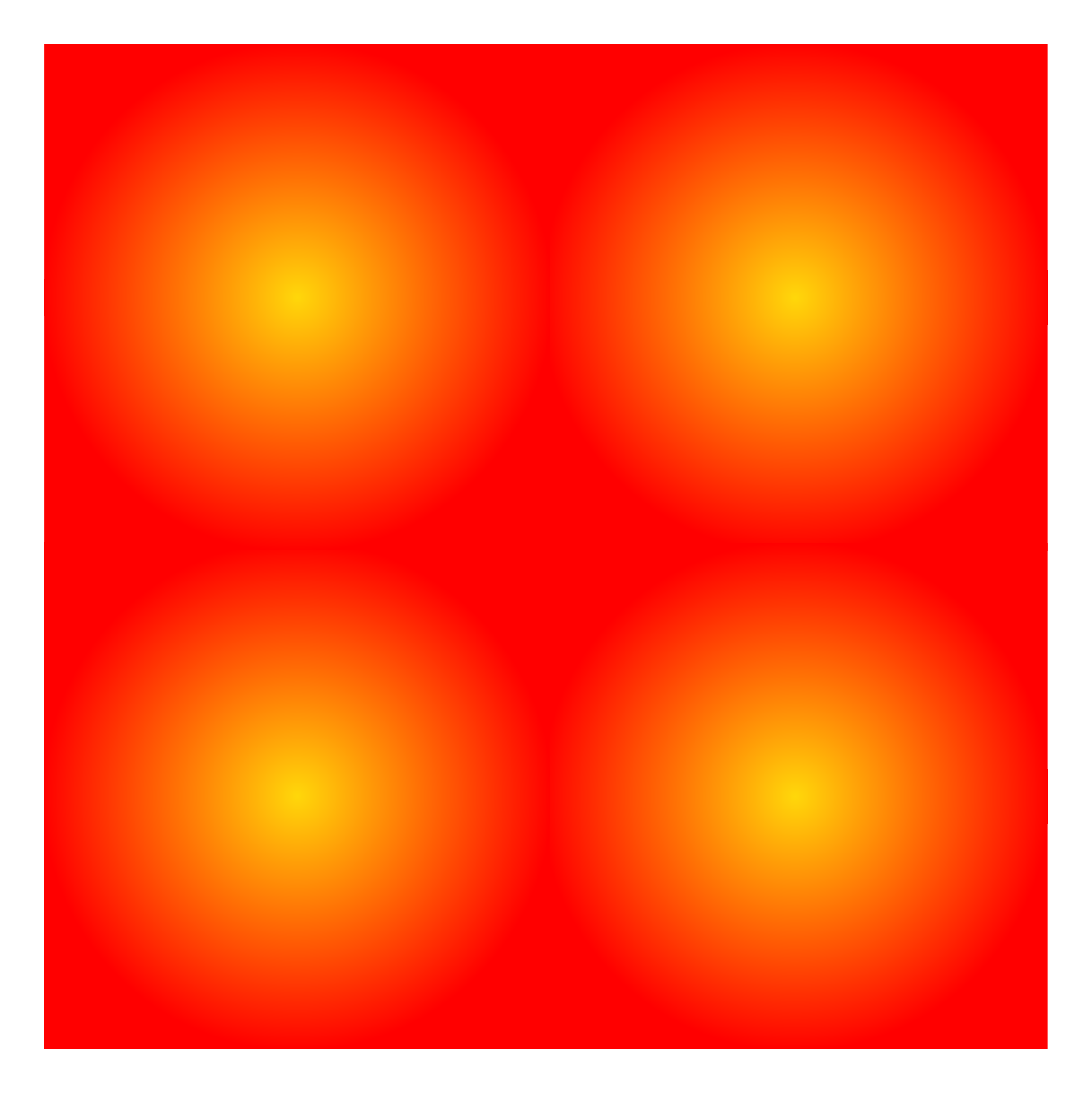}}

    \caption{(a) In a smooth state, the lattice spacing and the correlation length $\xi$ are much less than the unit cell size $a$ and the radius of spatial variation. (b) The topological response of a crystalline topological liquid is captured by a spatially-dependent TQFT that captures the spatial dependence within each unit cell but ``forgets'' about the lattice.}
    \label{fig:my_label}
\end{figure*}
In this section, we will briefly outline the arguments based on topological quantum field theory (TQFT) which lead to the Crystalline Equivalence Principle. The mathematical details are left to Section \ref{sec_space_dep_tqft}. The underlying physical concept is that of a \emph{smooth state}. A smooth state is a ground state of a lattice Hamiltonian that is defined on a lattice which is \emph{much} finer than the unit cell with respect to the translation symmetry, such that the lattice spacing $l$ and the correlation length $\xi$ are much smaller than the minimum radius $R$ of spatial variation within the unit cell. The condition $\xi, l \ll a$ (where $a$ is the unit cell size) was discussed as an assumption for classifying crystalline phases in Ref.~\cite{Huang2017}; our ``smooth state'' assumption is slightly stronger since we require $\xi,l \ll R$. This implies the condition of Ref.~\cite{Huang2017} since $R < a$, but the converse need not be true if there are regions in the unit cell where spatial variation happens rapidly (so that $R \ll a$).

A smooth state might not seem like the kind of system one would normally consider; a physical example would be a graphene heterostructure in which a lattice mismatch between two layers results in a Moire pattern with very large unit cell \cite{Hunt2013}.
Nevertheless, it is reasonable to expect that the \emph{classification} of smooth states would be the same as the classification of states in general. We will leave a rigorous proof for future work; presently, we merely state it as a conjecture and examine the consequences.

A very important property of a smooth state is that it can be coarse-grained while preserving the spatial symmetries. This is allowed only so long as the coarse-grained lattice is still small compared to the unit cell size, but given the assumption $\xi \ll a$ this still allows us to reach a ``topological limit'', by which we mean that $\xi$ becomes much smaller than the coarse-grained lattice spacing. Importantly, since the RG can take place in the neighborhood of any given point in the unit cell, the effective field theory that we obtain in this topological limit will still be spatially-dependent. (For this reason, we will avoid referring to the topological limit as an ``IR limit'', which would be misleading since the unit cell size -- but not the lattice spacing! -- is still an important length scale).

We expect that this topological limit will, as in the case of systems without spatial symmetries, be described by a topological quantum field theory (TQFT). In fact, given the afore-mentioned spatial dependence, it  should be described by a \emph{spatially-dependent TQFT}. We give the precise mathematical definition of this concept in Section \ref{sec_space_dep_tqft}.

Hence, we can define
\begin{defin}
A crystalline topological liquid is a phase of matter that is characterized by a spatially-dependent TQFT acted upon by spatial symmetries.
\end{defin}
We expect that this class of systems is quite large. Certainly, it includes ordinary topological liquids (which, by definition, have no explicit spatial symmetries and can be characterized by a spatially-\emph{constant} TQFT). Moreover, spatially-dependent TQFTs can capture a wide range of other topological crystalline phenomena, as we shall see.

In Section \ref{sec_space_dep_tqft}, we sketch a proof that on contractible spaces, spatially-dependent TQFTs with spatial symmetries are in one to one correspondence with spatially \emph{constant} TQFTs with internal symmetries. Since the latter are expected to characterize topological phases with internal symmetries, the Crystalline Equivalence Principle follows. In the following sections, we we will discuss how to understand this result in more concrete ways without resorting to the highly abstract formalism of TQFTs.

\section{Crystalline gauge Fields}
\label{sec_core_concepts}
\subsection{Gauge fluxes and crystal defects}
\label{sec_spatial_gauge_fields}
In order to understand crystalline topological phases, we want to study what it might mean to couple to a background gauge field for a symmetry group $G$ involving some transformation of space itself. More generally, we believe a framework exists where one can also consider symmetries that transform space-\emph{time}. However, for simplicity and to maintain contact with Hamiltonian models we will focus on purely spatial symmetries. We call our object of study the crystalline gauge field.

A special case of a background gauge field is an isolated \emph{gauge flux}. Isolated gauge fluxes are familiar objects for \emph{internal} symmetries. They are objects in space of codimension 2 (i.e. points in 2-D, curves in 3-D) which are labelled by a group element $g \in G$, and a particle moving all the way around one is acted upon by $g$. Actually, for a non-Abelian group only the conjugacy class of $g$ is gauge-invariant.

\begin{figure}
    \centering
\includegraphics[width=7cm]{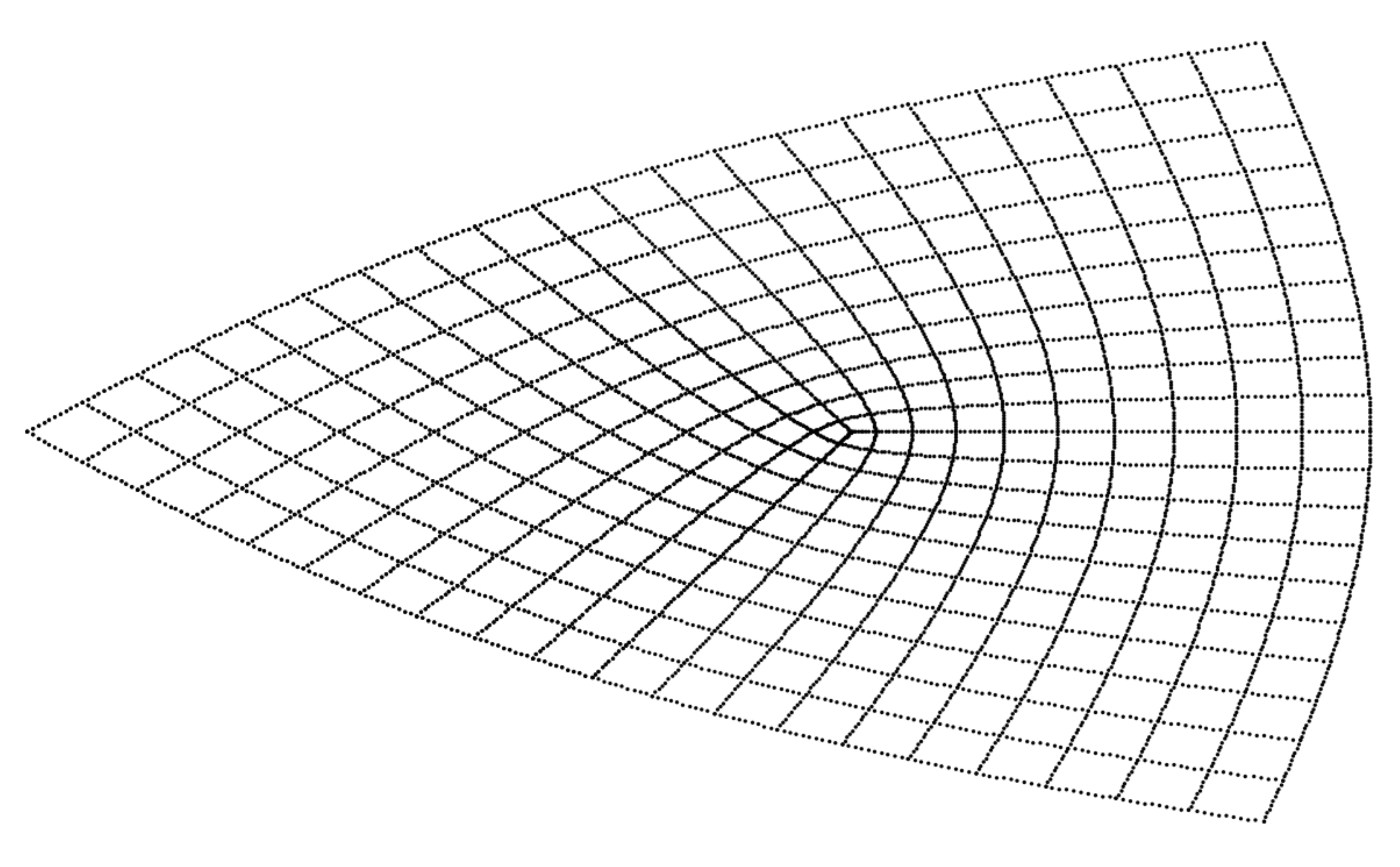}
\includegraphics[width=8cm]{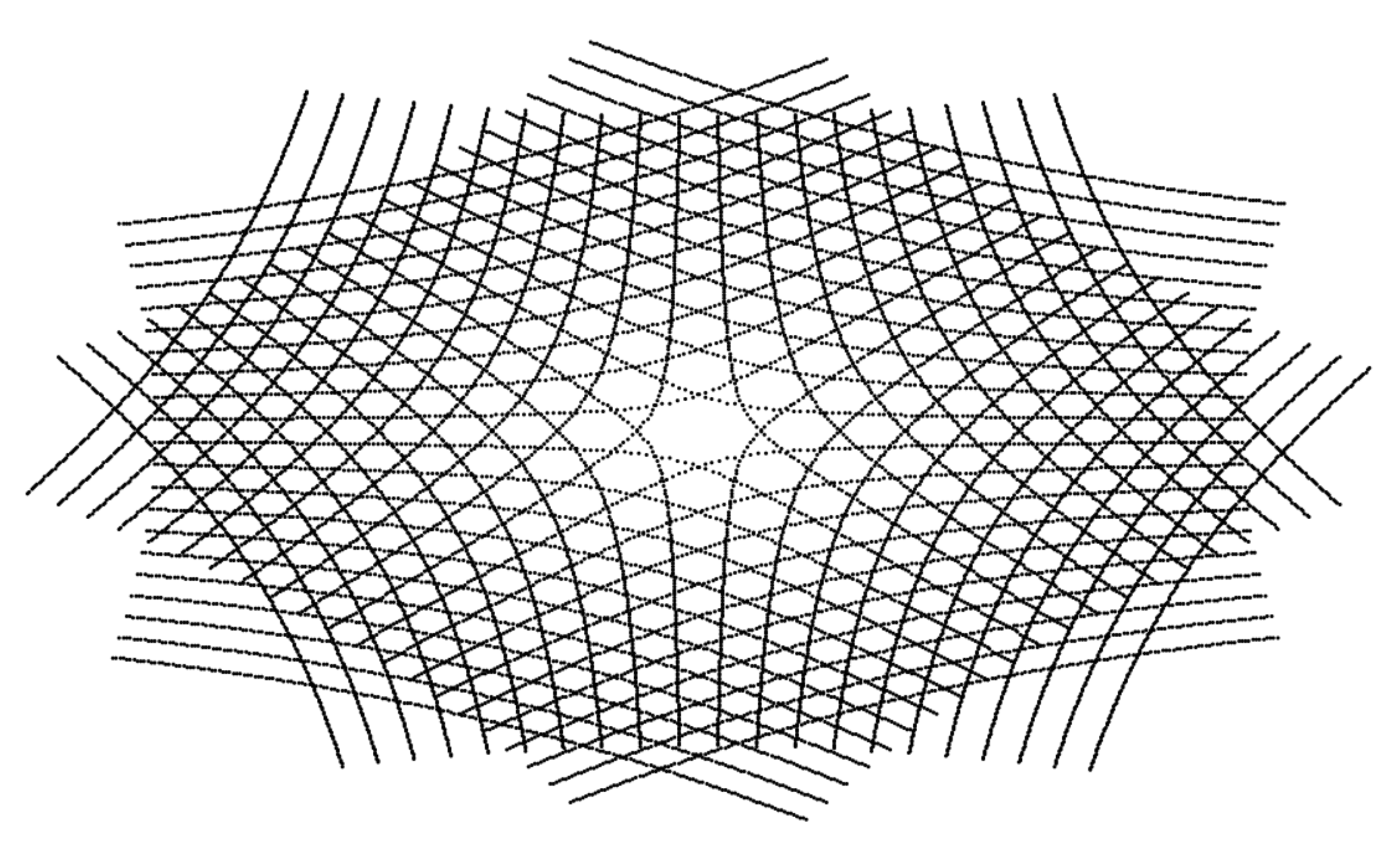}
    \caption{An angular defect of 90 degrees in a vertex-centered square lattice and an angular excess of 120 degrees in a face-centered kagome lattice.}
    \label{fig:deformeddefects}
\end{figure}

Gauge fluxes for spatial symmetries are also labelled by conjugacy classes of $G$. They are also well-known, but not under that name; they are more commonly referred to as \emph{crystal defects}. For example, a gauge flux for translational symmetry is a \emph{dislocation} and a gauge flux for a rotational symmetry is a \emph{disclination} (Fig \ref{fig:deformeddefects}). In 3d, the direction of dislocation does not have to be in the plane perpendicular to the defect, as in a screw dislocation. A defect for reflection symmetry is like the M\"obius band (a cross cap). For a glide reflection we also insert a shift in the lattice as we go around the band. We will see how this zoo of  defect configurations is tamed by topology.

\begin{figure}
    \centering
    \includegraphics[scale=0.35]{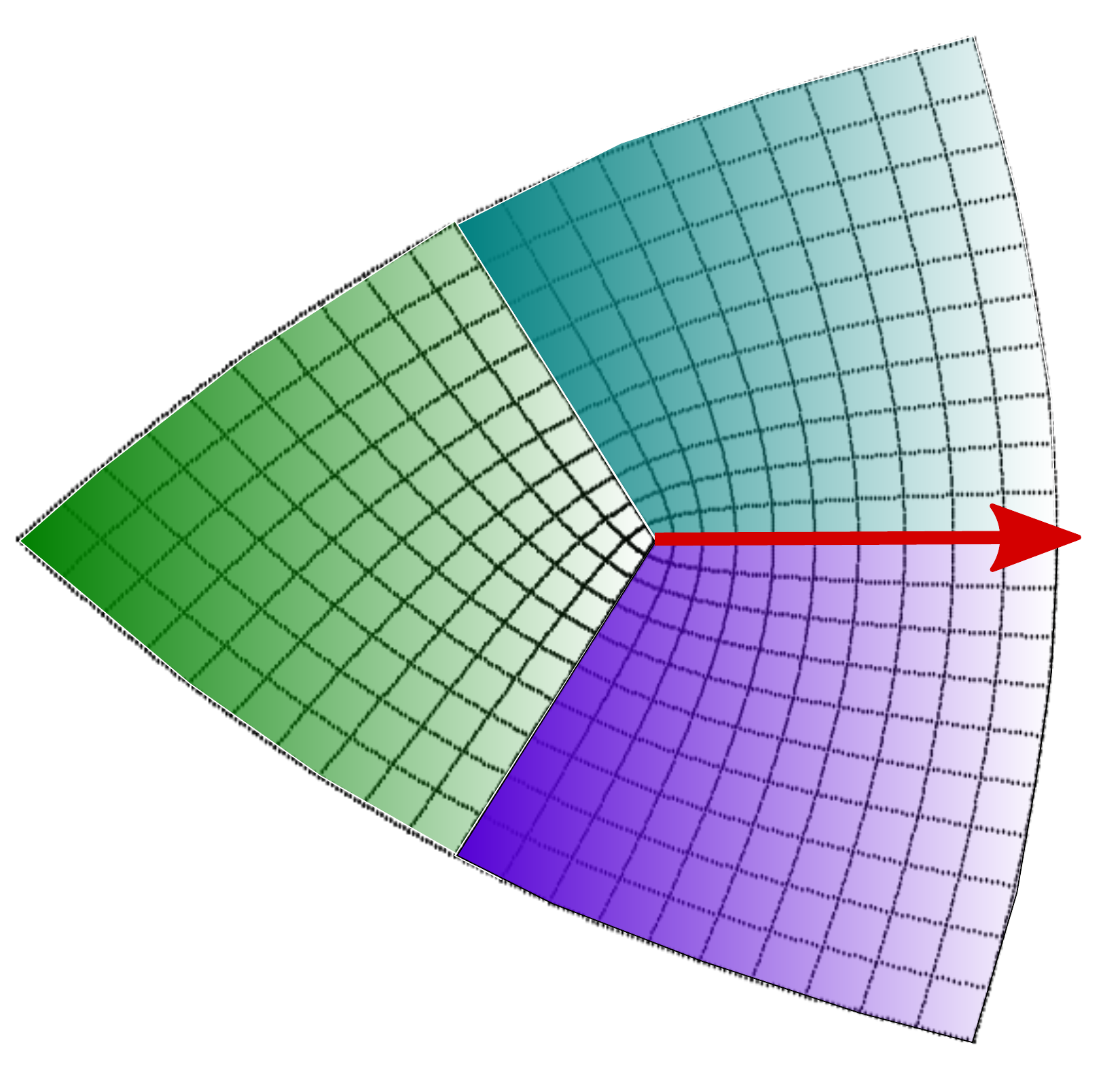}
    \includegraphics[scale=0.35]{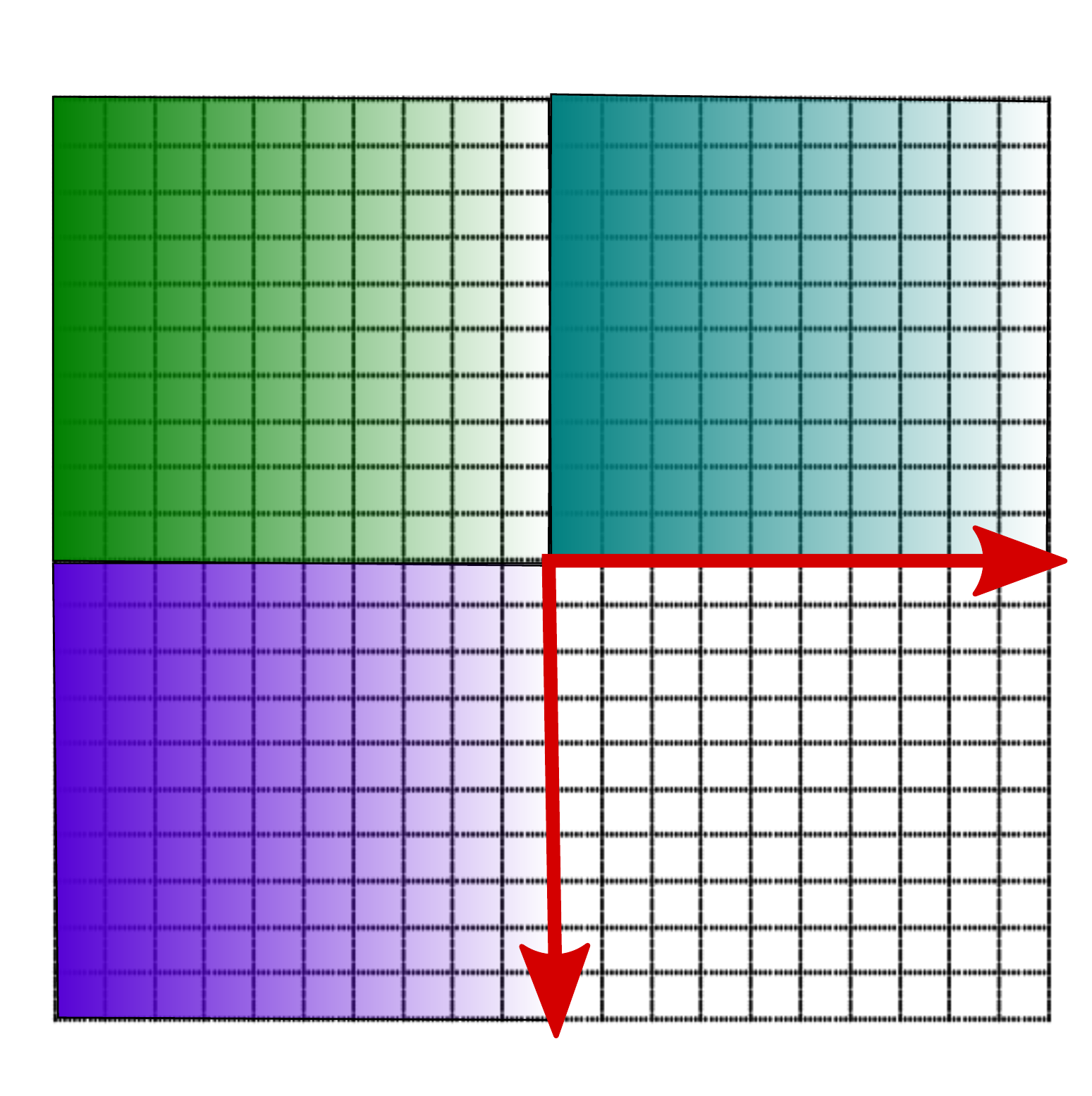}
    \caption{A 90 degree disclination maps discontinuously to the square lattice, as indicated with the colored quadrants. The red line is the branch cut across which the image rotates by 90 degrees. Because the discontinuity is by a rotation in $G$, this map descends to a continuous map from the disclination to the quotient of the square lattice by $G$.}
    \label{fig:Gtwistedmap1}
\end{figure}

Generalizing these examples, we can give a systematic definition of crystalline gauge flux, and more generally of a crystalline gauge field. For motivation, one can look again at Fig \ref{fig:deformeddefects}. The original lattice $\Lambda$ is a regular square or kagome lattice. The crucial property the defect lattice $\Sigma$ is that away from the singular point in the middle, it looks \emph{locally} the same as $\Lambda$, meaning that in a neighborhood of every face except the central one there is an invertible map sending $\Sigma$ to $\Lambda$. However, there is no \emph{global} map sending $\Sigma$ to $\Lambda$. Indeed, if we try to extend the domain of our map, we will eventually create a discontinuity after encircling the singularity. This is shown in Fig \ref{fig:Gtwistedmap1}. For the 90 degree angular defect, the discontinuity is a branch cut such that the limits on either side are related by a 90 degree rotation. For a crystal defect, this discontinuity is always by a $G$ transformation and labels the symmetry flux of the defect.

\begin{figure}
\subfloat[][]{\includegraphics[scale=0.5]{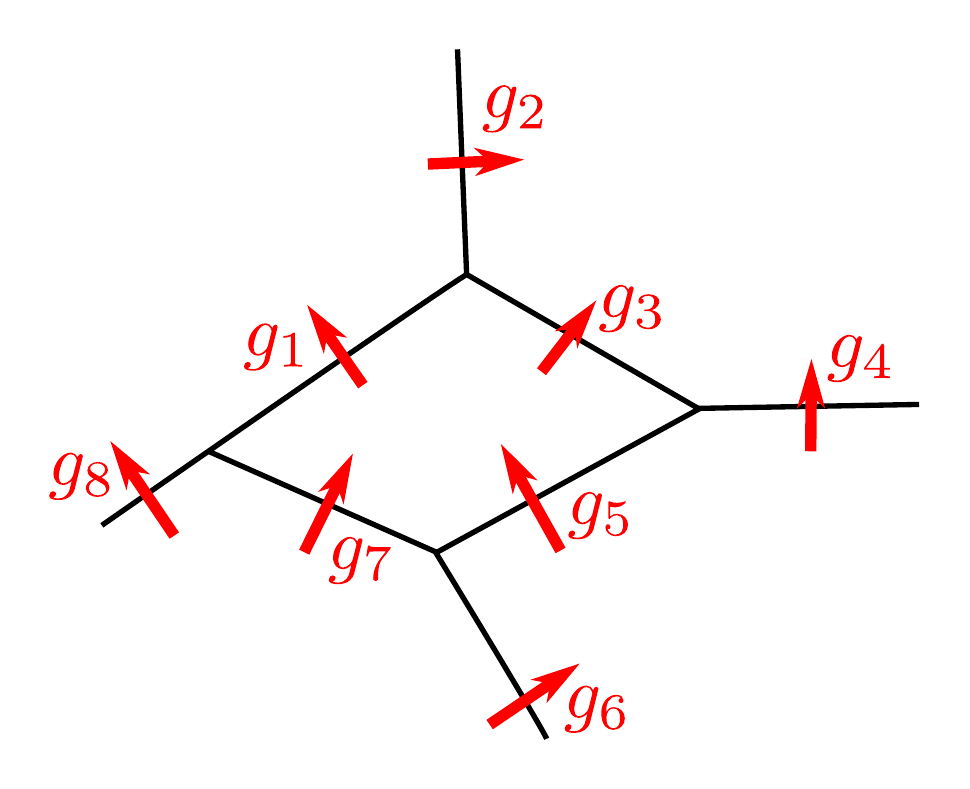}}
\subfloat[][]{\includegraphics[scale=0.5]{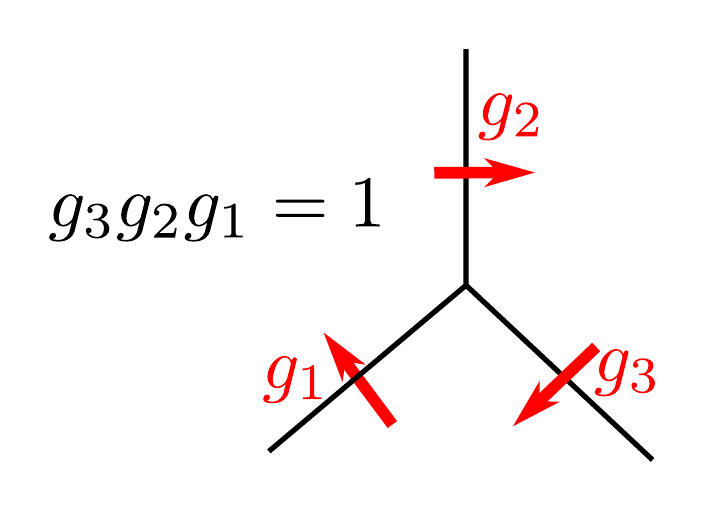}}

\subfloat[][]{\includegraphics[scale=0.5]{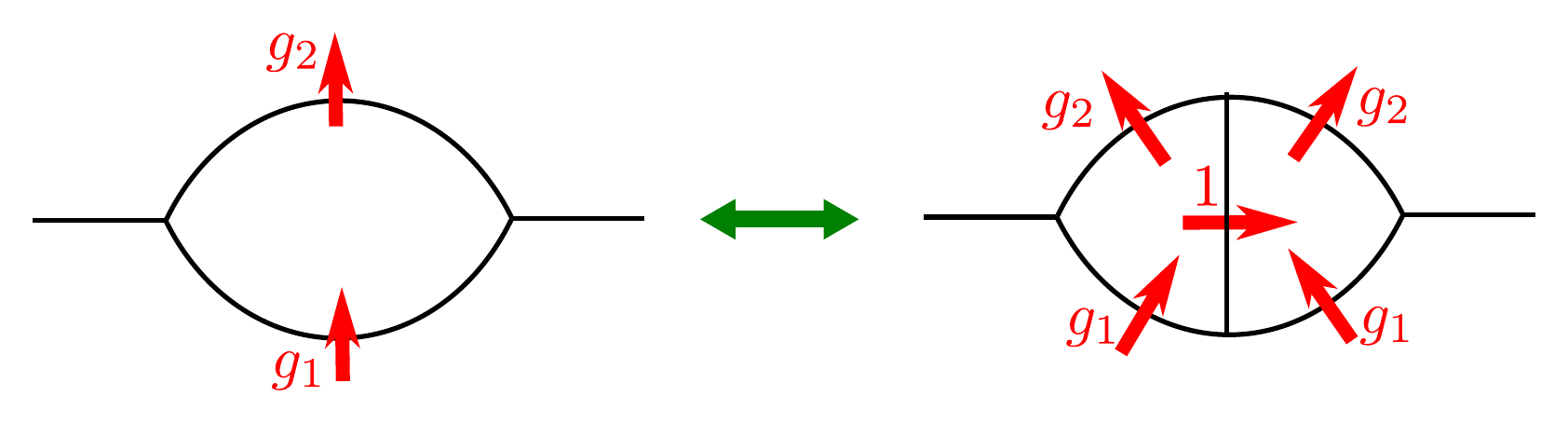}}

\subfloat[][]{\includegraphics[scale=0.5]{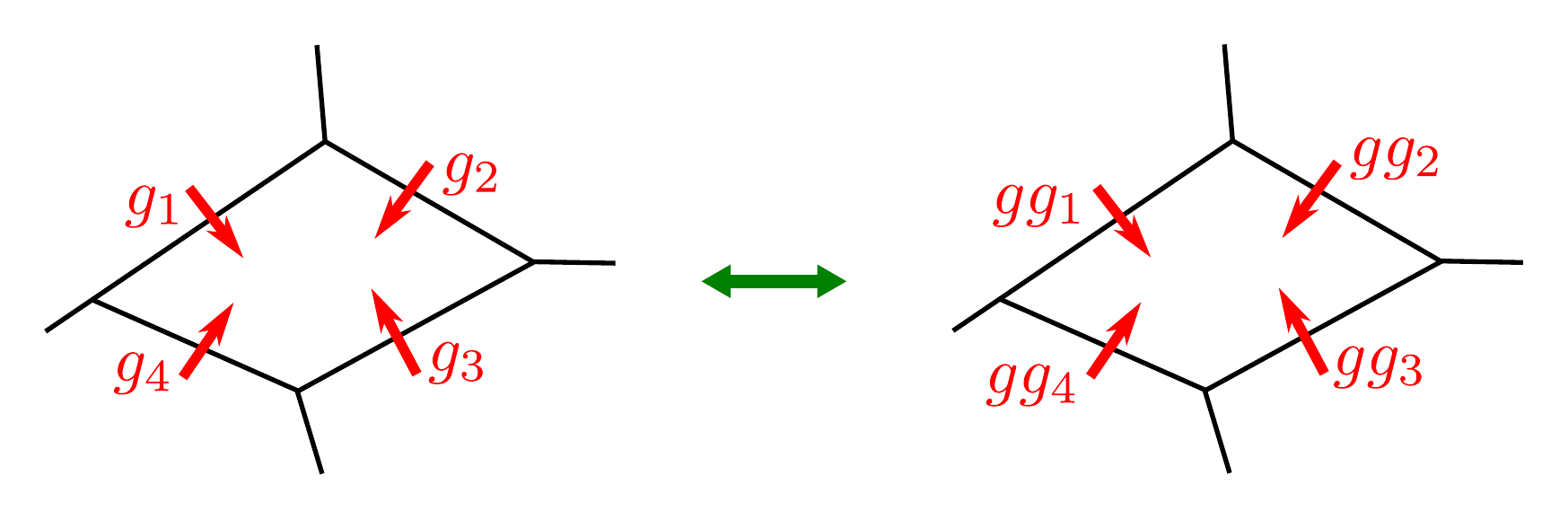}}
\caption{\label{fig:patches}The ``patches'' picture of a gauge field for an internal symmetry. (a): The manifold $M$ is divided up into patches, and the boundaries between patches are twisted by a group element $g \in G$. (b): The flatness constraint implies that the holonomy around a vertex must be trivial. (c) and (d): We identify configurations that differ by dividing patches or by acting on a patch with some $g \in G$.}
\end{figure}

To further motivate the definition, let us recall the definition of a gauge field for an \emph{internal} (discrete) symmetry. Gauge fields for discrete symmetries are somewhat more esoteric than gauge fields for continuous groups (like the familiar electromagnetic vector potential $A_\mu$). One way to think about them is that they encode ``twisted boundary conditions''. For example, threading a non-trivial gauge flux for an Ising symmetry through a system living on a circle means that we make a cut and identify spin-up on one side of the cut with spin-down on the other side of the cut (``anti-periodic boundary conditions''). In general, to specify a gauge field on a manifold $M$ we can build $M$ up out of ``patches''. The boundaries between patches (``domain walls") are ``twisted'' by an element $g \in G$ of the symmetry group (``transition functions"), which tells us how to identify the patches. A discrete gauge field must be ``flat'', which is to say there can be no non-trivial holonomy around a vertex where several patches intersect, as shown in Figure \ref{fig:patches}. This is to say there is no $G$-flux through the vertices (or along such line-like junctions in a 3d picture). There is some inherent gauge freedom: firstly, we can merge or split patches, provided that the boundaries thus created or destroyed are twisted by the trivial element $1 \in G$; secondly, we can apply an element $g_p \in G$ of the symmetry group to a given patch $p$, which has the effect of multiplying the twist carried by the boundaries of this patch by $g_p$. This gauge freedom relates two different representations of the same gauge field. More abstractly (but equivalently), we can define a gauge field as a principal $G$-bundle over $M$ \cite{nakahara}.

As an example, we can consider a $g$-flux at the origin of the plane. This $g$-flux is defined as a $G$ gauge field on the plane minus the origin. It may be defined using a single (simply-connected) patch which meets itself along a domain wall extending from the origin to infinity. This domain wall is labeled with the transition function $g$, indicating that a point charge taking along a path encircling the origin will return to its original position with any internal degrees of freedom transformed by the symmetry $g$. The similarity between the internal symmetry flux and the crystal defect is striking. It leads us to identify the role of the branch cut in the latter with the domain wall of the former. 

With this identification in hand, we are ready to state our definition of crystalline gauge field, by directly generalizing the patches picture of internal symmetry gauge fields. An important novelty will be that the lattice geometry is defined by the crystalline gauge background. That is, we fix our physical space $X$ containing the lattice $\Lambda$. $X$ is usually $\bR^d$, a torus, or some related spacetime. $G$ acts on $X$ preserving $\Lambda$. The lattice with defects $\Sigma$ will be embedded in a different space $M$. For example, in the disclination, $M$ is the plane minus the origin.
 
 To specify a crystalline gauge field, we will start with the same data we had before: a collection of patches $U_i$ dividing $M = \bigcup_i U_i$, with domain walls between intersecting patches $U_i \cap U_j \neq 0$ labelled by elements $g_{ij} = g_{ji}^{-1} \in G$, with the flatness condition $\prod_{i} g_{i,i+1} = 1$ imposed over all contractible loops. This is the definition of an internal symmetry $G$ gauge field, but it is not the end of the story, because as we saw in the examples above, there is an extra feature of crystalline gauge fluxes which needs to be captured: a \emph{map} $f : M \to X$. This represents the (continuum limit) of the identification between the lattices $\Sigma$ embedded in $M$ and $\Lambda$ embedded in $X$. Inside each patch $U_i$, this map $f:U_i \to X$ is continuous, but on the boundaries between intersecting patches $U_i \cap U_j \neq 0$ we impose the twisted continuity condition that for any $m \in U_i \cap U_j$, the limit of $f(m')$ as $m' \to m$ in $U_i$ and the limit of $f(m')$ as $m' \to m$ in $U_j$ are related in $X$ from the former to the latter by application of $g_{ij}$. For example, in Figure \ref{fig:Gtwistedmap1}, the different colored quadrants are patches on $M$ (which in this case is the punctured plane $\mathbb{R}^2\setminus\{0\})$, and the thick red line denotes a boundary between patches which is twisted by a $90$ degree clockwise rotation as we pass from the teal patch to the violet patch. We impose the same gauge freedom as before [Figure \ref{fig:patches}(c) and \ref{fig:patches}(d)], except that when we act on a patch by $g$, as shown in Figure \ref{fig:patches}(d), then inside the patch we replace the function $f$ according to $f(m) \to g f(m)$.
 
 There is a final condition we need to impose, related to the orientation (or lack thereof) of the manifold $M$. It is standard lore that a topological phase that is not reflection invariant cannot be put on an unorientable manifold, and moreover, that for a reflection invariant system, putting it on a unorientable manifold is essentially threading a ``flux'' of the reflection symmetry. So in order to enforce compatibility with these notions, we define $\mu(g) = -1$ if $g$ acts in an orientation-reversing way on $X$, and $\mu(g) = 1$ otherwise. For any closed loop $\gamma$ in $M$, we can define the ``flux'' $g_\gamma$, which is the product of the twist over each boundary crossed by $\gamma$. We also define $\lambda(\gamma) = \pm 1$ depending on whether going around the loop $\gamma$ would reverse the orientation on $M$. We require that $\lambda(\gamma) = \mu(g_\gamma)$.
 
For completeness, we will also formulate a more abstract mathematical definition. Basically we are specifying some extra data on top of a principal $G$-bundle. Formally, we have
 \begin{defin}
\label{def:crystalline}
A \emph{crystalline gauge field} is a pair $(\pi,\hat{f})$, where $\pi : P \to M$ is a principal $G$ bundle, and $\hat{f} : P \to X$ is a continuous map satisfying satisfying $\hat{f}(gp) = g \hat{f}(p)$ for all $p \in P$, $g \in G$. We require that the homomorphism $\mu : G \to \mathbb{Z}_2$ (where $\mu(g) = -1$ if $g$ has orientation-reversing action on $X$) reduces $\pi$ to the orientation bundle of $M$.
We say that two pairs $(\pi,\hat{f})$, $(\pi', \hat{f}')$ represent the same crystalline gauge field if the principal $G$-bundles $\pi : P \to M$ and $\pi' : P' \to M$ are isomorphic by a map $\sigma : P \to P'$ such that $\hat{f}' \circ \sigma = \hat{f}$.
\end{defin}
The map $\hat{f}$ in the definition above always induces a map $g$ from $P/G = M$ into $X/G$. Hence, we have the following commutative diagram:
\begin{equation}\label{diagram}
\begin{tikzcd}
P \arrow{r}{\hat f} \arrow{d}{\pi} & X \arrow{d}{\mod G} \\
M \arrow{r}{g} & X/G \\
\end{tikzcd}.
\end{equation}

It should be clear, from the disclination example, that crystalline gauge fields can describe the crystal defects which were our original motivation. However, now that have given a general definition, we had better ask whether \emph{all} crystalline gauge fields admit such a physical interpretation. In particular, there ought to be a well-defined sense of what it means to \emph{couple} to a general crystalline gauge field.

For internal symmetries it is familiar how to couple to a gauge field, at least when that gauge field lives on $M = X$. Given a gauge field $A$ for a (discrete) internal symmetry $G$, described using patches and transition functions, and given a Hamiltonian $H$ that commutes with the symmetry, we can define a Hamiltonian $H[A]$ that describes the system coupled to the gauge field. To do this, we assume that $H$ can be written as a sum of local terms. Then, $H[A]$ contains a local term for each local term in $H$. The terms in $H$ which act only within a patch carry over to $H[A]$ without change, while for terms in $H$ which act in multiple patches, we must first perform a gauge transformation so that the term acts in a single patch, add it to the Hamiltonian, and then reverse that gauge transformation. See, for example \cite{BBCW}.

Now suppose that we want to do the same thing for crystalline gauge fields. For crystal defects (for example, the disclination in Figure \ref{fig:Gtwistedmap1}) it should be clear how to do this; locally, the defect lattice looks the same as the original lattice, so we just pull local terms in $X$ back into $M$. On the other hand, this construction doesn't necessarily work for a general crystalline gauge field. We have to impose a condition which we call \emph{rigidity}.

\begin{defin}
A crystalline gauge field (expressed in terms of patches, twisted boundary conditions, and a map $f : M \to X$) is \emph{rigid} if near any point $m \in M$ that maps into a lattice point in $X$ under $f$, there exists a local neighborhood $U$ containing $m$ such that, after making a gauge transformation such that $U$ is contained in a single patch, $f$ is injective (one-to-one) when restricted to $U$; and, moreover, the image of $U$ under $f$ contains all lattice points that are coupled to $f(m)$ by a term in the Hamiltonian.\footnote{For certain applications, this last condition may be relaxed near a boundary of $M$. Terms in the Hamiltonian which fall of the edge may need to be discarded or modified in some arbitrary manner.}
\end{defin}

This somewhat technical definition is best understood by considering examples of crystalline gauge fields which are \emph{not} rigid. An extreme example is the case where $f : M \to X$ is the constant function: there is some $x_* \in X$ such that $f(m) = x_*$ for all $m \in M$. In other words, every point in $M$ gets identified with a single point in $X$. If the Hamiltonian in $X$ has terms coupling $x_*$ with some other nearby point, then there is no way to define corresponding terms acting in $M$, since the nearby point does not correspond to any point in $M$. More generally, rigidity fails when there are points at which $f$ is not locally invertible; if $f$ is a smooth map between manifolds, this is equivalent to saying that there are points at which its Jacobian vanishes.

For a rigid crystalline gauge field, on the other hand, there is always a well-defined procedure to couple it to the Hamiltonian. The idea is that rigidity guarantees that the local neighborhood is always sufficiently well-behaved that it makes sense to pull terms in the Hamiltonian from $X$ back into $M$. This is illustrated in Appendix \ref{appendix:gaugecoupling}

Finally, let us remark on a interesting property of the the definition of crystalline gauge field: in the case that the whole symmetry group acts internally (that is, the action of $G$ on $X$ is trivial), we might have expected the definition to reduce to the usual notion of a gauge field for an internal symmetry. However, this is evidently not the case, because there is still the map $f : M \to X$ (which in this case must be globally continuous). We believe that, in fact, this may be a more complete formulation of a gauge field for an internal symmetry.

\subsection{Crystalline topological liquids}
\label{sec_tcl}
From the discussion in the preceding discussion, it might seem that we should only consider \emph{rigid} crystalline gauge fields. Now, however, we want to argue that this is too restrictive. One indeed should require a crystalline gauge field $A$ to be rigid if one wants to go from a Hamiltonian $H$ to a Hamiltonian $H[A]$ coupled to $A$. But such a microscopic lattice Hamiltonian is a property of the system \emph{in the ultra-violet (UV)}. On the other hand, when classifying topological phases, what we actually care about is the \emph{low-energy limit}. The central conjecture of this work is that it is well-defined to discuss the low-energy topological response to \emph{any} crystalline gauge field (not just a rigid one). 

One reason for this is that a spatially-dependent TQFT that is invariant under a spatial symmetry can be expressed as a single TQFT coupled to a background field which is precisely our crystalline gauge background of Def \ref{def:crystalline} (with no rigidiy constraints)! This should be compared with the result for internal $G$ symmetry which says that a $G$ action on a (single) TQFT is equivalent to a TQFT with an ordinary background $G$ gauge field. In other words, topological field theories can be gauged and the resulting topological gauge theory retains all the information of the original theory and its symmetry action\cite{Drinfeldetal}\footnote{In the mathematics literature, this is often stated ``equivariantization is an equivalence".}. We discuss this further in section \ref{sec_space_dep_tqft}.

Such considerations provide the mathematical basis for our conjecture about the gauge response. Nevertheless, since these arguments are very abstract and potentially unappealing to readers not familiar with TQFTs, we will also give a more concrete prescription for coupling smooth states (recall that we introduced this concept in Section \ref{sec_toplimit}) to a general crystalline gauge field. For simplicity, we will only consider the case where there are no orientation-reversing symmetries, although we expect that this restriction can be lifted.

The idea is that there is a simple set of data which one can use to specify a smooth state. Firstly, in the neighborhood of every point in space, we need to specify the orientation of the fine lattice; this can be specified through a \emph{framing} of the manifold $M$ (i.e. a continuous choice of basis for the tangent space at every point). Moreover, in the neighborhood of every point in space, the state looks like it respects the (orientation-preserving) spatial symmetries of the fine lattice (globally, of course, this is not the case). Hence, there is a map $\psi : M \to \Omega$, where $\Omega$ is the space of all ground states invariant under the spatial symmetries of the fine lattice.
(For our arguments, it won't be important to characterize $\Omega$ precisely). For a smooth state, we require this map to be continuous.

As a warm-up, we will first show how to define coupling to a gauge field for an \emph{internal} discrete unitary symmetry $G$ in terms of smooth states. Let $\Omega$ be a space of ground states, with $G$ acting on $\Omega$ as a tensor product over every site, with the action at a given site described by the representation $u(g)$. Let $\psi \in \Omega$ be a $G$-invariant state. Now, given a framed manifold $M$ and a $G$ gauge field $A$ (i.e. collection of patches on $M$ with $G$-twisted boundary condition; alternatively, a principal $G$-bundle over $M$), we will show how to define a smooth state $\psi[A] : M \to \Omega$. For each $g$ we define a continuous path $u(g;t), t \in [0,1]$ such that $u(g;0) = \mathbb{I}$ and $u(g;1) = u(g)$. Given that $\psi$ is $G$-invariant, acting with $[u(g;t)]^{\otimes N}$ on $\psi$ defines a \emph{loop} $\psi_g(t) \in \Omega$, such that $\psi_g(0) = \psi_g(1) = \psi$.
Then, inside each patch we just set $\psi[A](m) = \psi$. But we decorate patch boundaries twisted by a group element $g \in G$ by the corresponding loop. That is, we require that, as $m$ crosses such a boundary, $\psi[A](m)$ goes through the loop described by $\psi_g(m;t)$. One might wonder whether this procedure is well-defined at the intersections between patch boundaries. For example, an obstruction would occur if the composition of the paths $\psi_{g_1}$, $\psi_{g_2}$ and $\psi_{(g_1 g_2)^{-1}}$ defines a non-contractible loop, i.e.\ a non-trivial element in the fundamental group $\pi_1(\Omega)$. In Appendix \ref{appendix_smooth}, we show that such obstructions can never arise, provided that we sufficiently enlarge the on-site Hilbert space dimension. We also give a more rigorous formulation in terms of the classifying space $BG$.

Now we return to the case of a crystalline gauge field, but by way of simplification we first consider the case where there is no symmetry. Then a crystalline gauge field $A$ on a manifold $M$ is simply a continuous map $f : M \to X$. In general, there is no way to define the Hamiltonian $H[A]$. But for a smooth state $\psi : X \to \Omega$ there is a well-defined way to define a corresponding smooth state $\psi[A] : M \to \Omega$ which describes $\psi$ coupled to $A$. Indeed, we just define $\psi[A](m) = \psi(f(m))$. (To completely specify the state, we also have to choose a framing on $M$). This should be compared with Kitaev's ``weak symmetry breaking" paradigm\cite{Kitaev2006}, where our $\Omega$ plays the role of Kitaev's $Y$.

Finally, we can combine the ideas from the previous two paragraphs to give a prescription for coupling a smooth state to a crystalline gauge field for a symmetry $G$ acting on $X$, living on a manifold $M$. The crystalline gauge field is specified (according to the discussion in Section \ref{sec_spatial_gauge_fields}) by a collection of patches on $M$ with twisted boundaries, and a function $f : M \to X$ respecting the twisted boundary conditions. We assume the symmetry action takes the form $U(g) = S(g) [u(g)]^{\otimes N}$, where $S(g)$ is a unitary operator that simply permutes lattice sites around according to the spatial action, and $[u(g)]^{\otimes N}$ is an on-site action. Then we define a path $u(g;t)$ for $t \in [0,1]$ such that $u(g,0) = \mathbb{I}$, $u(g,1) = u(g)$. By acting with $[u(g;t)]^{\otimes N}$ we obtain a path $\psi_g(x;t)$ in $M$. It's not a loop this time, though; instead $G$-invariance of $\psi$ implies that $\psi_g(x;0) = \psi(x)$, $\psi_g(x;1) = \psi(gx)$. Now we can define the coupled state $\psi[A]$ as follows. Inside each patch, we have $\psi[A](m) = \psi(f(m))$. Then, for patches connected by boundaries twisted by $g \in G$, we connect up the $\psi[A]$ in the respective patches by means of the paths $\psi_g(x;t)$. The previously noted endpoints of these paths are consistent with the fact that $f(m)$ jumps to $g f(m)$ as one crosses the boundary.
Again, we defer the proof that this procedure is well-defined at the intersection of boundaries to Appendix \ref{appendix_smooth}.

At this point, the careful reader might raise an objection. In our statement of the conjecture about coupling to a crystalline gauge field, we did \emph{not} require the manifold $M$ to be framed, only orientable (the orientability condition comes from our stipulation that there are no orientation-reversing symmetries, and from the compatibility condition between the orientation bundle of $M$ and the crystalline gauge field discussed in Section \ref{sec_spatial_gauge_fields} and again in Section \ref{sec_space_dep_tqft}). But so far, our smooth state arguments only showed how to couple to crystalline gauge fields on framed manifolds. There are two questions that still need to be addressed:
\begin{itemize}
    \item Question 1. Does the topological response depend on the choice of framing?
    \item Question 2. Can the topological response be defined on oriented manifolds that do not admit a framing?
\end{itemize}
These questions need to be addressed in any formulation of continuum limit. For bosonic systems we expect that the continuum limit, if it exists, can be defined on any oriented manifold and doesn't depend on any extra structure. For fermionic systems it also can depend on a spin or spin$^c$ structure. There are of course systems which, while gapped, still exhibit some metric or framing dependence in the IR, eg. Witten's famous framing anomaly of Chern-Simons theory \cite{witten}. We will later approach these questions in the TQFT framework of section \ref{sec_space_dep_tqft}. For now let us think about these questions from the perspective of smooth states.

For Question 1, we observe that that changing the framing corresponds to changing the fine lattice, and generally speaking, most topological phases have a ``liquidity'' property that ensures that the ground states on different lattices can be related by local unitaries. Since the states live on different lattices, this requires bringing in and/or removing additional ancilla spins that are not entangled with anything else, as is standard protocol when defining local equivalence of quantum states. Such a liquidity property will be necessary for the crystalline topological liquid condition to be satisfied. There are some notable exceptions, such as fracton phases \cite{Vijay2016}, of which a simple example is a stack of toric codes. We do not expect such fracton phases to be crystalline topological liquids.

As for Question 2, we believe that the answer is probably yes. To illustrate the issues at play, consider the 2-sphere. This is an orientable 2-manifold which does not admit a framing. As a consequence, there is no way to put a regular square lattice on a 2-sphere; there must be at least a singular face which is not a square or a singular vertex which is not 4-valent. So one cannot strictly define a smooth state. But we expect that there are ways to ``patch up'' such singular points so that they don't affect the long-range topological response. For example, the toric code is usually defined on a square lattice, which cannot be placed onto the sphere, but it is easy to put a toric-code-like state on the sphere by allowing a few non-square faces.f

We emphasize that coupling to non-rigid crystalline gauge fields is what allows us to establish the crystalline equivalence principle. For example, for internal symmetries one could consider braiding symmetry fluxes around each other. Does this make sense in the case of, for example, disclination defects? If the disclinations were interpreted strictly as lattice defects this would not be possible, since there is no continuous deformation of a lattice containing two disclinations such that the two disclinations move around each other with the lattice returning to its original configuration. But if we interpret disclination defects as special cases of (generally non-rigid) crystalline gauge fields, then this braiding process is allowed. The physical interpretation is that in the course of the braiding process, additional sites get coupled to, and superfluous sites decoupled from, the system by means of local unitaries (as discussed above in the context of the framing dependence). That is, the lattice geometry changes along the path.

In conclusion, this discussion motivates our terminology of ``crystalline topological liquid'': although such systems are ``crystalline'' in the sense that they have spatial symmetries, they are also ``topological liquids'' in the sense that the lattice is not fixed but can be transformed into other geometries by means of local unitaries (with ancillas). This is also consistent with our picture from Section \ref{sec_toplimit} that the topological response of crystalline topological liquids ``forgets'' about the lattice.


\subsection{The Crystalline Equivalence Principle}
\label{sec_crystalline_equivalence}

Most of the time, we will be interested in topological crystalline phases in Euclidean space $X = \mathbb{R}^d$. Moreover, the topological response should only depend on the deformation class of the crystalline gauge field.
It turns out that for $X = \mathbb{R}^d$ there is a very simple characterization of the collapsible homotopy classes of crystalline gauge fields:

\begin{thm}
\label{G_connection_equivalence}
If $X$ is contractible (e.g. $X = \mathbb{R}^d$), then
the deformation classes of crystalline gauge fields are in one-to-one correspondence with \emph{internal} gauge fields.
\end{thm}
That is, in the ``patches'' formulation of crystalline gauge fields, the deformation classes remember only the twisted boundary conditions and not the function $f : M \to X$. This theorem is a corollary of the more general classification theorem for crystalline gauge fields. See Thm \ref{homotopyquotient}. However, here we remark on an elementary way to see one part of Thm \ref{G_connection_equivalence}: namely, that homotopy classes can only depend on the twisted boundary conditions. (For the moment we will not attempt to prove the other part, namely that \emph{any} configuration of twisted boundary conditions has at least one function $f$ respecting it). Although the proposition holds more generally, for simplicity we consider the case where $X = \mathbb{R}^d$ and where the $G$ action on $X$ is affine linear:
\begin{equation}
\label{affine}
gx = A_g x + b_g,
\end{equation}
where $A_g$ is a $(d\times d)$ matrix and $b_g$ is a length $d$ vector. We then observe that given a patch configuration on $M$ with twisted boundary conditions, and two maps $f_0 : M \to X$ and $f_1 : M \to X$ respecting the same twisted boundary conditions, then there is a continuous interpolation
\begin{equation}
f_s = (1-s) f_0 + sf_1,
\end{equation}
which respects the same twisted boundary conditions all the way along the path.

Thm \ref{G_connection_equivalence} allows us to deduce the most important result of this paper. Thm \ref{G_connection_equivalence} shows that deformation classes of crystalline gauge fields are in one-to-one correspondence with principle $G$-bundles. On the other hand, deformation classes of gauge fields for an \emph{internal} symmetry also correspond to principal $G$-bundles. Topological phases are distinguished by their response to background gauge fields. Therefore we conclude the
\begin{framed}
\textbf{Crystalline Equivalence Principle}: The classification of crystalline topological liquids on a contractible space with spatial symmetry group $G$ is the \emph{same} as the classification of topological phases with \emph{internal} symmetry $G$.
\end{framed}
To be precise, the orientation-reversing symmetries on the spatial side are identified with the anti-unitary symmetries on the internal side.  Further, in fermionic systems, reflections with $R^2 = 1$ correspond to time reversal with $T^2 = (-1)^F$ and vice versa. (These statements are not clear from the above treatment since we haven't discussed gauging anti-unitary symmetries. However, they follow from the general TQFT picture, as discussed in \ref{space_dep_fermions}).

\subsection{Beyond Euclidean space}
\label{sec_beyond}
Before we delve into the details of how to classify crystalline topological liquids by their topological response to gauge fields, we recall that the above considerations refer to topological phases that exist in Euclidean space $\mathbb{R}^d$. In principle one can consider the more exotic problem of classifying topological phases on non-contractible spaces; for example, the $d$-sphere, the $d$-torus, or a Euclidean space with holes\footnote{We emphasize that, in the absence of translation symmetry, it does not make sense to relate a topological phase defined on one compact space to one defined on another space with different topology. That is, the classification can depend on the background space.
}.
The practical relevance of this problem may be a bit obscure, but from a theoretical point of view we find it more enlightening to formulate the problem we are interested in -- Euclidean space -- as a special case of the more general problem. It also illustrates an important conceptual point, because, as we shall see, the Crystalline Equivalence Principle does \emph{not} hold on non-contractible spaces (see, for example, Section \ref{noncontractibleproperties}). Thus, the Crystalline Equivalence Principle is not something that \emph{a priori} had to be true. Rather, it is a consequence of the fact that systems of physical interest live in Euclidean space.

On contractible spaces, we had the classification Theorem \ref{G_connection_equivalence} for crystalline gauge fields. This classification theorem is a special case of the more general result (see Appendix \ref{latticegaugefields} and Theorem \ref{homotopyquotient}) that deformation classes of crystalline gauge fields $M \to X$ are classified by homotopy classes of maps from $M$ into the ``homotopy quotient'' $X//G$, pronounced ``$X$ mod mod $G$". For $X$ contractible, $X//G$ is homotopic to the ``classifying space'' $BG$, so we recover Theorem \ref{G_connection_equivalence} if we invoke the well-known fact that principal $G$-bundles over $M$ are classified by homotopy classes of maps $M \to BG$.


\section{Exactly solvable models}
\label{sec_bootstrap}
It is of course important to show that we can explicitly construct Hamiltonians realizing topological crystalline phases classified in this work. We do this using a ``bootstrap'' construction. This is really a meta-construction, in the sense that it is a prescription for going from a construction for an SPT or SET phase with \emph{internal} symmetry to a construction for a topological crystalline phase. A similar idea was used by one of us to construct phases of matter protected by time-translation symmetry in Ref.~\onlinecite{DN}.

For simplicity we consider the case where the entire symmetry group $G$ acts spatially, i.e.\ the internal subgroup is trivial. We will also consider the case where $G$ does not contain any orientation-reversing transformations, and we work in Euclidean space, $X = \mathbb{R}^d$. First of all, let $\varphi$ be a surjective homomorphism from the symmetry group $G$ to a finite group $G_f$. We use one of many approaches to construct a topological liquid with an \emph{internal} symmetry $G_f$. In most of these approaches, there is no obstacle to construct the Hamiltonian to \emph{also} have a spatial symmetry $G$, which commutes with $G_f$ so that the full symmetry group is $\widetilde{G} = G \times G_f$ (for example, in the case of bosonic SPTs, this can be shown explicitly using the construction of Ref.~\cite{CGLW}, as detailed in Appendix \ref{appendix:CGLW_bootstrap}). We then can imagine deforming Hamiltonian to break the full symmetry group $\widetilde{G}$ down to the \emph{diagonal subgroup}
\begin{equation}
G' = \{ (g, \varphi(g)) \in \widetilde{G} \} \cong G.
\end{equation}
We expect that this model will be in the topological crystalline phase that corresponds to the internal symmetry-protected phase we started with via the crystalline equivalence principle. Indeed, we can do this construction on a lattice with lattice spacing much less than the unit cell size (thus giving a smooth state), and verify that, for the original model (without the $\widetilde{G}$-breaking perturbation), following the prescription given in Section \ref{sec_spatial_gauge_fields} to couple to a crystalline gauge field for the diagonal subgroup $G'$ gives the same result as coupling to an internal gauge field for the internal subgroup $G$. (A similar argument can be given in the spatially-dependent TQFT picture of Section \ref{sec_space_dep_tqft}).

Let us briefly sketch how to extend the above construction to symmetry groups $G$ containing orientation-reversing transformations. A general topological phase is not reflection-invariant, so the above argument needs to be modified. We expect that a topological liquid can always be made invariant under a spatial symmetry $G$ if we make the orientation-reversing elements of $G$ act \emph{anti-unitarily}; we can call this suggestively the ``CPT princple''\footnote{This is related to, but not a consequence of, the CPT \emph{theorem}, because here we are talking about lattice models, not relativistic quantum field theories. The CPT principle doesn't claim that \emph{every} lattice model is CPT invariant, which would be demonstrably false; rather, it posits that in any topological phase there is at least one CPT-invariant point.} We prove this explicitly for bosonic SPT phases in Appendix \ref{appendix:CGLW_bootstrap}. We then proceed as before, starting from a $(G \times G_f)$-symmetric topological phase, where the internal symmetry $\varphi(g) \in G_f$ acts anti-unitarily if $g$ was orientation-reversing. Then eventually the symmetry gets broken down to the diagonal subgroup $G'$, which contains spatial symmetries, possibly orientation-reversing, but all acting unitarily (since the orientation-reversing elements of $G$, which we have taken to act anti-unitarily, get paired with anti-unitary elements of $G_f$).
We expect that this gives the topological crystalline phase corresponding to the original internal symmetry-protected phase via the crystalline equivalence principle, but explicitly determining the topological response would involve explaining what it means to gauge an anti-unitary symmetry, which we will not attempt to do (but see Ref.~\onlinecite{Chen2015b}.)

\section{Topological Response and Classification}\label{sec_classification}

In this section, we will discuss how our understanding of what it means to gauge a spatial symmetry allows us to classify topological phases by their topological responses. Basically, any approach to understanding topological phases with \emph{internal} symmetries which relies on gauging the symmetry, can be applied equally well to space-group symmetries by coupling to crystalline gauge fields. Moreover, in Euclidean space, Theorem \ref{homotopyquotient} should imply that we obtain the same classification as for internal symmetries, in accordance with the Crystalline Equivalence Principle. In non-contractible spaces we may obtain a different classification.

There are two main approaches to thinking about topological response. The first is a bottom-up approach where one starts with a Hamiltonian in a lattice model and one attempts to work out all the topological excitations. For example in 2+1D, one has anyons and symmetry fluxes and one can ask about how they interact. This is tabulated mathematically in a $G$-crossed braided fusion category \cite{ENO,BBCW} and one can try to work out a classification of these objects or at least find some interesting examples and then look for lattice realizations.

The second approach is a top-down one where one first assumes the existence of a low energy and large system size ("IR") limit of the gapped system. This is a topological quantum field theory (TQFT) of some sort and one can just try to guess what it is from the microscopic symmetries, entanglement structure (short-range vs. long-range), and so on. One can make a bold statement that all possible IR limits are of a certain type of TQFT and then try to classify all of those. Despite its obvious lack of rigor, this approach has proven successful.

One reason for this is that it is often possible to bridge the two perspectives. For example, a $G$-SPT can be understood in terms of an effective action $\omega \in H^{D+1}(BG,\bZ)$\cite{DW,CGLW} leading ultimately to a TQFT. But considering the fusion of symmetry fluxes also leads to an element of $\mathcal{H}^D(G, U(1))$ through a higher associator of symmetry fluxes (in $2+1$-D, it is the F symbol). These are equivalent under the isomorphism $H^{D+1}(BG,\bZ) = \mathcal{H}^D(G,U(1))$ (see Appendix \ref{topphases} for more explanation of this isomorphism). In general, defects such as anyons and symmetry fluxes can be described  in the TQFT framework through the language of ``extended TQFT''.

Let us now discuss how these methods can be extended to the case of spatial symmetries.

\subsection{Flux fusion and braiding for SET phases in (2+1)-D with spatial symmetry}\label{sec_flux_fusion}
If we want to classify symmetry-enriched phases in (2+1)-D phases we can consider the ``bottom-up'' approach of Ref.~\onlinecite{BBCW}. There, one has a topological phase with an internal symmetry $G$, and one envisages coupling to a classical background gauge field. In particular, one can consider gauge-field configurations in which the gauge fluxes are localized to a discrete set of points. One can then consider the algebraic structure of braiding and fusion of such gauge fluxes, which is an extension of the braiding and fusion of the intrinsic excitations (anyons) that exist without symmetry. This structure is argued to be described by a mathematical object called a ``G-crossed braided tensor category''. For a crystalline topological liquid on Euclidean space, we expect that the equivalence between crystalline gauge fields and $G$-connections allows the arguments to carry over without significant change. (We will leave a detailed derivation for future work.) On non-contractible spaces, presumably a generalization of the arguments of Ref.~\onlinecite{BBCW} should be possible, but we will not explore this.


\subsection{Topological Response as Effective Action}
\label{sec_topological_actions}

Another way to compute topological response, which does not involve braiding or fusing fluxes is by computing twisted partition functions. That is, given a background gauge field (ordinary or crystalline) $A$ on a spacetime $M$, we can compute the partition function of  $Z(M,A)$ and compare it to the untwisted partition function $Z(M)$. The assumption is that
\[Z(M,A)/Z(M)\]
tends to a complex number of modulus 1 in the limit that $M$ becomes very large compared to the correlation length. In favorable situations, such as a crystalline topological liquid, the limiting phase is a topological invariant of $M$ and its gauge background $A$. We call this the topological response of our system to $A$ and its log the effective action for the gauge background $A$. In some cases, like $M = Y \times S^1$, $Z(M,A)$ can be interpreted as some kind of ``twisted trace'' of symmetry operators, as we soon discuss. In general there is such an interpretation but it involves topology-changing operators \cite{BaezDolan}. \footnote{Indeed, on a general spacetime, a generic choice of time direction defines a Morse function and a foliation of spacetime by spatial slices. At critical points of this Morse function, the spatial slice is singular and we have a topology changing operator that gets us from the Hilbert space just before the critical point to the Hilbert space just after. These are all handle attachments and can be thought of as generalized flux fusion processes.}. What is most important for classification of phases is that it is a number that captures some (or all) of the data in a ``spatially-dependent TQFT'', which we introduce in Section \ref{sec_space_dep_tqft} as the mathematical way to describe a ``crystalline topological liquid'' phase of matter.

For internal symmetries of bosonic systems, we know that in this case, the limiting ratio can be written
\begin{equation}\label{usualtopresp}
Z(M,A)/Z(M) \to \exp \left(2\pi i\int_M \omega(A)\right),
\end{equation}
where $\omega(A)$ is a gauge-invariant top form made out of the gauge field. In the case of a crystalline gauge field $A = (P,M,\pi,\hat f)$, we will also assume that the topological response is an exponentiated integral:
\begin{equation}\label{topresp}
Z(X,A)/Z(X) \to \exp \left(2\pi i \int_M \omega(\alpha,\hat f)\right),
\end{equation}
where $\omega(\alpha, \hat f)$ is a top form on $M$ made of the twisting field $\alpha \in H^1(M,G)$ which classifies the cover $P$ and the map $\hat f$, used to pull back densities from $X$. In the case that $G$ is purely internal, $\alpha$ plays the role of $A$ in \eqref{usualtopresp}.

As discussed in Ref \onlinecite{DW}, responses of the form \eqref{usualtopresp} are the same thing as cocycles in group cohomology, defined as cohomology of the classifying space $H^D(BG,U(1))$\footnote{Actually we should use only measurable cohomology or use different coefficients in a different degree: $H^{D+1}(BG,\bZ)$. We discuss this subtlety in Appendix \ref{topphases}.}, where $D$ is the dimension of spacetime $X$. This reproduces the classification of internal symmetry bosonic SPTs in Ref \onlinecite{CGLW}. To construct the effective action of $A$, we use the fact that the gauge field $A$ itself is the same as a map $A:X \to BG$, and given a $D$-cocycle on $BG$, we can pull it back along this map to get $\omega(A)$ over $X$.

Analogously, we can think of our crystalline gauge field as a map $A:M \to X//G$ (see Appendix \ref{latticegaugefields}) and take any form in $H^D(X//G,U(1))$, pull it back along this map to $M$ to get a $\omega(\alpha,\hat f)$ and integrate it (see Appendix \ref{cocycles}). We just need to be a little careful with coefficients. We intend to integrate $\omega(\alpha)$ over $M$, but if $G$ contains orientation-reversing elements like mirror and glide reflections (or time reversal), then $M$ may likely be unorientable. Integration on an unorientable $M$ is done by choosing a local orientation: orienting $M$ away from some hypersurface $N$ and performing the integration on $M - N$ with its orientation. To ensure the integral does not depend on this local orientation, we need our top form $\omega(\alpha)$ to switch sign with the local orientation is reversed. Mathwise, this means that $\omega(\alpha)$ should live in cohomology $H^D(M,U(1)^{or})$ with twisted coefficients $U(1)^{or}$. Luckily, if $X$ is orientable, then the unorientability of $M$ is entirely due to orientation-reversing elements of $G$, so if we use twisted cohomology $H^D(X//G,U(1)^{or})$ where orientation-reversing elements of $G$ act on $U(1)$ by $\theta \mapsto - \theta$, then the coefficients will pull back properly. This cohomology group is well known in algebraic topology as the equivariant cohomology of $X$, and is written
\[H^D_G(X,U(1)^{or}) := H^D(X//G,U(1)^{or}).\]

Another subtlety comes from considering the identity map $M = X \to X$ as a crystalline gauge field. Any non-trivial topological response to the identity cover is equivalent to a shift of all the partition functions by a phase. We may as well consider only the subgroup of all equivariant cohomology classes which pulled back along the identity map are trivial. This is called reduced cohomology and is denoted with a tilde $\tilde H$.

Summarizing (and recalling the subtlety about replacing $U(1) \to \mathbb{Z}$ increasing the degree by 1, as discussed in Appendix \ref{topphases} ), we find:
\begin{thm}\label{classification}
Homotopy-invariant effective actions in $D=d+1$ spacetime dimensions for crystalline gauge fields $A:M \to X//G$ which may be written as integrals over $M$ are in correspondence with ``twisted reduced equivariant cohomology":
\[\tilde H^{D+1}_G(X,\bZ^{or}) .\]
\end{thm}
In the following section, we will give examples of crystalline SPT states and how to compute the topological response as a class in equivariant cohomology.



Finally, even though these are all the effective actions, from what we've learned in the case with time reversal symmetry\cite{K} and consideration of thermal Hall response, we know these are very unlikely to be all the phases. There are some criteria, like homotopy invariance, that pick out these phases based on their effective action, but we don't know a microscopic characterization of which phases come from group cohomology and which phases are from the beyond. We say ``bosonic" because we have learned the importance of including spin structure in a careful way \cite{KTTW}. We discuss the relationship between topological actions and phases in Appendix \ref{topphases}.

\subsection{Examples of Topological Response}\label{sec_examples}

Let us explain in some examples how the topological response \eqref{topresp} manifests itself physically and how it can be computed starting from an SPT state. These examples were constructed using the techniques in Appendix \ref{appendix:CGLW_bootstrap}.

\subsubsection{Reflection SPT in 1+1D}

We consider a system of spin-$1/2$'s lying along the $x$-axis at integer coordinates $x=j$. We will use the $X$ basis for these spins and consider the state 
\[\cdots |\leftarrow\rangle\otimes |\leftarrow\rangle \otimes \left(|\rightarrow\rangle - |\leftarrow\rangle\right) \otimes |\rightarrow\rangle \otimes |\rightarrow\rangle \cdots.\]
There is no reflection-symmetric perturbation (keeping the gap) which can take that central minus sign to a plus. This is because it can be understood as an odd charge for an \emph{internal} $\bZ_2$ symmetry induced by reflection at the reflection center. This odd charge is the signature of this SPT phase. Let us see how to compute it as a topological response.

Observe that this odd charge can be detected using a trace
\begin{equation}\label{Rtrace}
{\rm charge\ at\ reflection\ center} = \lim_{\beta \to \infty} \Tr\ \mathcal{R}\ e^{-\beta H'} = -1,
\end{equation}
where $H$ is a gapped Hamiltonian with ground state as above and $\mathcal{R}$ is the reflection operator. Traces are computed by path integrals with a periodic time coordinate. The insertion of $\mathcal{R}$ means that as we traverse this periodic time, we come home reflected. This means that the geometry of the spacetime whose path integral computes this trace is a M\"obius strip.

We can represent this geometry as a crystalline gauge field over
\[X = \bR_x \times S^1_t = \{(x,t)|x \in \bR, t \in [0,1], (x,0) = (x,1)\},\]
the usual domain for background gauge fields used to compute twisted traces. We write the M\"obius strip
\[M = \{(m,s) | m \in \bR, s \in [0,1], (m,0) = (-m,1)\}.\]
We get a continuous map
\[
f(m,s) = (|m|,s): M \to X/\mathcal{R},
\]
where
\begin{multline*}
X/\mathcal{R} = \{(\bar x,t)|x \in \bR, t\in [0,1], (\bar x,0)\\  = (\bar x,1), (\bar x,t) = (-\bar x,t)\}.
\end{multline*}
There is no continuous lift of this map to $X$, so we insert a branch cut along $s = 0$ in $M$. This defines a covering space
\[P =\{(m,s') | m \in \bR, s' \in [0,2], (m,0) = (m,2)\}\]
with covering map $\pi:P \to M$ defined by
\[\pi(m,s') = \begin{cases} 
      (m,s') & 0 \leq s' \leq 1 \\
      (-m,s'-1) & 1\leq s'\leq 2
   \end{cases}\]
This has a map $\hat f:P \to X$ defined by
\[\hat f(m,s') = \begin{cases} 
      (m,s') & 0 \leq s' \leq 1 \\
      (m,s'-1) & 1\leq s'\leq 2
   \end{cases}\]
We summarize with a diagram (cf. Defn \eqref{diagram})
\begin{equation}
\begin{tikzcd}
P \arrow{r}{\hat f} \arrow{d}{\pi} & X \arrow{d}{(|x|,t)} \\
M \arrow{r}{f} & X/\mathcal{R} \\
\end{tikzcd}.
\end{equation}

The trace \eqref{Rtrace} is therefore interpreted as a topological response $-1$ to this crystalline gauge field is $-1$ a la \eqref{topresp} and we would like to write it in the form $\int \omega(\alpha)$ for some $\omega \in H^2(B\bZ_2,U(1)^{or})$, where $\alpha$ is our double cover $\pi:P \to M$ interpreted as a $\bZ_2$ gauge field which tells us where the branch cuts are. It turns out there is a unique non-trivial class $\omega(\alpha) = \frac{1}{2} \alpha^2$ and indeed if we compute (with particular boundary conditions)
\[\exp\left( 2\pi i \frac{1}{2} \int_M \alpha^2\right) = -1\]
we reproduce the trace \eqref{Rtrace}.

Before we move on to richer examples, let us make some comments for the mathematically inclined on the evaluation of this integral. To get the claimed answer, we used the one-point-compactification of $M$, where we add a single point at infinity, collapsing the boundary to a point. The one-point-compactification of $M$ is the projective plane $\mathbb{RP}^2$ and the usual integral $\int_{\mathbb{RP}^2}\alpha^2 = 1$. There is, however, no rigid crystalline gauge field over $X$ with $M = \mathbb{RP}^2$ since $X$ is noncompact. However, if we impose $\mathcal{R}$-symmetric, time independent boundary conditions on our crystal, then our cylindrical spacetime $\bR_x \times S^1_t$ gets each end collapsed to a point and becomes a sphere. The reflection group continues to act on this sphere and there is a rigid crystalline gauge field with $M = \mathbb{RP}^2$ over $S^2$.

In computing more complicated examples of this same SPT phase (examples which are not already disentangled), indeed one finds it necessary to choose some boundary conditions in \eqref{Rtrace} to get a nonzero trace. What if we use periodic boundary conditions? In that case, there is always a reflection center at $\infty$, and periodicity implies that the reflection center will also carry an odd charge. Therefore, the trace with periodic boundary conditions will receive a contribution from both reflection centers and be $(-1)^2 = 1$. We can see this with our crystalline gauge fields. Indeed, with periodic boundary conditions spacetime becomes a torus $S^1_x \times S^1_t$ and if we insert a reflection twist in the time direction we obtain $M$ as a Klein bottle. We identify $\alpha$ with the orientation class in $H^1(M,\bZ_2) = \bZ_2 \oplus \bZ_2$ and for this class $\alpha^2 = 0$, as expected! On the other hand, one can use M\"obius bands centered at either reflection center to see the odd charge at each one. Gluing them along their overlap gives us the Klein bottle again and the partition functions multiply\footnote{This is a form of cobordism invariance.}. We will see this in more detail in the following example.

\subsubsection{Reflection and Translation in 1+1D}\label{reftrans}

Next we consider a one-dimensional system with a translation and a reflection symmetry. In particular, the state
\[\bigotimes_{j \in \bZ} (|\leftarrow\rangle + (-1)^j |\rightarrow\rangle).\]
This state is symmetric under reflection around $0$: $\mathcal{R}_0(x) = -x$, and also under reflection around $1$: $\mathcal{R}_1(x) = 2-x$. The product $\mathcal{R}_1 \mathcal{R}_0 = T_2$ is a translation by two units. In the language of the previous example, even sites carry even charges and odd sites carry odd charges. These charges are detected by computing traces
\[\lim_{\beta \to \infty} \Tr\ \mathcal{R}_j\ e^{-\beta H} = (-1)^j.\]
The spacetime geometries of the two traces correspond to two different crystalline gauge backgrounds over $X = \bR_x \times S^1_t$. Depending on whether we twist by $\mathcal{R}_0$ or $\mathcal{R}_1$, our test manifold $M$ is a M\"obius band centered over $x = 0$ or $x = 1$.

Because we have a translation symmetry, we can also consider a trace in periodic space. If the length of the spatial circle is an odd number of unit cells (for a total length $4N+2$), then every reflection symmetry passes through an odd and an even site, so all the traces
\begin{equation}\label{circleRtrace}
\lim_{\beta \to \infty} \Tr_{S^1} \mathcal{R}_j e^{-\beta H} = -1.
\end{equation}
The spacetime geometry $M$ of this trace is a torus with a reflection twist as we go around the time circle, ie. a Klein bottle.

We can describe this trace as topological response to a crystalline gauge background over $X = \bR_x \times S^1_t$. We will need to insert branch cuts along both cycles of the Klein bottle $M$. Along the spatial direction, this because we are trying to map $S^1 \to \bR_x$. If we coordinatize $S^1$ using $y \in [-2N-1,2N+1]$, we can consider the map $f(y) = y$ with a branch cut from $2N+1$ to $-2N-1$ where we translate by $T_2^{2N+1}$. Denoting the compatible twisting for translations by $\tau$, a $\bZ$ gauge field on $M$, we therefore have $\int \tau = 2N+1$ around the spatial cycle. As before, we also have to twist around the time direction by a reflection, say $\mathcal{R}_0$. If we denote by $\alpha_0$ the corresponding compatible twisting, a $\bZ_2$ gauge field on $M$, then we have $\int \alpha_0 = 1$ (mod 2) around the time cycle. Summarizing, we have
\begin{equation}
\begin{tikzcd}
P = {\rm cylinder} \arrow{r}{\hat f} \arrow{d}{\pi} & X = {\rm cylinder} \arrow{d}{(|x| \mod 2,\ t)} \\
M = {\rm Klein\ bottle} \arrow{r}{f} & X/\mathcal{R}_0 \times T_2 \\
\end{tikzcd}.
\end{equation}

We wish to write the trace \eqref{circleRtrace} as a topological response \eqref{topresp} of the form
\[\exp 2\pi i \int_M \omega(\alpha_0,\tau) = -1.\]
There are two choices that work, namely
\[\omega_1(\tau,\alpha_0) = \frac{1}{2} \tau \alpha_0 \qquad \omega_2(\tau,\alpha_0) = \frac{1}{2}(\tau \alpha_0 + \alpha_0^2).\]
On the other hand, because of the $\alpha_0^2$ term in $\omega_2$, the second describes a non-trivial response to a M\"obius band centered over 0. We have argued that the partition function in this background computes the sign of the trace of $\mathcal{R}_0$, which for our SPT state above is positive. Therefore, our state must correspond to the class $\omega_1 \in H^2_G(\bR,U(1)^{or}) = \bZ_2 \oplus \bZ_2$.

Let us show as a consistency check that this cocycle correctly produces the negative trace over $\mathcal{R}_1$. Using the formula $\mathcal{R}_1 = T_2 \mathcal{R}_0$, we see that the M\"obius strip over $1$ has the twisting $\int_t \alpha_0 = \int_t \tau = 1$ (mod 2). Then adding the proper boundary conditions we indeed find
\[\exp \left(2\pi i \frac{1}{2} \int_{M} \tau \alpha_0\right) = -1 = \lim_{\beta \to \infty} \Tr \mathcal{R}_1 e^{-\beta H}.\]
Note that the same caveats about this integral and boundary conditions on the trace we discussed in the previous example apply.

From what we have computed so far, we see that $\omega_2$ corresponds to having odd charges on even sites and even charges on odd sites and $\omega_0 = \frac{1}{2}\alpha_0^2$ corresponds to having an odd charge on every site, the simplest translation-symmetric extension of our state in the previous example.

\subsubsection{Rotation SPT in 2+1D}

Now we consider another simple system, this time on the square lattice with a $C_2$ rotation symmetry $R_\pi$. This system has an odd $C_2$ charge, eg. $|\leftarrow\rangle - |\rightarrow\rangle$ at the rotation center and is a symmetric product state elsewhere. As with the reflection examples, we can see the odd charge at the rotation center using a trace:
\[\lim_{\beta \to \infty} \Tr\ R_\pi\ e^{-\beta H} = -1.\]
The geometry of this trace is a mapping cylinder $M = \bR^2_{x,y} \times S^1_\pi$ where we transform by $R_\pi(x,y) = (-x,-y)$ as we go around the circle. The unique non-trivial effective action $\omega(\alpha) = \frac{1}{2} \alpha \frac{d\alpha}{2}$ indeed has
\[\exp \left( 2\pi i \int_M \frac{1}{2} \alpha \frac{d\alpha}{2}\right) = -1,\]
so this is our phase.\footnote{If we use $C_2$-symmetric, time independent boundary conditions in our trace, corresponding to the one point compactification of $M$, we get real projective 3-space $\mathbb{RP}^3$.}

The equivalent internal $\bZ_2$-symmetry SPT is well known to be characterized by flux fusion: two $\pi$ fluxes fuse to an odd charge. This can be easily read off from the Chern-Simons form of its effective action $\frac{1}{2} A \frac{dA}{2}$, where $A$ is the (ordinary) background $\bZ_2$ gauge field. Indeed, if we read $dA/2$ as the density of $2\pi$ fluxes, we can read the effective action as a source term for $A$ saying precisely that $2\pi$ fluxes carry odd charge.

\begin{figure}
    \centering
    \includegraphics[width=9cm]{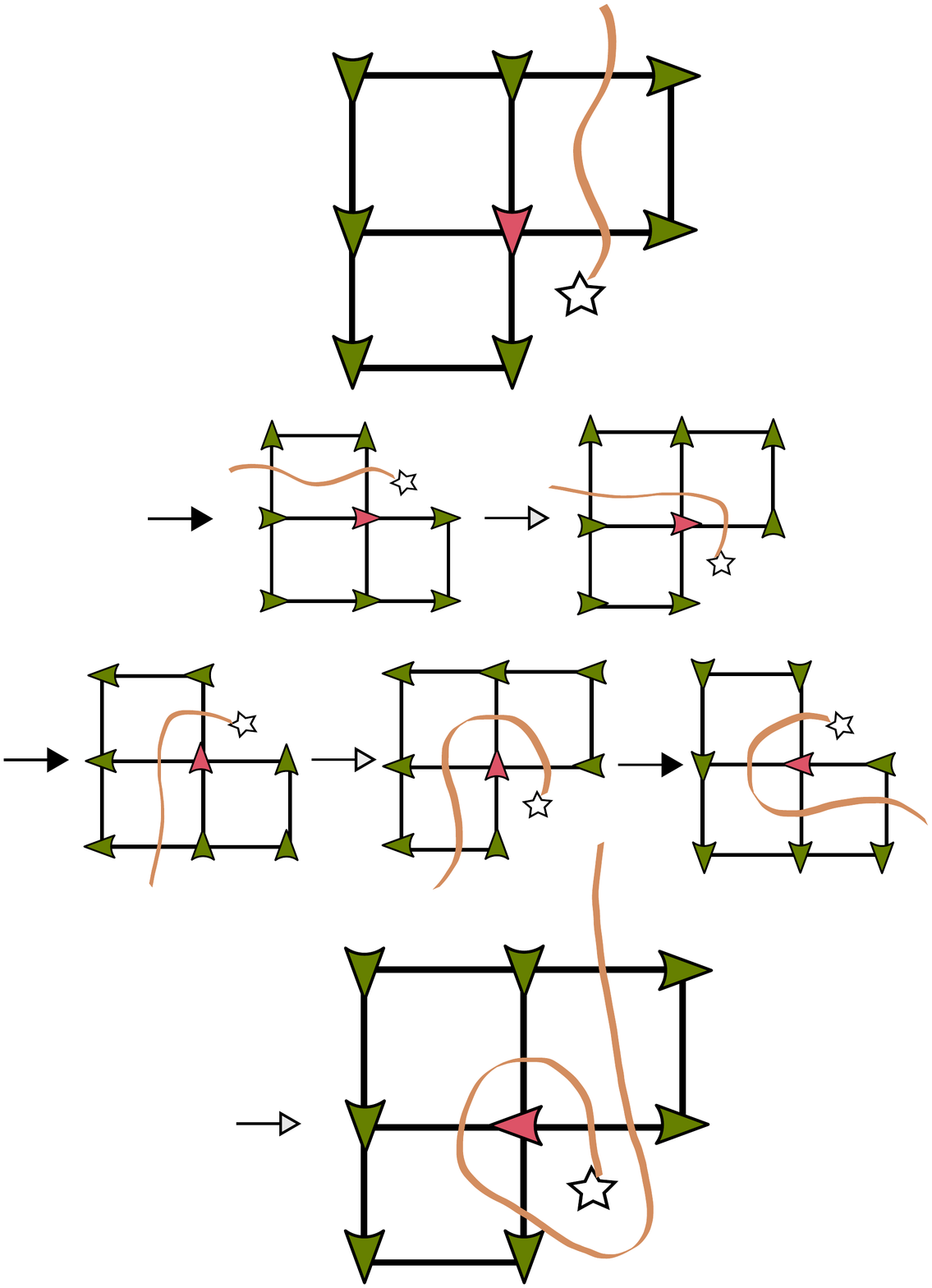}
    \caption{An example of a $C_4$ disclination is shown undergoing a full rotation. Our Hilbert space is a product of $C_4$ spins on each site and depicted is the transformation of a particular basis element. The star indicates the ``missing quadrant" of section \ref{sec_spatial_gauge_fields}, across which spins (green) away from the rotation center (red) are glued by a 90 degree rotation (see also Appendix \ref{appendix:gaugecoupling}). For convenience, we have used an orange ``domain wall" to indicate the boundaries between regions of homogeneous spin (compare Fig 6). The rotation itself is a two step process which must be performed three times. The first step (black arrows) is to simply rotate the picture around the rotation center counterclockwise by 90 degrees. The second step (white arrows) is a ``gauge transformation" (compare Fig 7) that moves the missing quadrant to its original position. At the end of the process, all green spins have returned to their original configuration while the spin at the rotation center has been rotated one unit. If there is a charge at the rotation center, the disclination thus picks up that charge as a phase after a full rotation.}
    \label{fig:disclinationsemion}
\end{figure}

The crystalline equivalent $\frac{1}{2} \alpha \frac{d\alpha}{2}$ has the same form, so can we read it in the same way? It turns out we can if we identify a gauge flux with the $C_2$ disclination and use the careful definition we gave in \ref{sec_spatial_gauge_fields}. We expect to find a half $C_2$ charge of the disclination, so we will rotate it twice by 180 degrees and see if we pick up a minus sign. As shown in Fig \ref{fig:disclinationsemion}, indeed we do.

Note that for an internal $\mathbb{Z}_2$ symmetry (it's less clear how it would work for a spatial symmetry), we can promote the gauge field to a dynamical quantum variable and then the gauge fluxes become deconfined excitations with semionic statistics \cite{Levin2012}. A semion has a topological spin (phase picked up under $2\pi$ rotation) of $i$, One might ask how this is consistent with the above statement that the symmetry defect picks up a phase of of $-1$ under two 180 degree rotations. However, we note that this is a rotation in $X$, whereas the rotation that defines topological spin does not take place in $X$ but rather in $M$.

\subsubsection{Crystalline Topological Insulators}

Now let us discuss 3+1D phases protected by time reversal symmetry or reflection symmetry and a $C_m$ rotation symmetry (typically $m=2,4,$ or $6$, odd $m$ has no nontrivial phase). For bosons, these have a $\bZ_2$ classification, with topological response resembling a $\theta = \pi$ topological term
\[\omega(\alpha) = \frac{1}{2} \left(\frac{d\alpha}{m}\right)^2,\]
where $\alpha$ is the $C_m$ twist of the crystalline gauge background. These phases are interesting because this topological term is only non-zero on non-orientable manifolds.\footnote{Indeed, the term may be written $\frac{1}{2} w_2 Sq^1 \alpha = \frac{1}{2} w_3 \alpha$, and $w_3 = 0$ for all orientable 4-manifolds.} This is where time reversal or reflection symmetry comes in. We will consider the reflection symmetry example, which acts across the x-y plane: $z \mapsto -z$. We will combine this with a $C_2$ subgroup of $C_m$, which we may write $x,y \mapsto -x,-y$, while $z \mapsto z$. The combined symmetry is a ``parity" symmetry:
\[P: x,y,z \mapsto -x,-y,-z.\]
Our topological response will be a trace of $P$. To describe this as a path integral, we begin with a cube $[-L,L]^3_{x,y,z} \times [-T,T]_t$ and glue $t = -T$ to $t = T$ with a $P$-twist. Then we choose $P$-symmetric, $t$-independent boundary conditions at $x,y,z = -L,L$. The resulting path integral is over a spacetime $\mathbb{RP}^4$, with $\alpha$ the generator of $H^1(\mathbb{RP}^4,\bZ_m)$. We therefore expect for these special states the topological response
\[\lim_{\beta \to \infty} \Tr P e^{-\beta H} = \exp\left(2\pi i  \int_{\mathbb{RP}^4} \omega(\alpha)\right) = -1.\]
Let us give an example of a state with this topological response. We can actually obtain it from dimensional induction from our $C_2$ symmetric state we discussed above. We place this state along the x-y plane. It is pinned there by the reflection symmetry across that plane. The above trace reduces to a trace of the rotation symmetry $x,y \mapsto -x,-y$ on this state, which we have computed sees an odd charge at the rotation center, yielding $-1$.




\subsubsection{Sewing Together a Pair of Pants and Internal Symmetry SPT}

So far we have discussed how to consider 1+1D twisted traces as crystalline backgrounds over either $X = S^1_x \times \bR_t$ or $X = \bR_x \times \bR_t$ in the case that translation is an explicit symmetry. Other partition functions of interest must be computed on higher genus surfaces and a basic building block of these is the pair of pants. Indeed, every orientable closed surface is glued together from discs and pairs of pants. Physically, the path integral over the pair of pants computes a sort of fusion process from $\mathcal{H}(S^1) \otimes \mathcal{H}(S^1)$ to $\mathcal{H}(S^1)$. Let us discuss how  the pair of pants is realized as a crystalline gauge background over $X = \bR_x \times \bR_t$ in a system with a unit translation symmetry $T_x$.

We construct $M$ starting with $[-L,L]_x \times [-T,T]_t$ mapping by inclusion into $X$ with a cut along the negative $t$ axis from $t = -T$ to $t = 0$ which doubles the $t$ axis into $(0^\pm,t)$ for $t<0$. We glue the $x \to 0^-$ side to $x = -L$ and the $x \to 0^+$ side of the branch cut to $x = L$ also with $T_x^L$. For $t \ge 0$, we glue $x = -L$ to $x = L$. This gives $M$ the topology of the pair of pants. We build a crystalline gauge field $M \to X//G$ by mapping the open domain $(-L,L)_x \times (-T,0) \cup (0,T) \subset M$ into $X = \bR^2_{x,t}$ by inclusion. We extend this to a $T_x$-twisted map on all of $M$ by inserting branch cuts so that $x = \pm L$, $t<0$ is glued to $x = 0^\pm$, $t<0$ with a twist $T_x^L$ and $x = -L$, $t>0$ is glued to $x = L$, $t>0$ with a twist $T_x^{2L}$. In terms of the translation twisting field $\tau$, we thus have $\int \tau = L$ on the two ``incoming" circles at $t = -T$ and $\int \tau = 2L$ on the ``outgoing" circle at $t = T$.

The path integral over the pair of pants is computed by stitching together propagators from the two legs into the waist. These propagators are computed on the cylinder with translation-twisted boundary conditions. For example, on the incoming circle from $x= -L$ to $x = 0$, we restrict the Hamiltonian from $\bR_x$\footnote{We assume this Hamiltonian is ultralocal to the lattice. In 1D this means it only couples neighbouring sites. Any finite-range Hamiltonian may be coarse-grained until it satisfies this.}, and use boundary conditions so that in the product state basis of the on-site Hilbert space $\mathcal{H}_{-L} \otimes \mathcal{H}_{-L+1} \otimes \cdots \otimes \mathcal{H}_0$, we restrict to the subspace spanned by product states such that the state at $\mathcal{H}_0$ is the same as $T_x^L$ applied to the state at $\mathcal{H}_{-L}$. Translation symmetry ensures that the Hamiltonian preserves this subspace. We compute $e^{-T H}$ as an operator from this subspace to $\mathcal{H}_{-L} \otimes \cdots \otimes \mathcal{H}_0$. We do the same for the other incoming circle, as an operator landing in $\mathcal{H}_0 \otimes \cdots \otimes \mathcal{H}_L$. Then we concatenate the two states and project so that the $\mathcal{H}_0$ parts agree. Then we are in the subspace of $\mathcal{H}_{-L} \otimes \cdots \otimes \mathcal{H}_L$ where the $-L$ part agrees with $T_x^{-2L}$ applied to the $+L$ part. This is the Hilbert space of the outgoing circle and we can apply $e^{-T H}$ on this subspace to obtain the complete pair-of-pants operator from the Hilbert space of the two incoming circles to the Hilbert space of the big outgoing circle.

\begin{figure}
    \centering
    \includegraphics{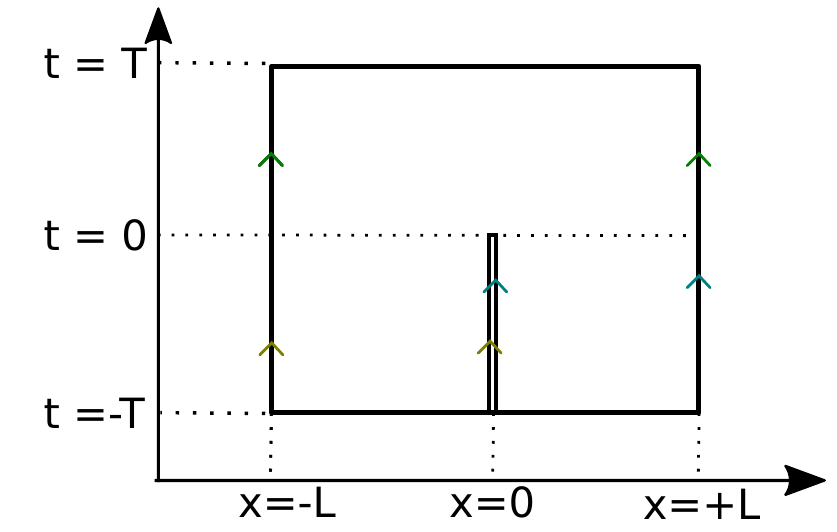}
    \caption{The pair of pants as a crystalline gauge background. The branch cuts gluings are indicated with colored arrows. $(x,t) = (0,0)$ is a singularity of the smooth structure but not the continuous structure. It can be smoothed out into a high curvature region but does not affect our calculations.}
    \label{fig:pairofpantsgluing}
\end{figure}

If we also have an internal symmetry $G$ with associated background gauge field $A$, then we can also have $G$ twists around these circles encoded in the $G \times T_x$ crystalline gauge background. We denote the twists around the two incoming circles as $\int_{1,2} A = g_1, g_2$ and around the outgoing circle as $\int_3 A = g_3$. We find that for continuity they must satisfy $g_1 g_2 = g_3$. We can imagine this is describing a $G$-flux fusion process occurring inside the pair of pants (see FIG. \ref{fig:pairofpantsgluing}. If we choose representative ground states $|g\rangle$ in each sector $g \in G$, the path integral over $M$ (topological response \eqref{topresp} with boundary conditions) will be some phase $c(g_1,g_2)$. The crystalline topological liquid assumption implies that if we glue two such pairs of pants together in two different ways, we get equal response, at least in the large $M$ limit. This implies that $c(g_1,g_2)$ is a group 2-cocycle, encoding the possibility of projective flux fusion. On the other hand, since the phases of our states are unphysical, if we rephase them each $|g\rangle \mapsto e^{i\phi(g)}|g\rangle$, then $c \mapsto c + \delta \phi$ changes at most by an exact cocycle, so $c \in \mathcal{H}^2(G,U(1))$ is well defined in group cohomology.

Let us show how this is computed in an example. We consider a very simple $G = \bZ_2 \times \bZ_2$ symmetric SPT. There are two associated on-site $\bZ_2$ degrees of freedom we denote $\phi_{1,2}$. We consider the state
\[|0\rangle = \sum_{\phi_1,\phi_2 \in C^0(\bR_x,\bZ)} (-1)^{\sigma(\phi_1,\phi_2)} |\phi_1,\phi_2\rangle\]
The sum is over all labelings of vertices $j \in \bZ \subset \bR_x$ by a pair $(\phi^j_1,\phi^j_2) \in \bZ_2 \times \bZ_2$ and the relative phase factor $\sigma(\phi_1,\phi_2)$ is the number of edges $j \to j+1$ where $\phi_1^j = 1$ (mod 2) and $\phi_2^{j+1} - \phi_2^j = 1$ (mod 2). The unit translation symmetry $T_x:j \mapsto j+1$ is manifest. Less manifest but still a symmetry is the $G = \bZ_2 \times \bZ_2$ which acts by shifting the respective $\bZ_2$ variables $\phi_{1,2}$ by a global constant. It can be seen by writing \[\sigma(\phi_1,\phi_2) = \int_{\bR_x} \phi_1 d\phi_2,\]
which is invariant under a constant shift $\phi_2 \mapsto \phi_2 + 1$ and invariant up to boundary terms under $\phi_1 \mapsto \phi_1 + 1$.

We must determine the propagator for which this state is the unique ground state. This can be done using tensor network techniques or in this case merely by inspection. We find this state is computed by the path integral over a strip $\bR_x \times [-T,0]_t$ with fixed boundary conditions $\phi_{1,2}(x,-T) = 0$ and free boundary conditions at $t = 0$, using the path integral weight $e^{iS}$ which is $-1$ to the number of intersection points between $\phi_1$ and $\phi_2$ domain walls.

\begin{figure}
    \centering
    \includegraphics{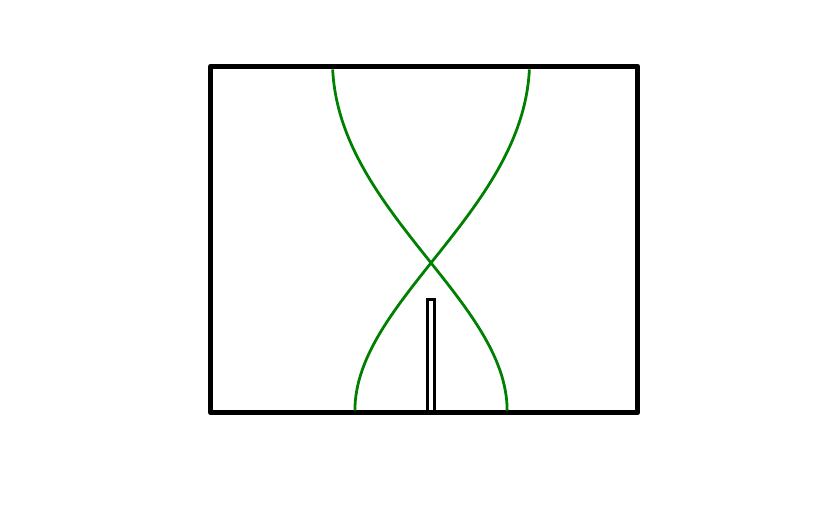}
    \caption{A $\phi_2$ flux and a $\phi_1$ flux fuse with a minus sign, computed as a path integral over the pair of pants with an unavoidable domain wall crossing.}
    \label{fig:pairofpantscrossing}
\end{figure}

Now for each $g \in G = \bZ_2 \times \bZ_2$, we choose a gauge-fixed reference state for the twisted ground state on $S^1_g$. This just means we have to decide where the domain walls go. Let's agree there is at most one domain wall for $\phi_1$ at $x = -L/2$ and one for $\phi_2$ at $x = L/2$, where the circle is coordinatized by $x \in [-L,L]$. Then we see all the pairs of pants have trivial phase factor except for the one where the left incoming circle has $\phi_2$ twist and the right has $\phi_1$ twist. In this case, the domain walls have to cross (see FIG. \ref{fig:pairofpantscrossing}) so we get a phase $-1$. This gives us a function $c: G \times G \to \bZ_2$ which turns out to be a group 2-cocycle and classifies the nonabelian extension
\[\bZ_2 \to D_4 \to \bZ_2 \times \bZ_2,\]
where $D_4$ is the dihedral group of the square. Extending coefficients to $U(1)$ we get the corresponding SPT cocycle $c \in \mathcal{H}^2(G,U(1))$. A similar construction of a translation-symmetric state can be made for any $G$-SPT phase.

\subsubsection{Weak SPTs and Lieb-Schultz-Mattis}

Now we consider a 2+1D system with internal symmetry $G$ and unit translation symmetry in one direction, say $T_x$. We fix a 1+1D $G$ SPT $c(A) \in H^2(BG,U(1))$ and consider infinitely many copies of this SPT laid side-by-side along $x \in \bZ$. The edge of this system has a projective $G$ symmetry $c$ per unit cell. We would like to assemble a crystalline gauge background that distinguishes the bulk from the trivial phase. This way, we derive the Lieb-Schultz-Mattis (LSM) theorem as the anomalous edge constraint of this crystalline SPT.

As we discussed, any 1+1D SPT cocycle $c(A)$ is detected using the pair of pants (FIG. \ref{fig:pairofpantsgluing}), interpreted as a projective fusion of fluxes. In our case, this flux threads a circle in the $y$-direction, so we must assume there is also a $T_y$ symmetry or at least an emergent one that we can use to roll up the system in that direction. There are various ways to make such a construction. Given a $G$-SPT state
\[|c\rangle_{\rm single\ layer} = \sum_{\phi \in C^0(\bR_y,G)} \exp\left( 2\pi i \int_{\bR_y} c_1(\phi)\right)|\phi\rangle,\]
where $c_1(\phi)$ is some 1-form density depending on $\phi$ (compare previous example)\footnote{This state is in SPT phase $c$ iff $c(d\phi) = dc_1(\phi)$ for all $\phi$. It is called the first descendant of $c$. See \cite{KT1}}, we can write the layered SPT state as
\[|c\rangle_{\rm many\ layered} = \sum_{\phi \in C^0(\bR^2_{x,y},G)}\exp\left(2\pi i \int dx c_1(\phi)\right)|\phi\rangle,\]
where $dx$ integrates to 1 across the $x$-direction of any unit cell. This is equivalent to
\[|c\rangle_{\rm many\ layered} = \bigotimes_{x \in \bZ} |c\rangle_{\rm single\ layer}.\]
We can use this to write a state on a torus $S^1_x \times S^1_y$ as
\[|g,c_1\rangle = \sum_{\phi \in C^0(S^1_x \times S^1_y,G)} \exp\left( 2\pi i \int_{S^1_x \times S^1_y} \tau_x\ c_1(A,\phi)\right)|\phi\rangle,\]
where $\tau_x$ is the $T_x$ twisting around $S^1_x$ with $\int \tau_x = L$ the length of $S^1_x$ and $\int_{S^1_y} A = g$ is the $G$ twisting around $S^1_y$, and $c_1(A,\phi)$ is a 1-form density encoding $c_1(\phi)$ and its coupling to the background $G$ gauge field $A$\footnote{This density will satisfy $c(A+d\phi) - c(A) = dc_1(A,\phi)$.}. There is a gauge where $\int \tau_x = 1$ across each unit cell in the $x$ direction. This most closely matches our $|c\rangle_{\rm many\ layered}$ above, with $L$ copies of $|c\rangle_{\rm single\ layer}$ arranged around the circle $S^1_x$. In a gauge with a single branch cut, all $L$ copies are piled up at the branch cut.

We see that the proper crystalline gauge background for computing $c(A)$ is given by taking $\int \tau_x$ (the size of $S^1_x$) coprime to the order of $c$ and assembling the pair of pants using the remaining coordinates $y,t$. With incoming twists $\int_{1,2} A = g_{1,2}$, one can compute using the state above the topological response $c(g_1,g_2)$. Essentially this crystalline gauge background is a kind of compactification along $S^1_x$. Of course, since we are already in the topological limit, this $S^1_x$ need not be small. Its size need only be coprime to the order of $c$. If $G$ is finite, then it suffices to be coprime to the order of $G$.

Generalizing to the case where the $d$-dimensional unit cell carries the projective $G$ representation with 2-cocycle $c$, the topological response is
\[ \int_M \tau_1 \cdots \tau_d\ c(A),\]
where $\tau_j$ are the twists corresponding to the unit translation $T_j$ in the $j$th lattice coordinate. To check for this topological response, we can take our test spacetime $M$ to be a $d$-torus (of size coprime to the order of $c$ or $G$) times a pair of pants with $G$-twists.

\section{Spatially-dependent TQFTs}\label{sec_space_dep_tqft}
Here, we will explain our proposal for the description of the low-energy limit of a crystalline topological phase in terms of a TQFT. In this setting, our results, such as the crystalline equivalence principle, and the fact that the low-energy limit can be coupled to an arbitrary crystalline gauge field, can be proven mathematically. We will focus here on the physical motivations; however, we give enough detail that the full mathematically rigorous treatment should be apparent to TQFT experts.


Recall that the starting point is that a phase of matter should have a spatially-dependent ``topological limit'', which we expect to be described by a \emph{spatially-dependent TQFT}. Indeed, we define

\begin{defin}
A \emph{$(d+1)$-dimensional spatially-dependent TQFT} on a space $X$ is a continuous map $\sigma : X \to \Theta$, where $\Theta$ is the space of all $(d+1)$-dimensional TQFTs.
\end{defin}

Now, what exactly do we mean by ``space of all TQFTs''? Familiar notions of TQFTs (at least in 2+1D) look quite rigid, suggesting that any such space would be discrete. However, we want to argue that there is a natural way to think about TQFTs as living in a richer topological space $\Theta$. First of all, we note that for classifying phases of matter it will not be necessary to specify $\Theta$ exactly, only up to homotopy equivalence. Let us discuss a physical motivation for the homotopy type of $\Theta$.

\begin{figure}
    \includegraphics[width=7cm]{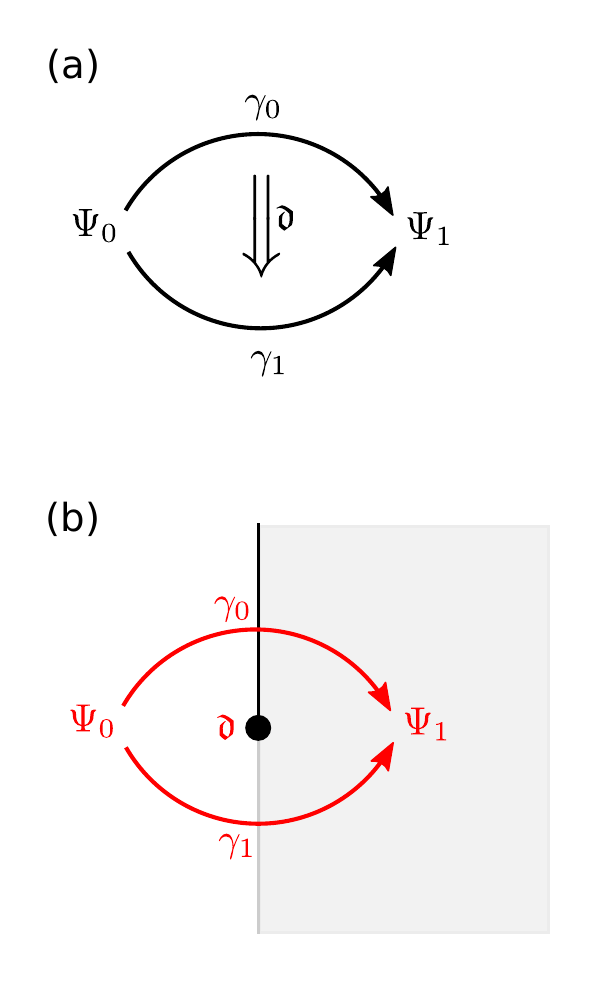}
    \caption{(a) Specifying the homotopy type of the space $\Theta$ of all TQFTs involves specifying points in this space, paths between arrows (single arrows), deformations between paths (double arrow), and so on. We want these to capture features of the space of quantum ground states.
    (b) These features can also be interpreted as interfaces. Depicted is a spatial configuration of interfaces in a 2-dimensional system, with two 1-dimensional interfaces separated by a junction of dimension 0. We can imagine that these interfaces are ``smoothed out'' such that the spatial variation occurs on scales large compared to the lattice spacing (thus, we have a a ``smooth state'' as discussed in Sections \ref{sec_toplimit} and \ref{sec_tcl}). Traversing a path in $\mathbb{R}^2$ from the left half-plane to the right half-plane, the local quantum state goes through the path $\gamma_0$ or $\gamma_1$ depending on whether the path in $\mathbb{R}^2$ goes through the upper 1-dimensional interface or the lower one. As one deforms the path in $\mathbb{R}^2$ through the 0-dimensional junction (black dot), the corresponding path in the space of quantum states goes through the deformation described by $\mathfrak{d}$.}
    \label{fig:paths}
\end{figure}

Generally, specifying the homotopy type of a topological space involves identifying points, paths between points, deformations between paths, and so on. The idea is that the structure of $\Theta$ should represent features of ground states of quantum lattice models. Thus, the points in $\Theta$ should correspond to ground states of quantum lattice models; the paths in $\Theta$ should correspond to continuous paths of ground states of quantum lattice models; and so on. There is another way to interpret these statements. A path in the space of ground states of quantum lattice models can also be implemented spatially, giving rise to an interface of codimension 1. Similarly, deformations between paths give rise to interfaces of codimension 2 between interfaces of codimension 1, and so on. (See Figure \ref{fig:paths}).

Roughly, therefore, the idea is that $\Theta$ should have the homotopy type of a cell complex with vertices $v$ labeled by $(d+1)$-dimensional TQFTs $T(v)$. Edges $e:v \to w$ are labeled by invertible $d$-dimensional topological defects $D(e)$ between $T(v)$ and $T(w)$. 2-Cells $f$ with $\partial f = v_1 \xrightarrow{e_{12}} \cdots v_n \xrightarrow{e_{n1}} v_1$ are labeled by invertible $d-1$-dimensional junctions between the defects $D(e_{12}) \cdots D(e_{n1})$. This continues all the way down to 0-dimensional defects, which for topological field theories with a unique ground state on a sphere is a copy of the complex numbers.
\footnote{Note that if two topological theories share an invertible topological defect, it means they are isomorphic, so in a formulation of TQFT up to isomorphism, eg. modular tensor category, each component of $\Theta$ will have a single vertex, perhaps with many other cells attached to it. In a state sum or tensor network formulation, on the other hand, there could be lots of state sums giving rise to the same TQFT with invertible MPO defects between them\cite{Williamson2016}.} In \cite{ENO}, this space was considered for $d=3$ in the tensor category framework and was referred to as the Brauer-Picard 3-groupoid.

A version of the bulk-boundary correspondence says that the set of boundary conditions and boundary operators determines the bulk topological field theory (see \cite{KICM} for some perspective on this in general dimensions and \cite{KS,FRS,Henriques} in 2+1D especially).  For theories admitting gapped (therefore topological in the IR) boundary conditions, this is the Baez-Dolan-Lurie cobordism theorem (sometimes ``hypothesis") \cite{BaezDolan,Lurie}, which characterizes possible boundary data as special objects in a $d+1$-category $\mathcal{C}$. This characterization can be used to construct $\Theta$ in a mathematically precise way. (Specifically, it is a space whose homotopy type is described by the core of the category $\mathcal{C}$).

Let us now consider the effect of symmetries. There is a natural way to define a $G$-action on a TQFT. From the Baez-Dolan-Lurie framework, one can show that a TQFT with symmetry $G$ is equivalent to TQFT coupled to a background $G$ gauge field. What we mean by the latter is the following. A $(d+1)$-dimensional TQFT assigns topological invariants to manifolds; for example, it assigns complex numbers (the partition function) to $(d+1)$-dimensional manifolds, and finite-dimensional Hilbert spaces (the state space) to $d$-dimensional manifolds. A $(d+1)$-dimensional TQFT coupled to a background $G$ gauge field assigns invariants to $G$-manifolds: manifolds decorated with $G$ gauge fields. Physically, this is supposed to describe response the topological response of the system to background gauge fields. We want to extend this result to systems with spatial symmetries.

Let us first review the case of a TQFT $\theta \in \Theta$ with an internal unitary $G$-action. Indeed, we define:
\begin{defin}
A $G$ action on a TQFT is a collection of isomorphisms $\phi_g : \theta \to \theta$ for each $g \in G$, with consistency data.
\end{defin}
In fact, in the Baez-Dolan-Lurie framework discussed above, isomorphisms are just paths in the space $\Theta$. These have the interpretation of defects of codimension 1. In fact, these are just symmetry twist branch cuts (e.g. see Ref.~\cite{BBCW}), such that particles moving through them get acted upon by the symmetry $G$. What we mean by ``consistency data'' is that the implementation of the relations of $G$ are also data in the $G$-action (see for instance \cite{KT2}). This data describes the codimension 2 junctions where domain walls fuse, the codimension 3 singularities where two junctions slide past each other, and so on. In fact, a more succinct way to formulate this definition is that a (anomaly-free, see below) TQFT with $G$ symmetry is a continuous map $\phi : BG \to \Theta$.
The statement about equivalence between TQFTs with $G$-action and TQFTs coupled to background gauge field then follows from the following general consequence of the Baez-Dolan-Lurie framework (see Thm 2.4.18 of \cite{Lurie}):
\begin{lemma}
\label{asdf_lemma}
For any space $W$, a continuous map $f : W \to \Theta$ is equivalent to a TQFT for manifolds equipped with maps into $W$.
\end{lemma}
Indeed, we set $W = BG$ and note that maps into $BG$ are the same as $G$ gauge fields.

Finally, we are ready to consider the general case of a spatially-dependent TQFT with a spatial symmetry $G$. We define:
\begin{defin}
A \emph{$(d+1)$-dimensional spatially-dependent TQFT with symmetry $G$} on a space $X$ is an action of the group $G$ on $X$ along with a $G$-equivariant map $\sigma : X \to \Theta$, meaning for all $x$ and $g$ we have a choice of isomorphism
\begin{equation}\label{equivcond}
\phi_{g,x}:\sigma(g \cdot x) \simeq \sigma(x).
\end{equation}
(with consistency data).
\end{defin}
Note that the isomorphisms should be taken to be unitary or anti-unitary for orientation-preserving or orientation-reversing symmetries respectively.

Once all the appropriate consistency data has been taken into account, we find that a spatially-dependent TQFT with an orientation-preserving spatial symmetry $G$ corresponds to a map from the homotopy quotient $X//G$ (discussed in section \ref{sec_beyond} and appendix \ref{latticegaugefields}) into $\Theta$. (We will discuss the orientation-reversing case later). Applying Lemma \ref{asdf_lemma}, we find

\begin{thm}\label{gaugethm}
A $(d+1)$-dimensional spatially-dependent TQFT on $X$ with symmetry $G$ is equivalent to a TQFT for $(d+1)$-manifolds $M$ equipped with a (homotopy class of) map $M \to X//G$, where $X//G$ is the homotopy quotient we have discussed in section \ref{sec_beyond}.
\end{thm}
This statement suggests that we can consider any map $M \to X//G$ as a crystalline gauge background, whereas in section \ref{sec_spatial_gauge_fields} we only showed how to couple a Hamiltonian to a rigid crystalline gauge background. Indeed, spatially-dependent TQFT mathematically formalizes our notion of smooth states in section \ref{sec_tcl} and appendix \ref{appendix_smooth}. Further, restricting to the case that $X$ is contractible, $X//G$ is homotopy equivalent to $BG$, so we find the same classification whether $G$ acts internally or on $X$.

\subsection{Spatially-dependent TQFTs for orientation-reversing symmetries and fermions}\label{space_dep_fermions}
Let us now discuss how to extend the above results to systems with orientation-reversing symmetries and/or fermions.
First we need to be more specific about the nature of the TQFTs we are discussing. TQFTs come in different flavors (framed TQFTs, spin TQFTs, and so on), depending on what structures we impose on the manifolds to which it assigns invariants. For bosonic systems with orientation-preserving symmetries, the natural choices are TQFTs for either framed manifolds or oriented manifolds. Choosing the latter amounts to assuming that the low-energy limit of our physical system has an emergent Lorentz symmetry, which becomes an $SO(d+1)$ Euclidean symmetry after Wick rotating to Euclidean spacetime)\footnote{This is because, in the Baez-Dolan-Lurie framework, one can show that oriented TQFTs arise from an action of $SO(d+1)$ on the space of framed TQFTs. See Corollary 2.4.10 of \cite{Lurie}.}.
Henceforth, we will always assume that such an emergent Lorentz symmetry is present, though we do not know how to justify this microscopically.

The situation gets interesting when there is also the microscopic $G$ symmetry. Indeed, besides the simplest possibility of a total $SO(d+1) \times G$ symmetry, there is also the possibility that $G$ acts on $SO(d+1)$ or there is some kind of extension. The former happens especially when $G$ contains orientation-reversing symmetries. Let us suppose $X$ is orientable. Whether $G$ preserves or reverses the orientation of $X$ describes a group homomorphism $G \to \bZ_2$. We expect this $\bZ_2$ to act non-trivially on any emergent $SO(d+1)$ symmetry generators so that the total symmetry algebra contains $O(d+1)$.

Let us see this in a simple example, a glide reflection acting on $X = \bR^2\times \bR_t$, $G = \bZ$. We assume that the infrared limit of such a system will be a field theory of a map $M \to X//G$. As we have discussed, in the case of group actions without special Wyckoff positions (ie. free actions), the homotopy quotient $X//G$ is equivalent to the ordinary quotient $X/G$, which in this case is a M\"obius band (cross the time coordinate). Our claim is that the continuum limit of this system is a ``sigma model", a 2+1D field theory whose only field is a map $\phi:M \to X/G$. What does it take to write an action for this field theory? The usual sigma model action is the volume of the image of $\phi$, or
\[S(\phi) = \int_M \phi^* vol(X/G),\]
where $vol(X/G)$ is a volume form on $X/G$. The issue is that $X/G$ is not orientable, so the volume form cannot be globally defined. Rather it is a 3-form valued in the orientation line bundle $\mathcal{L}_{or}$, a real line bundle whose sections switch sign when one follows them around the M\"obius band. When we pull back the volume form using $\phi$, the result is a 3-form valued in $\phi^*\mathcal{L}_{or}$. In order to integrate this over $M$, the fundamental class $[M] \in H_3(M)$ must be also valued in $\phi^*\mathcal{L}_{or}$. This occurs iff $\phi^*\mathcal{L}_{or}$ is isomorphic to the orientation line of $M$. This means that if a 1-cycle of $M$ encircles the M\"obius band, $M$ must be unorientable around that cycle. In the gauge theory description, these are the cycles with odd $G = \bZ$ holonomy, which have non-trivial image under the map $G \to \bZ_2$ we have just discussed. We discuss more about sigma models in Section \ref{sigma}.

Another way of saying this, which makes sense in general, is that the tangent bundle of $X$ gives rise to a bundle $T_G X$ over $X//G$, and rather than $TM$ being oriented, the orientation is on $TM \oplus A^*T_G X$. This sort of phenomenon is familiar also in the internal symmetry case, and is especially crucial once fermions are considered \cite{KTTW}. For this reason, we expect this to be the form emergent Lorentz symmetry takes in crystalline topological liquids. \footnote{In the case that there is no emergent Lorentz symmetry and $M$ must be framed, the orientation-reversing elements will act on the framing, and so it must be considered a framing not of $TM$ but rather $TM \oplus A^*T_G X$.}

To proceed, let us recall more from the case where $G$ acts trivially on $X$ \cite{KTTW}. The data that specifies the group of fermionic SPT phases is a map $w_1:G \to \bZ_2$ encoding which elements of $G$ are time-reversing elements, and also a group extension $w_2 \in H^2(G,\bZ_2)$ encoding how the relations of $G$ may be extended by fermion parity $(-1)^F$. As the notation suggests, both may be encoded in a bundle $\xi$ over $BG$, with $w_1 = w_1(\xi)$ and $w_2 = w_2(\xi)$. We call this the characteristic bundle. It may be interpreted in field theory as the $G$ representation of the fermion bilinears. The low energy limit of a gapped theory with such a symmetry is expected to be a theory defined on manifolds $M$ with a gauge field $A:M \to BG$ and a spin structure on $TM \oplus A^*\xi$. The cobordism group of such manifolds only depends on $\xi$ through $w_1(\xi)$ and $w_2(\xi)$. Since $X//G$ is itself a bundle over $BG$, any characteristic bundle $\xi$ also defines a bundle over $X//G$. This leads us to:
\begin{thm}
A bosonic (fermionic) crystalline topological liquid with microscopic symmetry $G$, characteristic bundle $\xi$, and emergent Lorentz symmetry has a topological limit defined on manifolds $M$ with a map $A:M \to X//G$ and an orientation (spin structure) on $TM \oplus A^*T_G X \oplus A^*\xi$.
\end{thm}
As an example, if $G = C_2$ acts on $X = \bR^2 \times \bR_t$ by $\pi$ rotation of space, $T_G X$ is a plane bundle over $B\bZ_2$. In terms of the tautological line bundle $\lambda$ (corresponding to the sign representation), this bundle is $\lambda \oplus \lambda$. One computes $w_1(\lambda \oplus \lambda) = w_1(\lambda) + w_1(\lambda) = 0$ and $w_2(\lambda \oplus \lambda) = w_1(\lambda)^2$, which generates $H^2(B\bZ_2,\bZ_2)$, corresponding to the extension $\bZ_4$ of $\bZ_2$. In the absence of any nontrivial $\xi$, this means that the $\pi$ rotation squares to the fermion parity. This is what we typically expect of rotations of fermions.

On the other hand, if there is an internal symmetry squaring to the fermion parity, such as a charge conjugation (unitary) symmetry, $C^2 = (-1)^F$, then we can combine this with the rotation symmetry to obtain a unitary $\bZ_2$ symmetry squaring to 1. By the crystalline equivalence principle, or by Thm \ref{gaugethm}, we expect such phases to give rise to topological field theories for manifolds $M$ with a spin structure on $TM$ and a $\bZ_2$ gauge field $A:M \to B\bZ_2$. Let's see how the conjecture above encodes this. As explained in \cite{KTTW}, a unitary $\bZ_2$ representation with $C^2 = (-1)^F$ corresponds to the characteristic bundle $\xi = \lambda \oplus \lambda$ over $B\bZ_2$ (recall $\lambda$ is the bundle of the sign representation $\bZ_2 \circlearrowright \bR$). Since $\xi$ of the $C_2$ rotation symmetry was trivial, this is also $\xi$ of the composite symmetry. The conjecture above says that we will obtain in the infrared a theory of manifolds $M$ with $A:M \to B\bZ_2$ and a spin structure on $TM \oplus A^*T_G X \oplus A^*\xi = TM \oplus A^*(\lambda \oplus \lambda \oplus \lambda \oplus \lambda)$. It turns out $T_G X \oplus \xi = \lambda \oplus \lambda \oplus \lambda \oplus \lambda$ has both $w_1 = 0$ and $w_2 = 0$. This means that we may choose a spin structure on this bundle and thereby form an isomorphism between spin structures on $TM$ and spin structures on $TM \oplus A^*(T_G X \oplus \xi)$ and there is no contradiction.

Given this conjecture, we can state the crystalline equivalence principle more precisely:
\begin{thm}\label{CEP}
If $X = \bR^d$, then any $G$ action on $X$ defines a vector bundle $T_G X$ over $X//G = BG$. The $G$ action on internal degrees of freedom defines another bundle $\xi$ over $BG$. SETs with this sort of crystalline symmetry are isomorphic to SETs with internal symmetry $G$ and characteristic bundle $T_G X \oplus \xi$.
\end{thm}
The most important aspect of this theorem is the identification of orientation-reversing symmetries like reflection with anti-unitary symmetries like time-reversal in the classification of topological phases. Indeed, for bosonic systems with $G = \bZ_2$, all that matters of $T_G X \oplus \xi$ is the determinant of the $G$ representation (ie. $w_1$), which is multiplicative over direct sum. Thus, a time reversal symmetry, which has $det(\xi) = \lambda$ and $T_G X$ trivial, is equivalent to a reflection symmetry, which has $det(T_G X) = \lambda$ and $\xi$ trivial.

An important caveat about this in fermionic systems is that reflection with $R^2 = 1$ corresponds to $T^2 = (-1)^F$. Indeed, the former leads us to consider TQFTs for manifolds $M$ with $A:M \to B\bZ_2$ and a spin structure on $TM \oplus A^*\lambda$. This is equivalent to a Pin$^+$ structure on $M$, which in \cite{KTTW,FH} is what needs to be considered for the classification of $T^2 = (-1)^F$ phases. This is because our spacetimes $M$ are Euclidean, and Wick rotation changes the behavior of orientation reversing symmetries.

\subsection{Comments on Anomalies}

In the beginning of this section, we made an identification between TQFTs with a $G$ symmetry and TQFTs parametrized by $BG$. This identification is actually only possible when the $G$ symmetry is anomaly free.

0+1D is an instructive example. Such TQFTs describe the ground states of quantum mechanical particles and so the space of 0+1D TQFTs $\Theta$ can be described as the space of finite dimensional Hilbert spaces $\Theta = \bigsqcup_n BU(n)$ (here $n$ is the dimension of the ground state degeneracy). The ground states of a quantum mechanical particle with symmetry $G$ are characterized by a unitary representation of $G$. This representation defines a vector bundle over $BG$ whose fibers can be considered different 0+1D TQFTs, and has a classifying map $BG \to \Theta = \bigsqcup_n BU(n)$.

In quantum mechanics, however, there is the possibility that the $G$ action on the space of ground states is projective. In 0+1D this counts as an anomaly. In this case, we don't get a vector bundle over $BG$ but rather a projective vector bundle. Such a bundle does not have a classifying map $BG \to \Theta$. Rather, the anomaly $\alpha$ is characterized by a bundle $\Theta(\alpha)$ over $BG$ with fiber $\Theta$, and the anomalous theory can be described as a section of this bundle. For group cohomology anomalies, $\alpha \in H^2(BG,U(1))$ classifies bundles over $BG$ with fiber $BU(1)$. This defines a bundle $\Theta(\alpha)$ by the diagonal map $BU(1) \to \Theta$. The story in all dimensions is a direct generalization with no new ingredients, though in practice it is difficult to work out the details \cite{NSS,relativetqft}.

We expect that there could be an analogous phenomenon for the spatially-dependent TQFTs. That is, we defined a crystalline topological liquid as a system with a topological limit described by a map $X \to \Theta$. However, it could be that there are ``anomalous" crystalline topological liquids, characterized by a section of a bundle $\Theta(\alpha)$ over $X$ with fiber $\Theta$. With symmetries this would become a section of such a bundle over $X//G$.

The anomaly itself would be characterized by the bundle $\Theta(\alpha)$. The simplest examples would come from ordinary equivariant cohomology $H^{D+1}(X//G,U(1))$, where $D$ is the spacetime dimension of the anomalous theory. This suggests a kind of crystalline anomaly in-flow mechanism which would be interesting to study. We leave this to future work.

\section{Classification of phases in non-contractible space}\label{noncontractibleproperties}

As mentioned above in Section \ref{sec_beyond}, our framework in principle can be applied to the classification of topological phases with spatial symmetries on any space $X$, not just $X = \mathbb{R}^d$. In this section we discuss a few examples of this classification to indicate the general flavor, and state some general properties. We emphasize that, although our classification can be applied to any space $X$ (for example, a compact manifold), our results are only expected to be physically valid when the size of this manifold is much greater than the lattice spacing and the correlation length; otherwise, we cannot define the ``topological limit'' discussed in Section \ref{sec_toplimit} which underlies our arguments. 

\subsection{Properties of the classification}
 All these statements are derived from properties of the homotopy quotient $X//G$ discussed in Appendix \ref{latticegaugefields}.
 \subsubsection{Properties that hold in general}
 The following statements are valid for the classification of crystalline phases in full generality.
We let $\mathcal{C}_G(X)$ denote the classification of crystalline topological phases with symmetry $G$ acting on a space $X$. We let $\mathcal{C}_G(*)$ denote the classification of crystalline topological phases with \emph{internal} symmetry $G$.
\begin{itemize}
    \item{\bf Crystalline Equivalence Principle} If $X = \mathbb{R}^{D}$, then $\mathcal{C}_G(\mathbb{R}^d) \cong \mathcal{C}_G(*)$. For orientation-preserving symmetries, we find an isomorphic classification with those phases protected by an internal unitary symmetry. For orientation-reversing symmetries, we find for example that inversion-symmetric phases have the same classification as time-reversal-symmetric phases.
     \item{\bf Rolling and Unrolling} If $G$ contains a normal subgroup $H$ which acts freely on $X$, then we can quotient $X$ by $H$ and obtain an equivalent rolled up phase: $C_G(X) \cong C_{G/H}(X/H)$. We can also unroll a phase along any circular coordinates to get an equivalent phase with a translation symmetry in the unrolled direction, so these classifications are isomorphic. In particular, if $G$ is the space group of a crystal, then the translations are a normal subgroup which act freely and we may instead study the point group acting on the quotient torus. This works even for non-symmorphic symmetries, as observed in Ref \onlinecite{Hiller}: even if there is no symmetry center, the point group will act on the fundamental torus. However, this residual group $G/H$ can still contain things like glide reflections (and certainly will if we are in a non-symmorphic situation).
\end{itemize}

\subsubsection{Properties that hold for ``in-cohomology'' bosonic SPTs}.
The following statements hold for the ``in-cohomology'' bosonic SPTs discussed in Section \ref{sec_topological_actions}, which are classified by the equivariant cohomology $\mathcal{C}_G(X) = H^{D+2}_G(X, \mathbb{Z}^{\mathrm{or}})$.

\begin{itemize}
    \item{\bf No Translation SPTs} If $G$ acts freely on $X$, eg. translations on $\bR^d$, then by rolling up along all the translations we get an equivalent phase on a torus without any symmetries: $H^{D+1}_G(X,\bZ^{or}) = H^{D+1}(X/G,\bZ^{or})$. In particular, since $X/G$ is a manifold, the reduced cohomology in the top degree is always zero, so there are no non-trivial phases.
    
        \item{\bf New Internal SPTs Protected by Topology} On the other hand, if $G$ is an internal symmetry, meaning it doesn't act on $X$, then $X//G = X \times BG$. This means $H^{D+1}_G(X,\bZ^{or}) = \bigoplus_{j+k=D+1} H^j(BG,H^k(X,\bZ))$. If we require that $\omega$ has no dependence on $X$, we find the usual group cohomology classification for internal symmetry SPTs\cite{CGLW}. We see however the possibility for new equivariant cohomology SPT phases which are protected by the topology of $X$ as well as $G$ symmetry. These all look like lower dimensional $G$ SPT phases wrapped perpendicular to non-contractible cycles of $X$.

            \item {\bf Finitely Many Phases in Each Symmetry Class $(X,G)$} In fact, all phases originate in some element of $E_2 = \oplus_{j+k=D+1} H^j(BG,H^k(X,\bZ^{or}))$. We discuss how this works in Appendix \ref{cocycles}. Because we can roll up our phases to equivalent ones with finite symmetry $G$, the only piece of $E_2$ which can contribute infinite order elements is the $j=0$ piece $H^0(BG,H^{D+1}(X,\bZ^{or}))$. Happily, since $X$ is $D$-dimensional, this piece always vanishes, so $E_2$ is finite and therefore there are only finitely many phases in each symmetry class $(X,G)$.
\end{itemize}



\subsection{Examples of Phases on Non-contractible Spaces}\label{noncontractiblexamples}


\subsubsection{Reflection Acting on a Circle and Unrolling}

\begin{figure}
    \centering
    \includegraphics{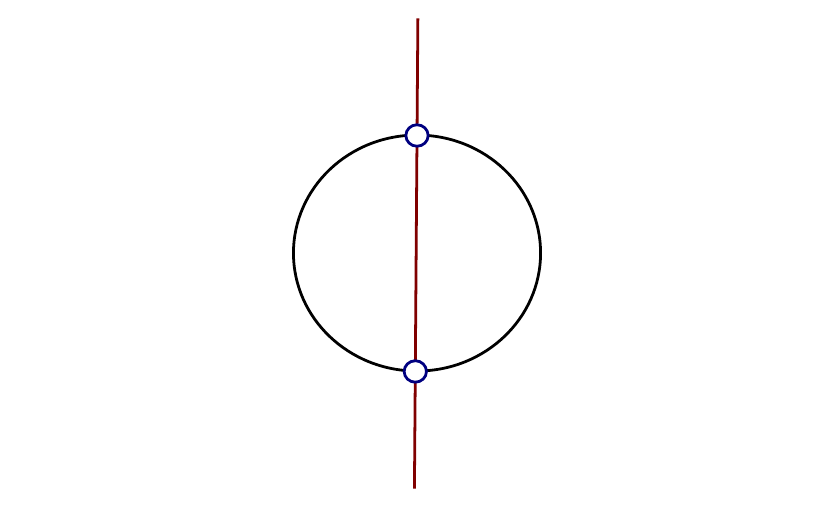}
    \caption{A reflection acting on a circle. The reflection axis is shown in red and the fixed points are marked with blue circles. SPT phases of this symmetry set up are classified by a $\bZ_2$ charge at each fixed point. There are more phases in this geometry than on a line with a single reflection center!}
    \label{fig:circlefixedpoints}
\end{figure}


Consider a system on a circle with a reflection symmetry $\theta \mapsto -\theta$. By arguments of Ref \onlinecite{SHFH}, there should be a $\bZ_2 \oplus \bZ_2$ classification corresponding to the residual $\bZ_2$ charges at the two fixed points (see Figure \ref{fig:circlefixedpoints}). For our classification, these phases live in $\tilde H^2_{\bZ_2}(S^1,U(1)^{or})$. The cohomology groups in the descent sequence (Appendix \ref{cocycles}) which contribute are
\[H^2(B\bZ_2,U(1)^{or}) = \bZ_2, \quad{\rm generated\ by\ } \frac{1}{2}\alpha^2,\]
\[H^1(B\bZ_2,H^1(S^1,U(1))^{or}) = \bZ_2, \quad{\rm generated\ by\ } \alpha \frac{d\theta}{\pi},\]
where $\alpha$ is the (degree one) generator of cohomology of $B\bZ_2$ and $d\theta/\pi$ is the element of $H^1(S^1,U(1)) = U(1)$ which integrates to $1/2$ over the circle. It is the unique fixed element under the $\bZ_2$ action.

We observe the similarity with $\frac{1}{2}\alpha_0^2$ and $\frac{1}{2}\tau \alpha_0$ in Section \ref{reftrans}, where we considered a system with a reflection (corresponding to $\alpha_0$) and translation (corresponding to $\tau$) symmetry. Indeed, taking the quotient of that system by the translation symmetry, we obtain the circle with reflection action. The translation part of the crystalline gauge field $\tau$ becomes the volume form $d\theta/2\pi$ of the quotient $S^1$. Identifying the topological response, we find that $\frac{1}{2}\alpha^2$ has an odd reflection charge at both fixed points, while $\alpha \frac{d\theta}{\pi}$ has a single odd reflection charge at a fixed point. Which fixed point in particular depends on the choice of cocycle representing $\frac{d\theta}{\pi}$ and can be traced to the choice of preferred reflection center of the unrolled system. The state with the odd reflection charge at the other fixed point will correspond to the topological response $\frac{1}{2}\alpha^2 + \alpha \frac{d\theta}{\pi}$.

\subsubsection{Orientable $G$ Actions on Spheres}

Another interesting case is when space is a sphere $S^d$ with $G$ acting via $\rho:G \to SO(d+1)$. The descent tells us that all such phases are a combination of those in $H^{d+1}(BG,U(1))$, the ordinary $G$ SPT phases, and $H^1(BG,H^d(S^d,U(1))) = H^1(BG,U(1))$, which correspond to a system whose ground state on $S^d$ has a (abelian) $G$ charge with character $\chi \in H^1(BG,U(1))$. These phases are not independent, but are related by the Gysin sequence \cite{Hatcher}. If we write $V$ for the volume form on $S^d$ with unit volume $\int_{S^d} V = 1$, then the classes in $H^1(BG,H^d(S^d,U(1)))$ may be written $\chi V$. These classes are not necessarily well-defined on $S^d//G$. This is because $dV$ is no longer necessarily zero, but instead can be
\[dV = w_{d+1} \in H^{d+1}(BG,\bZ).\]
This is the top Stiefel-Whitney class (or often ``Euler class") of the associated $\bR^{d+1}$ bundle over $BG$ of which $S^d//G$ is the unit sphere bundle. Physically, this equation says that the ``Skyrmion number" $\int_\Sigma A(t)^*V$ is not conserved in the presence of instantons $A^*w_{d+1}$, where $\Sigma$ is a (2d) time slice and $A:\Sigma\times\bR_t \to S^d//G$ is the crystalline gauge field. The topological response $\chi V$ tells us that Skyrmions carry $G$ charge $\chi$ so for $G$ charge to be conserved we need to satisfy the anomaly vanishing formula $d(\chi V) = \chi w_{d+1} = 0 \in H^{d+2}(BG,U(1))$. On the other hand one can show that $H^{d+1}(BG,U(1)) \to H^{d+1}(S^d//G,U(1))$ is always injective. We believe this to be related to the fact that one can obtain all SPT phases from sigma models on the sphere \cite{Bietal} (see also Section \ref{sigma} below). To summarize, the group of phases on $S^d$ sits in an exact sequence
\begin{multline}
0 \to H^{d+1}(BG,U(1)) \to H^{d+1}(S^d//G,U(1)) \\\xrightarrow{\int_{S^d} -} H^{1}(BG,U(1))  \xrightarrow{- \cup w_{d+1}} H^{d+2}(BG,U(1)).
\end{multline}
The first map is the inclusion of ordinary SPT phases, the second measures the $G$ charge of the ground state on $S^d$, and the third is the anomaly map.

\section{Generalizations}
\label{sec_generalizations}

\subsection{Floquet SPTs and gauged Floquet SPTs}
\label{sec_floquet}

The TQFTs defined by our topological actions do not have a preferred axis of time. In a sense there is no difference between a discrete space translation and a discrete time translation. This suggests that all of the crystalline topological phases we have discussed with a discrete spatial translation symmetry can be thought of as Floquet crystalline topological phases \cite{Khemani2015a,vKS,DN,Potter2016b,Roy2016b,Po2016,Potter2016c,Roy2016c} which appear in a driven system with time-periodic Hamiltonian $H(t)$, with $H(t+T) = H(t)$. (To make sense of this, we should take the space $X$ on which the symmetry $G$ acts to be space-time rather than just space). We will not, however, attempt to explain how precisely one arrives at a topological action describing such a driven system, nor even what it actually means to ``gauge'' a discrete time-translation symmetry.

For example, in 1+1D the $H^3(BG \times \bZ,\bZ) = H^3(BG,\bZ) \oplus H^2(BG,\bZ)$ phases can be interpreted as Floquet phases with internal $G$ symmetry. For finite $G$, this is $H^2(BG,U(1)) \oplus H^1(BG,U(1))$ and agrees with the classification in Ref. \onlinecite{vKS}. The first group represents bulk SPT order and the second group represents a charge being pumped each time step.

We mention that $H^2(BU(1),\bZ) = \bZ$ and so there seems to be a topological response analogous to these charge pumping phases but with symmetry $U(1)$. Though it is possible to construct finely tuned models which do this, they all seem to transport some non-trivial quantum information, so we remain skeptical that they exist in real systems.

There are also phases that mix Floquet and crystalline symmetry, for example in 2+1D on $X = \bR_{x,y} \times \bR_t$ with a translation symmetry $x \mapsto x+1$ and a Floquet symmetry $t \mapsto t + 1$ there is a phase in $H^2(BG,H^2(B\bZ \times B\bZ,\bZ)) = H^2(BG,\bZ)$. The interpretation is that the system pumps a $G$ charge per unit cell per time step on the boundary. This can be rolled up into a Floquet system on a cylinder which pumps a $G$ charge per time step to the boundary circle.

It is possible to also pump higher dimensional SPT phases to the boundary. In a sense this is because SPT phases are themselves generalized abelian charges. These phases live in $H^D(BG,H^1(B\bZ,\bZ))$, which is $H^{D-1}(BG,U(1))$ for finite $G$. For example, in 3+1D we are talking about $H^4(BG,\bZ)$. For connected Lie groups $G$ these classify Chern-Simons actions. These correspond to 3+1D phases which pump a $G$ Chern-Simons theory to the boundary each time step. For $G = U(1)$ these are integer quantum Hall states.

We pause here to appreciate that if we can build a Floquet system which pumps (internal) $G$ SPT phases to the boundary, then we can think about coupling to a \emph{dynamical} $G$ gauge field. Generically, this makes the $G$ SPT phase no longer invertible. For example, a trivial $\bZ_2$ SPT in 2+1D becomes the toric code, which is topologically ordered. This does not mean, however, that we pump any non-trivial degrees of freedom to the boundary every time step. Rather, we have a $G$ gauge theory at the boundary and at each time step we pump a \emph{topological term}. For example, the 3+1D Floquet phase with internal $\bZ_2$ which pumps the $\bZ_2$ group cohomology SPT to the boundary each time step has in the gauged picture a $\bZ_2$ gauge theory on the boundary which at one time step is a toric code and at the next is a double semion and then toric code again. It is as though half a $\bZ_2$ charge is pumped \emph{to the fluxes} and they become semions. If one models this with a Walker-Wang model, then the fluxes on the boundary are themselves the boundaries of strings that reach into the bulk. It is as though these strings are decorated with half (!) the non-trivial 1+1D $\bZ_2$ Floquet SPT, something that pumps half a $\bZ_2$ charge each time step.

Finally, 
let us note that, according to the ``unrolling'' principle in Section \ref{noncontractibleproperties}, a TQFT in a space-time with discrete time-translation symmetry is formally equivalent to a TQFT in a space-time with a compactified time dimension. One might wonder how this differs from a system at finite temperature, which also can be interpreted in terms of a compactified time dimension. The difference is that the latter system still has a \emph{continuous} time-translation symmetry, whereas Floquet systems do not.

\subsection{Fermions and Beyond Group Cohomology}\label{fermions}

Although we have mainly been talking about bosonic systems, the general framework of our paper is applicable to fermionic systems as well. Indeed, the notion of crystalline gauge field is independent of whether the system is bosonic or fermionic, and in Section \ref{space_dep_fermions} we explained how to define fermionic versions of  ``spatially dependent TQFTs''. Here we will think some more about fermionic crystalline SPT phases.

We note that this section is not really a ``generalization'', since it falls within our general framework. It is, however, a generalization of the equivariant cohomology formulation of bosonic SPT phases discussed in Section \ref{sec_topological_actions}.
We can go beyond equivariant cohomology by considering equivariant cobordism. This is where the invariants discussed in Ref \onlinecite{SSR} live. The definition of the most general equivariant cobordism is delicate, but by the Crystalline Equivalence Principle, we can assume that fermion phases for spacetime symmetries are like fermion phases for internal symmetries. By what we've learned from Ref \onlinecite{KTTW}, the data we need to define spin cobordism is a homomorphism $w_1:G \to \bZ/2$ telling us which elements of $G$ are orientation-reversing and a group extension $w_2 \in \mathcal{H}^2(G,\bZ)$ telling us how $G$ is extended by fermion parity. Then there is a corresponding equivariant spin cobordism group $\Omega^d_{\rm spin}(G,w_1,w_2)$. These two classes can be nicely encoded in a single $G$-representation $\xi$ called the characteristic bundle (see Section \ref{sec_space_dep_tqft}). Then we write $\Omega^d_{\rm spin}(G,\xi)$. These are equivariant cobordism invariants of $d$-manifolds $X$ with a spin structure on $TX \oplus A^*\xi$, where $A$ is the $G$ gauge field. For example, the ordinary topological insulator in 3+1D lives in $\Omega^4_{\rm spin}(U(1) \rtimes \bZ_2,\xi)$ where $\xi$ is the fundamental 2d representation of $U(1) \rtimes \bZ_2 = O(2)$ on $\bR^2$ plus two copies of the sign representation $O(2) \xrightarrow{\rm det} \bZ_2 = O(1)$ on $\bR$. This corresponds to the $\bZ_2$ being orientation reversing and squaring to $(-1)^F$. Details are in Ref \onlinecite{KTTW}. It so happens $\Omega^4_{\rm spin}(U(1) \rtimes \bZ_2,\xi) = \bZ_2$.

Let us give a concrete example of a 2+1D fermionic crystalline SPT with $C_4$ symmetry (not extended by fermion parity), which is related to an internal $\bZ_4$ fermionic SPT which carries a Majorana zero mode at a symmetry flux \cite{KTTW,Tarantino2016,Gaiotto2016,Bhardwaj2016}. This model is very simple. It exists on a square lattice with $p4$ symmetric Kastelyn orientation and four Majorana operators per site, each associated with one of the 4 edges incident there (see figures). The Kastelyn orientation is used to define a Kitaev-wire type Hamiltonian that dimerizes the Majoranas and will be at least $p4$ symmetric. One sees that at a disclination, there is a vertex with just three bonds incident, so one Majorana is left unpaired, so in a sense, the disclinations will carry \emph{nonabelian} statistics.

\begin{figure}
    \centering
    \includegraphics[width=10cm]{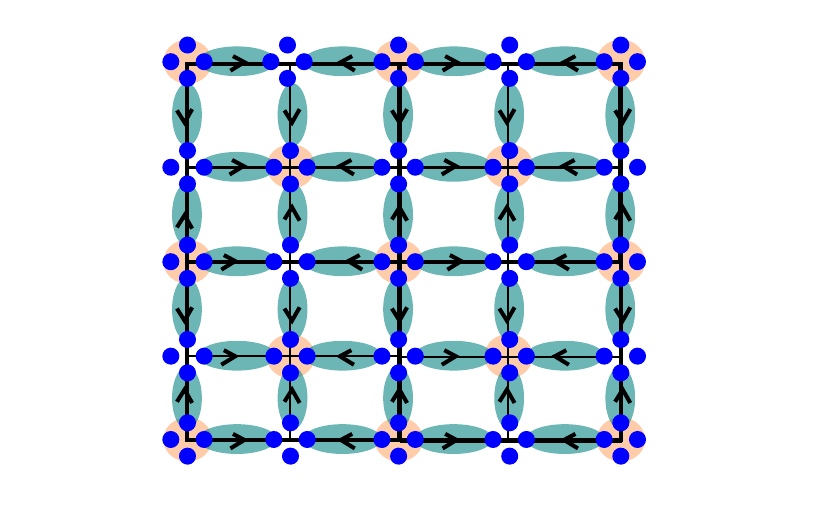}
    \caption{A $p4$ (actually $p4m$!) symmetric Majorana Hamiltonian. At each site there are four Majorana operators (blue dots) which are hybridized along the teal bonds. An arrow from Majorana $c_i$ to Majorana $c_j$ is a term $-\frac{i}{2} c_i c_j$ in the Hamiltonian. The arrows are chosen to be a $p4$ symmetric Kastelyn orientation, which is necessary for the ground state to be free of fermion parity $\pi$-fluxes. See \cite{Tarantino2016,Braydenetal}. The $C_4$ rotation centers are highlighted in beige. The other sites are $C_2$ rotation centers.}
    \label{fig:majoranalattice}
\end{figure}

\begin{figure}
    \centering
    \includegraphics[width=10cm]{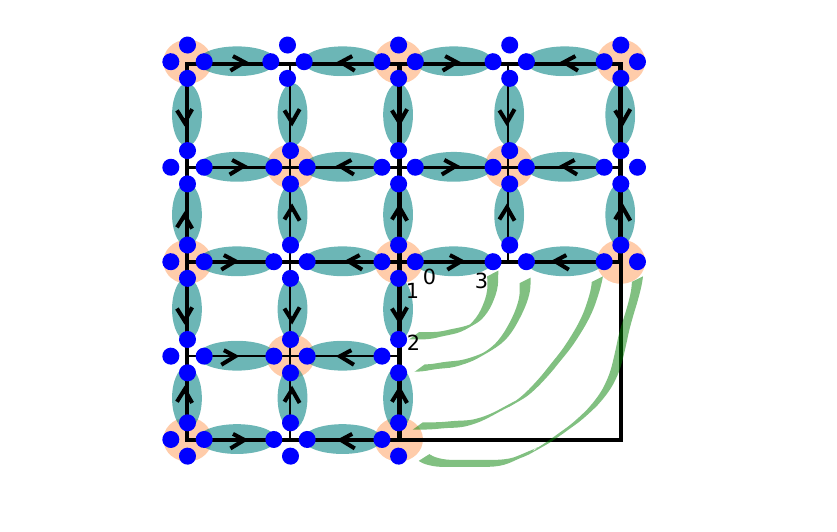}
    \caption{A 90 degree disclination defect in the Majorana lattice above. The green curves denote Majoranas that are equivalent in the projected Hilbert space (see Appendix \ref{appendix:gaugecoupling}). In particular, $c_2 = c_3=c'$ on the defect Hilbert space. This means that the rightward and downward bond from the singular vertex combine into a single term $-\frac{i}{2} (c_0+c_1)c'$ in the Hamiltonian, meaning $c_0 - c_1$ is an unpaired Majorana zero mode. Likewise one can show a 180 degree disclination carries two unpaired Majorana zero modes. These can't be paired in a symmetric way. However, we expect that a fusion of four 90 degree disclinations will have four unpaired Majorana zero modes which can be paired in a $C_4$ symmetric way.}
    \label{fig:majoranalatticedefect}
\end{figure}

If we add to the ordinary TI a reflection symmetry $R$, then the new cobordism group we need to compute is $\Omega^4_{\rm spin}(U(1) \rtimes \bZ^T_2 \times \bZ^R_2,\xi \otimes \sigma)$, where $\sigma$ is the sign rep of $\bZ_2^R$ acting on $\bR$. This group contains at least a $\bZ_2 \oplus \bZ_8$, with the first being the ordinary TI invariant and the second being the $\eta$ invariant of the $Pin^{c+}$ structure made out of the reflection symmetry. Indeed, a system with a $U(1)$ symmetry and a reflection symmetry is a lot like the $U(1) \times \bZ^{CT}_2$ TI. Indeed, for our classification there is no difference in the group data. As is well known, this has a $\bZ_8$ invariant. So we know these are cobordism invariants, we just have to determine whether they're non-trivial. From Ref \onlinecite{BG}, we know that in Pin$^{c+}$ bordism, $\mathbb{CP}^2$ and $\mathbb{RP}^4$ are independent. Further, they each also have a Pin$^{\tilde c+}$ structure, so they define independent classes in $\Omega^4_{\rm spin}(U(1) \rtimes \bZ^T_2 \times \bZ^R_2,\xi \otimes \sigma)$. We know from, eg. Ref \onlinecite{KTTW} that the $U(1)\rtimes T$ invariant is order 2 on both $\mathbb{CP}^2$ and $\mathbb{RP}^4$ and the $U(1) \times CT$ invariant is order 2 and order 8 on them respectively. This proves that $\Omega^4_{\rm spin}(U(1) \rtimes \bZ^T_2 \times \bZ^R_2,\xi \otimes \sigma) \supset \bZ_2 \oplus \bZ_8$. In particular, since all the free fermion phases are classified by cyclic groups (see Ref. \onlinecite{SS}), some of these must be inherently interacting fermionic phases.

We can of course also consider situations where the space group $G$ is a non-trivial extension of time reversal (there are 1651 magnetic space groups in 3d\cite{Litvin}) and charge symmetries. This gives the possibility of an endless zoo of topological superconductors and insulators, most of which will likely be beyond free fermions!

We can also go beyond cohomology in just bosonic phases. For phases with orientation-reversing symmetries this is very important. As we learned from Ref \onlinecite{K}, we need to consider the unoriented bordism group. If $K$ denotes the subgroup of $G$ whose symmetries are orientation-preserving, then by the Crystalline Equivalence Principle the most general bosonic phases on spacetime $\bR^D$ will be classified by $\Omega^D_O(K)$. For these we are allowed to use the Stiefel-Whitney classes in our topological terms \cite{K,R}.

\subsection{Coupling a QFT to TQFT and sigma models}\label{sigma}

One very important use of (invertible) topological field theories besides classifying (short-range-entangled) gapped Hamiltonians is in forming twisted versions of dynamical theories. Indeed, the partition function of a quantum field theory often splits into a sum over topological sectors
\[Z = \sum_\alpha Z(\alpha).\]
A usual example is when $Z$ is a gauge theory and $\alpha$ is the topological class of the gauge bundle. In this situation, the theory may be twisted by an SPT $\omega$ to obtain a new theory
\[Z \mapsto \sum_\alpha Z(\alpha) \exp \int \omega(\alpha).\]
In \cite{ReadingBetweenTheLines,KapustinThorngren, KapustinSeiberg} it was described how these twists change the categories of line and surface operators in the gauge theory. In \cite{ScaffidiParkerVasseur} this technique was employed to symmetry breaking critical points to define interesting CFTs.

Another example of a class of theories with topological sectors are the ``sigma models." A sigma model on a $D$-dimensional spacetime $M^D$ has an $n$-dimensional target space $X^n$ and counts among its degrees of freedom a map $\sigma:M \to X$. The partition function splits as a sum over homotopy classes of this map. When $n>0$, such models are typically gapless and depend on a choice of metric on $X$. Indeed, a typical Lagrangian is proportional to the volume of $\sigma(M)$. However, it is also possible to study \emph{topological} sigma models, which depend only on the homotopy class of $\sigma$. An invertible such theory, ie. one whose partition function is just a phase $\Omega([\sigma]) \in U(1)$ can be used to define the twisted sigma model partition function
\[Z \mapsto \sum_{[\sigma]} \Omega([\sigma]) \int_{\sigma \in [\sigma]} D\sigma \exp i S,\]
where $[\sigma]$ denotes a homotopy class of map $\sigma:M \to X$ and $S$ is the action of the dynamical sigma model (and we have suppressed other degrees of freedom).

The theories $\Omega([\sigma])$ are among the ones we have already studied in the case $G = 1$. We have described how those $\Omega([\sigma])$ which can be writted as an exponentiated integral of a local density $\exp\left(\int_M \sigma^*\omega\right)$ come from $\omega \in H^D(X,U(1))$. However, in light of the previous section, we should expect that to account for all possible twists, we should use the cobordism of $X$ instead: either $\Omega_{SO}^D(X)$ in the bosonic case or $\Omega_{Spin}^D(X)$ in the fermionic case.

For example, if $X = S^2$ and $D = 3$, there is a $\bZ_2$ possibility of topological terms for a fermionic system, classified by $\Omega^3_{spin}(S^2) = H^2(S^2,\Omega^1_{spin}) = H^2(S^2,\bZ_2) = \bZ_2$, while for bosons there is no non-trivial topological term. This topological term was discussed in \cite{AbanovWeigmann,FKS} and references therein. When $M = S^3$, it equals minus one to the Hopf index of the map $\sigma:S^3 \to S^2$.

It is interesting to include $G$ symmetry in the picture as well. Suppose first that $G$ is an internal symmetry, in that it doesn't act on the spacetime $M$, but does act on the target space $X$. When we couple to a background $G$-gauge field, $\sigma$ becomes a map to $X//G$. Recall this is a fiber bundle over $BG$ with fiber $X$. The gauged model will be topological in the $BG$ directions but may be dynamical in the fiber direction. Anyway, it can be twisted by a topological sigma model with target $X//G$. These are the theories we have studied, and are classified by $\Omega_{SO}^D(X//G)$ in the bosonic case and $\Omega_{spin}^D(X//G)$ in the fermionic case.\footnote{There is also the possibility of adding a characteristic bundle $\xi$ to modify the tangent structures involved, \`a la Section \ref{space_dep_fermions}.}

For example, if $X = S^2$, and $D=2$, with $G = \bZ_2$ acting by the antipodal map of $S^2$, $S^2//\bZ_2 = \mathbb{RP}^2$ and $H^2(\mathbb{RP}^2,U(1)) = \bZ_2$. This can be understood beginning from the ungauged model, which has an integer topological invariant, the degree of the map $M \to S^2$. The only $G$-symmetric values of the corresponding $\theta$ angle are $\theta = 0, \pi$, corresponding to this $\bZ_2$ cohomology. The special point $\theta = \pi$ coincides with the spin-1/2 Haldane chain\cite{HALDANE1983464}.



\section{Beyond crystalline topological liquids}\label{sec_beyond_tcl}

In section \ref{sec_space_dep_tqft}, we discussed a picture of a limit of crystalline topological liquids which looks like cells occupied by topological orders and boundaries between cells carrying invertible topological defects. It is interesting to consider the generalization where these domain defects are not necessarily invertible. For instance, a chiral Chern-Simons theory forms a topological defect between a Walker-Wang model and a trivial vacuum. The groundstate degeneracy prevents this defect from being invertible, indeed, its tensor product with its parity-inverse is a non-chiral Chern-Simons theory which carries some ground state degeneracy and therefore is not isomorphic to the trivial defect.

In order to account for the non-invertible of these defects, we will need to co-orient the boundaries between cells of $X$ (as well the higher codimension junctions). Along with an orientation of the ambient space(time) $X$, this defines an orientation of all the domain walls, allowing them to support parity-sensitive theories like Chern-Simons. We call a space $X$ with a cellular decomposition and co-orientations of all cell junctions a directed space. One can study spatially-dependent TQFTs over a directed space $X$.

A directed space can be most concisely discussed as a sort of category, although this involves passing to the dual cellulation with orientations. For example, a circle with a directed edge can be thought of as a category with one object and an endomorphism for every \emph{nonnegative} integer. This integer tells how many times one winds around the circle but we can only go one way. For example, if $X = S^1_t \times \bR^2$ with a preferred time direction, then it would be possible to describe a family of TQFTs which says the system pumps a toric code every time step, since now time evolution need not be invertible. Perhaps such a thing is possible in a system with an emergent arrow of time! We could also have the circular direction be a direction of space and this is a quotient of the stack of toric codes we've talked about before. So this exotic phase is not beyond TQFT afterall, but the directed space structure is very important. This tells whether we stack toric codes or the CPT conjugate of toric codes. These edge directions in space might be used to describe ``fracton" excitations \cite{Vijay2016,Haah} that can only move in certain directions. Whether more general fracton phases can be encoded this way, we don't know, but while we suspect it is possible, it is probably not enlightening.

There are some sorts of directed space structures which have already appeared in physics. An important one is the Kastelyn orientation of a surface. These are (relative to a dimer configuration) equivalent to a choice of spin structure. Perhaps spatially-dependent TQFTs on Kastelyn oriented surfaces have something to do with spin TQFTs. One can think about the maximum sort of this structure: a branching structure. It seems that a branching structure plays the role of a framing in the Baez-Dolan-Lurie cobordism theorem. Indeed, it gives one all the proper orientations to decorate the cell structure of $X$ with line operators, surface operators, ... and co-orient them properly so that they can be consistently fused.

\section{Open problems}
\label{open}
In this work we have presented a general framework for understanding the classification of interacting topological crystalline phases, for both bosons and fermions. An important question for future work is to understand the physical signatures of these phases.

The classic signature of an SPT phase is the protected gapless modes on the boundary (though in strongly interacting systems the boundary can also spontaneously break the symmetry or be topologically ordered). One would expect similar statements to hold for crystalline SPT phases, but there are some caveats. Firstly, of course, a boundary will in general explicitly break the spatial symmetry down to a subgroup, and one only expects protected modes when the phase is still non-trivial with respect to this subgroup. But even then there are exceptions. For example, an SPT protected in 1-D by inversion symmetry about $x=0$ does \emph{not} have a protected degeneracy when placed on the interval $[-L,L]$, even though the entire boundary (comprising two points) is in fact invariant under the symmetry \cite{Pollmann2010}. Another example is a phase in 2-D with a $C_4$ rotation symmetry, which can be constructed using the techniques of Ref.~\cite{SHFH}. A ground state in this phase is equivalent by a local unitary to a product state, with a $C_4$ charge pinned to the origin; therefore, there will not be any non-trivial edge states for any choice of boundary.
Thus, it is still an open question to determine what is the criterion which ensures protected boundary modes. A way to answer this would be to extend our spatial symmetry gauging procedure to systems with boundary. This is, however, beyond the scope of the present work.

Another question is the robustness of the topological crystalline phases that we have found to disorder, which explicitly breaks the spatial symmetries. There are some topological crystalline phases which have been argued to be robust to disorder, so long as the spatial symmetry is respected \emph{on average} \cite{Chiu2016}. It would be interesting to determine the general circumstances under which this happens.

\begin{acknowledgments}
We thank M. Barkeshli, P. Bonderson, N. Bultnick, X. Chen, L. Fidkowski, M. Hermele, M. Metlitski, C. Nayak, S. Galatius, B. Ware, D. Williamson and C. von Keyserlingk for helpful discussions. We also thank M. Hermele for drawing our attention to a mistake in a previous version of Table I. RT is supported by an NSF GRFP grant. DVE is supported by the Microsoft Corporation. This work was conducted in part at the Boulder Summer School for Condensed Matter and Material Physics, which is supported by the NSF and the University of Colorado; and at the Kavli Institute for Theoretical Physics, which is supported by NSF grant no.\ NSF PHY-1125915.
\end{acknowledgments}

\begin{appendices}

\section{Topological Terms vs. Topological Phases}\label{topphases}

Let us discuss Dijkgraaf and Witten's construction of the topological action in more detail. The important observation is the existence of a classifying space $BG$: homotopy classes of gauge fields for discrete internal symmetry group $G$ are in one-to-one correspondence with continuous maps from $M \to BG$. $BG$ may be constructed as a quotient $EG/G$ where $EG$ is any contractible space on which $G$ acts freely, ie. by translations. $EG$ and hence $BG$ are often infinite dimensional, in fact necessarily so if $G$ has any torsion. This allows there to be nontrivial SPT phases in infinitely many dimensions. From this definition, one easily verifies that the homotopy class of $BG$ is determined only by the group $G$. Moreover, the cohomology of the group can be defined to be the cohomology of the classifying space (but see below for some subtleties for non-finite groups) $H^*(BG)$. \footnote{This correspondence can be used to connect many algebraic facts about groups to topological properties of the classifying space and vice versa. For example, a representation $G \to O(n)$ corresponds to a $\bR^n$ vector bundle over $BG$. The characteristic class $w_1$ of this bundle determines whether it is an orientable bundle. The characteristic class $w_2$ determines whether it lifts to a spin representation $G \to Spin(n)$. Furthermore, computing the cohomology of $BG$ is often easier than attempting a direct calculation of the group cohomology of $G$.}

Given $\omega \in H^{D}(BG,U(1))$ on $BG$, we write the action
\begin{equation}
\label{dw_action}
S[A] = 2\pi \int_M A^*\omega,
\end{equation}
where $A$ is a gauge field on $M$ considered as a map $A:M \to BG$ and $A^*\omega$ is the pullback of forms. Because $\omega$ has coefficients defined only mod integers, $S[A]$ is defined modulo $2\pi$ times integers. This is because the physical quantity is really $e^{ i S}$.

For infinite discrete groups, one cannot subsequently sum over the gauge field in the path integral to obtain a gauge theory because there are infinitely many gauge equivalence classes of gauge field. However, it still makes sense to consider these gauge fields as backgrounds, if not dynamical, in which case again the topological terms which can be written as local integrals of just the gauge field are classified by 
$H^{D}(BG,U(1))$. 

An important caveat is that the topological terms in $H^{D}(BG,U(1))$ do not necessarily correspond to topological phases. For example, with $U(1)$ symmetry in 3+1d one can have a theta angle for the $U(1)$ gauge field, but this does not mean there is an entire circle of different topological phases. Instead, the existence of the theta term descends to the existence of the Chern-Simons term in 2+1d, whose coefficients are \emph{quantized}, as phases of matter should be. The famous relationship is
\begin{equation}\label{CSFF}
dCS(A) = F \wedge F,
\end{equation}
where $CS(A)$ is the Chern-Simons form of $A$.

For connected symmetry group $G$ like $U(1)$ or $SU(2)$ (and indeed for any group) the proper cohomology to consider actually lives \emph{one dimension higher than spacetime} and over the integers: $H^{D+1}(BG,\bZ)$. Let us remark that physicists often prefer\cite{CGLW} to write the classification as $\mathcal{H}^D(G, U(1))$, where $\mathcal{H}^{*}$ denotes the ``measurable'' or ``Borel'' group cohomology. This can be proven to be isomorphic to $H^{D+1}(BG, \bZ)$.

For finite $G$, $H^D(BG, \bZ)$ coincides with $H^{D}(BG,U(1))$. The problem is infinite order elements in the homology of $BG$ with which one can define theta angles. These were encountered in Refs \onlinecite{K} and \onlinecite{KTTW} and the solution there was to use only the torsion part of homology: $H_{D}(BG)^{tors}$ and consider maps from that group to $U(1)$: ${\rm Hom}(H_{D}(BG)^{tors},U(1))$. Since abelian groups always split between their torsion and free part, there is an inclusion of this into $H^{D}(BG,U(1))$ but more importantly also into $H^{D+1}(BG,\bZ)$ by measuring the winding number of the form, much like in Eq \ref{CSFF}. These are the sort of gappable-edge SPT phases encountered in studying finite symmetries. The non-torsion pieces in homology define things like Chern-Simons terms, which are descended from 4d integer characteristic classes like the first Pontryagin class $p_1$  or second Chern class $c_2$ and their corresponding theta terms $\theta p_1$, $\theta c_2$. Our belief is that the universal coefficient sequence \cite{Hatcher}
\begin{widetext}
\begin{equation}
\label{Pcomm}
\begin{tikzcd}
 H^{D+1}(BG,\bZ)\arrow{r}{} & {\rm Hom}(H_{D+1}(BG),\bZ) \simeq H_{D+1}(BG)^{free} \\
{\rm Hom}(H_{D}(BG)^{tors},U(1)) = {\rm Ext}(H_{D}(BG),\bZ) \arrow{u}{} &
\end{tikzcd}
\end{equation}
encodes a physical short exact sequence with the same functorial and splitting properties
\[{\rm \{torsion\ phases\}} \to {\rm \{topological\ phases\}} \to {\rm\{non-torsion\ response\}}.\]
\end{widetext}
This is discussed also in Ref \onlinecite{KTTW} where the inclusion of infinite order homology elements in shifted degree corresponds to thermal Hall-type gravitational response and Ref \onlinecite{FH} where it is related to the Baez-Dolan-Lurie cobordism theorem and the Anderson dual of the sphere spectrum, which in topology plays the role of $U(1)$, the Pontryagin dual of $\bZ$.

\begin{table*}
{\footnotesize
\input{classif_pg1.txt}
}
\caption{\label{bosonic3d} The ``230-fold way''. This table shows the classification of bosonic crystalline SPT phases in (3+1)-D for each of the 3-D space groups. For space groups 227, 228 and 230 the classification has not been computed.}

\end{table*}

\addtocounter{table}{-1}

\begin{table*}
{\footnotesize
\input{classif_pg2.txt}
\caption{(continued)}
}
\end{table*}

\section{Computing the bosonic classification}
\label{sec_bosonic_classif}
A nice feature of our results, at least in the case of bosonic crystalline SPTs (in Euclidean space) is that the classification is readily computable. According to the general discussion of Section \ref{sec_topological_actions}, we see that the classification in $d$ space dimensions for a given space group $G$ is given by $H^{d+2}(BG, \mathbb{Z}^{\mathrm{or}})$. Computing this object turns out to be within the capabilities of the GAP computer algebra program \cite{GAP4}. We show the results in Table \ref{bosonic2d} (in the introduction) for the (2+1)-D case and in Table \ref{bosonic3d} for the (3+1)-D case. There were 3 space groups in (3+1)-D for which the classification took too long to compute and is not shown.

We recall that this classification is expected to be \emph{complete} in (2+1)-D, and for the Sohncke groups (those not containing any orientation-reversing elements) in (3+1)-D. What about explicit constructions of these phases? Let us fix some element $\omega \in H^{d+2}(BG, \mathrm{Z}^{\mathrm{or}})$. Suppose that there exists a finite group $G_f$ and a group homomorphism $\varphi : G \to G_f$ such that $\omega$ is in the image of the map $\mathcal{H}^{d+1}(G_f, \mathrm{U}(1)^{\mathrm{or}}) \cong H^{d+2}(BG_f, \mathbb{Z}^{\mathrm{or}}) \to H^{d+2}(BG, \mathbb{Z}^{\mathrm{or}})$ induced by $\varphi$. Then indeed we have an explicit construction of the crystalline SPT correponding to $\omega$, using the bootstrap argument of Section \ref{sec_bootstrap} (leveraging, for example, the construction of Ref.~\onlinecite{CGLW} for the SPT protected by $G_f$ acting internally). We conjecture that there will always be some such $G_f$ for any element of $H^{d+2}(BG, \mathbb{Z}^{\mathrm{or}})$.

\section{Coupling a Hamiltonian to a rigid crystalline gauge field}\label{appendix:gaugecoupling}

In this appendix, we explain how to couple a finite range Hamiltonian to a crystalline gauge field. To fix notation, $X$ will be the physical space with $G$ action, $\Lambda$ the crystalline lattice therein, $M$ the test space, divided into patches $\bigcup_i U_i = M$ with local homeomorphisms $f:U_i \to X$ and transition functions $g_{ij} \in G$ such that for all $x \in U_i \cap U_j$, $f_i(x) = g_{ij} f_j(x)$. We will use the shorthand $A$ to denote the whole crystalline gauge field.

We begin by defining the Hilbert space on $M$, assuming that the Hilbert space of $X$ is local to the lattice $\Lambda$, that is, there is a space $\cH_x$ for every $x \in \Lambda$ and $\cH_X = \bigotimes_{x \in \Lambda} \cH_x$. We define the pulled-back lattice $\Sigma = \bigcup_j f_j^{-1}\Lambda$ and assign to each $m \in f_j^{-1}\Lambda$ the Hilbert space $\cH_m(A) := \cH_{f_j(m)}$. The total Hilbert space may be written $\cH(A) = \bigotimes_{m \in \Sigma} \cH_{m}(A)$.

Next we discuss (rigid) gauge transformations. These come in three sorts. The first are homotopies of the maps $f_j$ (fixing the boundary). We suppose that the patches are transverse to the lattice (this is generic) so that each $m \in \Sigma$ lies in a unique $U_j =: U_{j(m)}$. In the rigid case, these are simply continuous deformations of the lattice in $M$\footnote{In the non-rigid case, new lattice sites could appear or disappear in conjugate pairs by creating ``folds" of $f_j$.}.

The second type are given by the action of a group element $g_j \in G$ on a $U_j$ and are analogous to ordinary gauge transformations. To define these, we need to assume the symmetry action on $\cH_X$ is ``ultralocal", meaning that it is a tensor product operator $U(g) = \bigotimes_{x \in \Lambda} U(g)_x$ where $U(g):\cH_x \to \cH_{gx}$. Then we can isolate the part acting on $f_j(U_j)$, $U(g_j)_j = \bigotimes_{x \in \Lambda \cap f_j(U_j)} U(g)_x$ and apply this to $\cH(A)$. This takes us to a different Hilbert space $\cH(A^{g_j})$, where $A^{g_j}$ is the crystalline gauge field obtained from $A$ by replacing $f_j$ with $g_j f_j$ and $g_{ij}$ with $g_{ij} g_j^{-1}$ for all adjacent $U_i$ to $U_j$.

The third type involve moving the patches themselves. This is actually a combination of the previous type of gauge transformation as well as splitting or joining patches. A patch $U$ becomes split into $U_{1} \cup U_{2}$ with $f_{1}, f_{2}$ defined by restricting $f$ and $g_{12} = 1$. Likewise, if there are every any adjacent patches $U_{i,j}$ with $g_{ij} = 1$, then $f_i$ and $f_j$ can be joined to a continuous function across both patches which can then be considered a single patch $U_i \cup U_j$. In both cases the adjacent transition functions do not change. Moving a domain wall can then be achieved by first splitting a patch, applying a $G$ element to the new patch, and joining patches again.

Now we discuss how to couple a Hamiltonian to this crystalline gauge field. For each $m \in \Sigma$ and each term $h$ in the Hamiltonian $H$ acting on $f_j(m)$, we will have a corresponding term in the Hamiltonian $H(A)$ acting on $\cH(A)$. If the support of $h$ lies entirely inside $f_j(U_j)$, then it acts on $\bigotimes_{x \in f_j(U_j) \cap \Lambda} \cH_x = \bigotimes_{m \in U_j \cap \Sigma} \cH_m$, which is a tensor factor of $\cH(A)$ so we can include $h$ in $H(A)$ with no issue. 

Difficulty comes when the support of $h$ is not contained inside any one $f_j(U_j)$. This is where we have to use the rigidity assumption. We assume that it is possible to move the patch $U_j$ by a gauge transformation so that $h$ is contained in $f_j(U_j)$ (the Hamiltonian built so far comes along for the ride according to our gauge transformation operator). Then we add $h$ to the Hamiltonian and perform the inverse gauge transformation to return to the original gauge field configuration. Compare Appendix \ref{latticegaugefields}, especially Fig \ref{fig:Ghampullback}.

As a simple example of this technique, consider a 1+1D spin-1/2 Ising model, focusing on a specific edge $12$ with Hamiltonian term $X_1 X_2$ and global $\bZ_2$ symmetry $\bigotimes_j Z_j$, where $X, Z$ denote Pauli spin operators. Suppose that 1 and 2 belong to different patches with a non-trivial transition function. Then rather than adding $X_1 X_2$ to the Hamiltonian, we first perform a gauge transformation $Z_2$, which pushes the domain wall off to the right and we get the term $-X_1 X_2$. Note because $\bZ_2$ is a symmetry, it doesn't matter which way we push the domain wall off. Using $Z_1$ would result in the same term.

\begin{figure}
    \centering
    \includegraphics{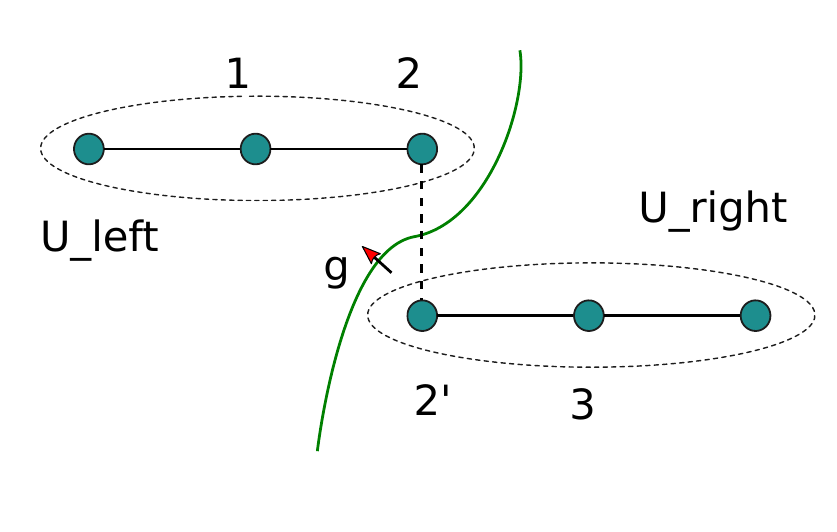}
    \caption{In this approach to defining the Hamiltonian coupled to crystalline gauge field, patches are allowed to overlap to include some vertices. In this particular example, $U_{left} \cap U_{right}$ includes vertex 2, which gets duplicated. Hamiltonian terms (denoted by solid edges) lying entirely inside $U_{left}$ or $U_{right}$ are taken to act on those Hilbert spaces. Then spurious degrees of freedom are eliminated by applying a projection operator which in a product state basis identifies the state at 2 with $g$ applied to the state at 2'. This is indicated by the green curve labelled by $g$ cutting the dashed vertical line from 2 to 2'.}
    \label{fig:splittingpatches}
\end{figure}

We end this appendix with a second method for describing the Hamiltonian coupled to a crystalline gauge field, which is equivalent but does not require one to perform gauge transformations to obtain all the terms in the Hamiltonian. In this version, the patches $U_j$ are taken to be an open covering of $M$ and are allowed to overlap. Then a lattice (hence a Hilbert space) is first defined on the disjoint union $\bigsqcup_j U_j$ by $\tilde\Sigma := \bigsqcup_j f_j^{-1}\Lambda$. We denote the associated Hilbert space $\cH_{\tilde M} = \bigotimes_j \cH_{U_j}$, where $\cH_{U_j} = \bigotimes_{m \in \Sigma \cap U_j} \cH_m$. Note that the map $\bigsqcup_j U_j \to \bigcup_j U_j  = M$ sends $\tilde \Sigma$ to $\Sigma$. Then rigidity means that for each $m \in \Sigma$, and for each term $h$ acting on $f(m)$, there is some $U_j \ni m$ such that the support of $h$ is contained in $f_j(U_j)$. We choose $h$ to act on the $U_j$ part of the Hilbert space $\cH_{\tilde M}$. Then we project everything to $\cH_M$ by identifying duplicated vertices $m \in U_j, m' \in U_k$ in the disjoint union by the transition maps $U(g_{ij}):\cH_{U_j} \to \cH_{U_k}$. A simple example is shown in Fig \ref{fig:splittingpatches}.

This method is particularly convenient for describing crystal defects. In the case of a single defect in $\bR^d$ supported along $\partial H$, where $H$ is a $d-1$-dimensional branch cut (which, fixing $\partial H$, is a choice of gauge), the defect space $M = \bR^d - \partial H$ can be covered with a single patch $U$ given by a thickening of $\bR^d - H$, which intersects itself in $M$ along a neighborhood of $H$. In other words, the degrees of freedom near the branch cut are doubled (see Figs \ref{fig:Gtwistedmap1}, \ref{fig:disclinationsemion}, and \ref{fig:majoranalatticedefect}), coupled to either side of the branch cut, and then reglued by a projection map twisted by the crystal symmetry.

\section{Coupling smooth states to gauge fields}
\label{appendix_smooth}
Here we prove the claims made in Section \ref{sec_tcl} about the well-definedness of the construction to couple smooth states to gauge fields. 
We first consider the case of an internal symmetry $G$. We adapt an argument due to Kitaev (Appendix F of Ref.~\onlinecite{Kitaev2006}). We assume that our original ground state $\psi$ lives on a lattice with a spin of Hilbert space dimension $d$ at each site. However, we will define the space $\Omega$ which our smooth states target to be the space of states with Hilbert space dimension $m > d$ per site. Of course, given a choice of isometric embedding $e : \mathbb{C}^d \to \mathbb{C}^m$, we could think of our original state $\psi$ as living in $\Omega$ too. The resulting state depends on $e$ and we call it $e(\psi)$.

Recall that the symmetry is assumed to act on-site, with the action on each site described by a representation $u(g) \in U(d)$. For each $g \in G$, we also considered a path $u(g;t)$, $t \in [0,1]$ such that $u(g;0) = \mathbb{I}$ and $u(g;1) = u(g)$. Then, (at least locally) we can reformulate the prescription in Section \ref{sec_tcl} for defining the smooth state $\psi[A] : M \to \Omega$ as follows in terms of a \emph{spatially-dependent} isometric embedding $e_m : \mathbb{C}^d \to \mathbb{C}^m$, according to $\psi[A](m) = e_m(\Psi)$. We then require that when passing over a patch boundary twisted by $g \in G$, $e_m$ goes through the continuous path obtained by acting with $u(g;t)$. But now we see that there will not be any obstructions to making this process well-defined due to non-contractible loops (or higher non-trivial homotopy groups)  at intersections between patch boundaries, provided that we take $m$ sufficiently large. This is because in the limit $m \to \infty$ the space $\mathrm{Emb}(d,m)$ of all isometric embeddings $\mathbb{C}^d \to \mathbb{C}^m$ is contractible, i.e. all its homotopy groups are trivial.

A more rigorous (and succinct) way to think about the above construction is obtained by thinking about the classifying space $BG$. Indeed, since $EG := \mathrm{lim}_{m \to \infty} \mathrm{Emb}(d,m)$ is a contractible space with a free action of $G$, it follows that $EG/G$ is a model for $BG$, and we find that there is a continuous map $BG \to \Omega$. A $G$ gauge field over $M$ is the same as a principal $G$-bundle over $M$, which can be represented by a a continuous map $M \to BG$. Hence, composing these two maps gives a smooth state $\psi[A] : M \to \Omega$.

The ``patch'' version of the argument for a crystalline gauge field proceeds similarly to above and we will not write it out again. Let us simply note that a rigorous version of the construction can be formulated in terms of the homotopy quotient $X//G$. Indeed, given a smooth state $\psi : X \to \Omega_d$ (where $\Omega_d$ is the space of ground states with Hilbert space dimension $d$ per site), there is a map from $X \times \mathrm{Emb}(d,m) \to \Omega$ defined by $(x,e) \mapsto e(\psi(x))$. This map is invariant under the diagional action of $G$. Therefore, taking the limit $m \to \infty$, we find a map from $(X \times EG)/G = X//G \to \Omega$. A crystalline gauge field on $M$ can be represented by a map from $M \to X//G$. By composing these two maps we obtain a smooth state $\psi[A] : M \to \Omega$.

\section{Lattice Crystalline Gauge Fields}\label{latticegaugefields}

The cellular description we give in this section is dual to the patch picture we gave in Section \ref{sec_spatial_gauge_fields}, where $g$ elements labelled codimension 1 walls between volumes in the crystal. Here in order to compare with the usual definition of a lattice gauge field, we label edges with $g$ elements.

Recall for a discrete group $G$ a lattice gauge field has a very nice description where each edge $e$ gets a group label $g_e \in G$ and any 2-face $\tau$ imposes a flatness constraint
\begin{equation}\label{flatness}
\prod_{e \in \partial \tau} g_e = 1,
\end{equation}
where the multiplication is performed in the order the edges are encountered in a circular traversal of the boundary. This conservation law allows us to express these labels as a configuration of domain walls running about our manifold. The conservation law says that a $g_1$ and a $g_2$ fuse to a $g_1 g_2$. The domain walls are codimension one so fusion can be non-commutative in this way.

\begin{figure}
    \centering
    \hspace*{-1cm}\includegraphics[width=8cm]{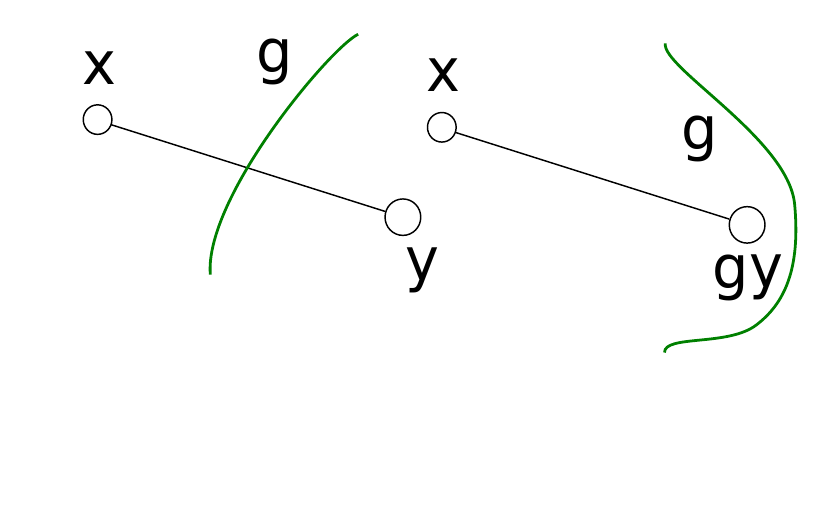}
    \caption{When the $g$ domain wall is pulled off of this edge, it is revealed to be an edge from $x \to gy$. Note the similarity with the Hamiltonian coupling procedure in Appendix \ref{appendix:gaugecoupling}.}
    \label{fig:edgecut}
\end{figure}

Let's imagine drawing a configuration like this on $X$ where the $G$ elements act non-trivially on $X$. Does this make sense? Let's look at a particular edge, Fig \ref{fig:edgecut}. It looks like an edge from $x \to y$, but if we push the domain wall out of the way, we see the actual data there is an edge (actually path; see below) from $x$ to $gy$! This means that while our underlying manifold has points labelled by points in $X$, it is perhaps a different space $M$! To see what data is assigned to a face or higher facet, one performs a similar procedure, pushing all the domain walls off and collecting $g$ labels. The flatness condition on $G$ implies that this is always unambiguous. At a symmetry defect like the core of a disclination, the flatness condition is violated and it is impossible to unambiguously assign a face of $X$ to the core of the defect. When this happens, the underlying space $M$ may have different topology from $X$! In fact, we may end up with a space $M$ whose labels don't even close up unto a map to $X$! In such a case, we end up with only a map $P \to X$, where $P$ is the $G$-cover corresponding to the $g$ labels (equivalently the $G$ gauge bundle).

Note that if $X$ is contractible the extra information beyond the $G$ gauge field, the $X$ labels, contributes no non-trivial data up to homotopies of this map. Indeed this is basically another proof of the Crystalline Equivalence Principle.

Let us try to be more systematic about the construction. We start with a warm-up, just describing cellular maps $\hat f:M \to X$ in a lattice gauge theoryish way. A cellular map means the $n$-skeleton of $M$ gets sent to the $n$-skeleton of $X$ for every $n$. This means every vertex $m\in M$ gets a vertex $\hat f(m) \in X$, every edge $e:m_1 \to m_2 \in M$ gets a \emph{path} $\hat f(e):\hat f(m_1) \leadsto \hat f(m_2) \in M$, every plaquette $\tau$ gets a chain $\hat f(\tau)$ with $\partial \hat f(\tau) = \hat f(\partial \tau)$, every volume gets a 3-chain with prescribed boundary and so on. This data describes a general partial covering $M \to X$ (i.e. a map which gives a rigid crystalline gauge field with trivial transition functions).

\begin{figure}
    \centering
\includegraphics[width=10
cm]{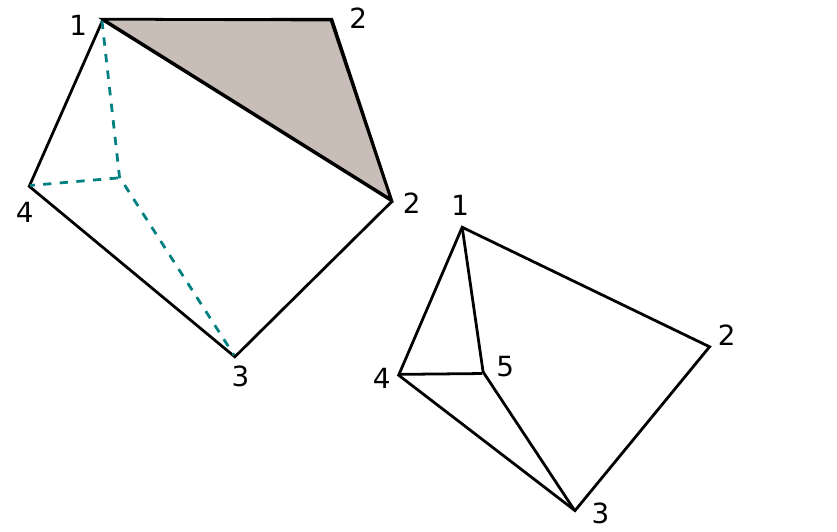}
    \caption{Here we depict of a piece of $M$ (northwest) mapping to a piece of $X$ (southeast). We have given the vertices of $X$ unique labels and labelled the vertices of $M$ with their image vertices in $X$. Note that vertex 2 has two adjacent preimages. This edge of $M$ is mapped to a degenerate edge and the triangle it lies on (grey) is mapped to a degenerate face 122 in $X$. Note also that vertex $5 \in X$ has no preimage and to map faces to faces we must refine the lattice of $M$, depicted by the dotted blue lines.}
    \label{fig:hampullback}
\end{figure}

To account for maps which are not locally homeomorphisms, we need to include in this definition the \emph{degenerate facets of $X$}. For example, if we had the constant map $M \mapsto x \in X$, this definition only makes sense if there is a hidden edge ${\rm id}:x \to x$, hidden faces $x\to x \to x$, $x \to x \to y$, and so on. All higher degenerate facets should be included as well.

This means that any map $\hat f:M \to X$ is homotopic to one given first by refinement of the lattice in $M$ and then by labelling vertices, edges, faces, ... of the refinement with vertices, edges, faces, ... (possibly degenerate ones) of $X$. This should be intuitive, since the cell structure in $M$ is not really physical. It's just a way to encode the topology of $M$ combinatorially.

\begin{figure}
    \centering
\includegraphics[width=10
cm]{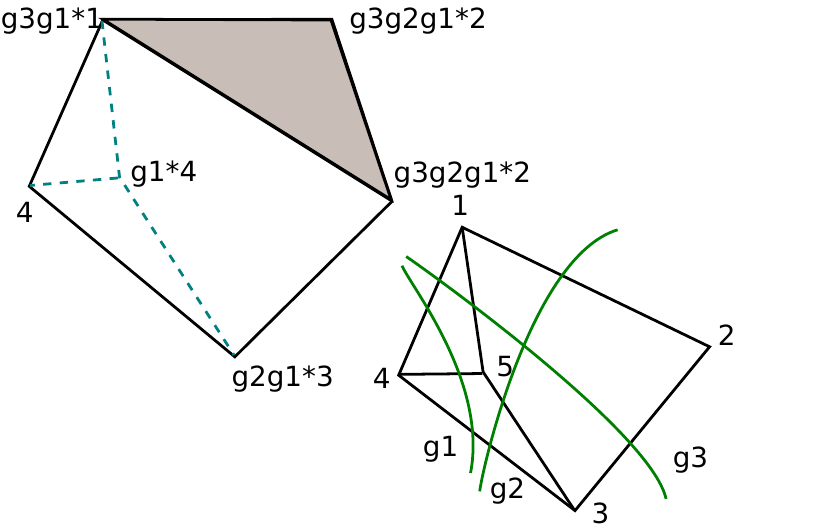}
    \caption{The conservation law for $G$ labels allows us to draw them as $G$ domain walls in $X$. Then in any contractible patch of $M$ we can describe our local map $M \to X$ by ``pushing off the domain walls". Then we look at the northwest picture of our patch in $M$. See how the vertices have been transformed; so have the edges. Then we fill in the transformed picture with faces of $X$ as we would in describing an ordinary map $M \to X$. This always requires a choice of basepoint. Here our basepoint is 4 and we have pushed all the domain walls (green) straight to the east. The choice of basepoint is like a local choice of gauge. It should be compared with the construction for coupling to Hamiltonians in Appendix \ref{appendix:gaugecoupling}.}
    \label{fig:Ghampullback}
\end{figure}

Now let us consider maps with $G$-twisted continuity conditions. As before we assign vertices of $X$ to vertices of $M$. Before to an edge in $M$ we would assign a path $x\leadsto y$ connecting the $X$ labels $x$ and $y$ of the endpoints. For $G$-fs, these paths can pass through domain walls, resulting in something we call a $G$-path:
\[x_1 \leadsto y_1 \xrightarrow{g_1} x_2 \leadsto \cdots \xrightarrow{g_k} y_k.\]
Around the boundary of a face $\tau \in M$, we get a $G$-path by concatenating the $G$-paths on each edge. Our conservation law
\[\prod_j g_j = 1\]
must be supplemented by the condition that the boundary $G$-path forms a $G$-loop:
\[y_k = x_1.\]
If this is the case, then we can push all the $g$'s to the right, acting on the paths as we do to obtain an honest path $x_1 \leadsto g_k^{-1}\cdots g_1^{-1} x_1$. If the $G$ conservation law holds then this path is a loop in $X$. This is just like pushing the domain walls off $\tau$ towards vertex $1$. We ask that $\tau$ be assigned a chain with boundary equal to this loop. A picture of this is depicted in Fig \ref{fig:Ghampullback}.

Now we discuss homotopies of this data (collapsible crystalline gauge transformations). Such a homotopy $A(0)\mapsto A(1)$ is itself a crystalline gauge field $A(t)$ but on the prism $M \times [0,1]$ with boundary conditions equal to $A(0)$ and $A(1)$ on each copy of $M$.

\begin{figure}
    \centering
\hspace*{-1.5cm}\includegraphics[width=10
cm]{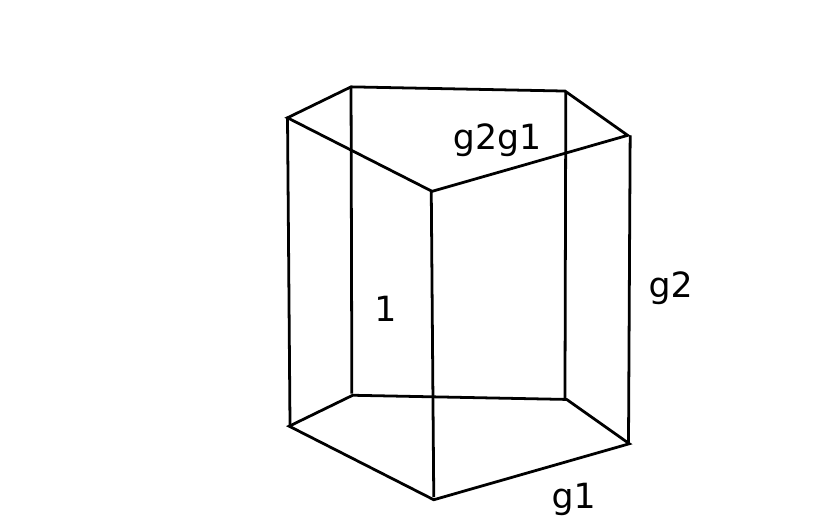}
    \caption{A prism $M \times [0,1]$ mapping to $BG$ means an assigning of $G$ labels also to the interior edges. These correspond with the vertices of $M$ so we can think of them as a function $g:M^0 \to G$ where $M^0$ is the set of vertices of $M$. Then the conservation law on the internal faces of the prism forces a constraint between corresponding edge labels in each $M$. The constraint reads that the top labels are the gauge transformation of the bottom labels by $g$. The direction is fixed by an orientation of the internal prism edges. If we reverse all of them, it takes $g \mapsto g^{-1}$ (and locally as well).}
    \label{fig:Gtransformation}
\end{figure}

As a first warm-up, let's just consider ordinary $G$ gauge fields. See Fig \ref{fig:Gtransformation}. There is a cell complex of $M \times [0,1]$ with one inner $p+1$-cell for every $p$ cell of $M$. These inner cells are the only ones where the boundary conditions do not fix the data. For an ordinary $G$ gauge field we must specify the $G$ labels on the inner edges. These correspond to vertices of $M$, so the data is like an element of $G$ for each vertex of $M$. The flatness condition on the inner faces determines how these must act on the edge variables.

A second warm-up, really getting going this time, is to consider homotopies of a map $M \to X$. This is the case with no symmetry, $G=1$. This gets quite complicated but it is possible to divide homotopies into elementary pieces, where all the inner $p$-cells but one are degenerate but one: $\tau_p$, meaning the map $M \to X$ does not change away from $\tau_p$. The map $h:M\times [0,1] \to X$ identifies $\tau_p$ with a $p$-chain $h(\tau_p)$ and because all other inner cells in $M\times [0,1]$ are degenerate, $\partial h(\tau_p)$ is divided into two $p-1$-chains in the image of the boundaries: $\partial h(\tau_p) = h(N_0) \sqcup h(N_1)$, where $N_j \subset M\times \{j\}$ are $p-1$-chains in $M$. In fact these are the same $p-1$-chains and $h(\tau_p)$ is telling us how they move inside $X$ during the homotopy $h$. A general gauge transformation of $A$ is essentially a combination of these two ingredients.

Just as the cellular description of $G$ gauge fields reflects a convenient cellular structure of $BG$, what we have described above amounts to a cellular structure on the homotopy quotient $X//G$. One can see what we've written as a simultaneous construction of $X//G$ and a proof of
\begin{thm}\label{homotopyquotient}
A crystalline gauge field is the same as a cellular map $A:M \to X//G$ with the cell structure induced by the action of $G$ on a compatible cell structure of $X$. Thus, gauge equivalence classes of crystalline gauge fields are the same as homotopy classes of maps $A:M\to X//G$.
\end{thm}
There is a nice way to get a handle on the homotopy type of $X//G$. Recall from, eg. Ref.~\onlinecite{Hatcher}, that $BG$, the classifying space for ordinary $G$-gauge fields and the special case of our construction when $X$ is a point, is itself constructed as an ordinary quotient $EG/G$, where $EG$ is some (usually very large) \emph{contractible} space on which $G$ acts freely. For discrete groups, $EG$ can be constructed as a simplicial complex where vertices are group elements $g \in G$, edges are pairs, triangles are triples, and so on. The gluing maps use the $G$ multiplication. For example, an edge $(g_0,g_1)$ is glued to $g_1$ and to $g_0g_1$; a triangle $(g_0,g_1,g_2)$ is glued to $(g_1,g_2)$, $(g_0,g_1g_2)$; and $(g_0g_1,g_2)$, and so on. This space has a $G$ action which acts on all the labels simultaneously. It's also contractible. The quotient structure is the usual structure on $BG$. Likewise, we can invent a cell structure on the space $EG \times X$ so that the quotient structure is the one we've described on $X//G$. This proves $X//G = EG \times X/G$ where $G$ acts diagonally. In fact, to preserve the homotopy type of $X//G$, we just need any space $EX$ which is homotopy equivalent to $X$ and on which $G$ acts freely. $EG \times X$ is an example, but if $G$ already acts freely on $X$, then $X$ itself is an example and the homotopy quotient reduces to the ordinary quotient $X//G = X/G$. In the other extreme, which $G$ is a purely internal symmetry, $X//G = BG \times X$.

\section{Explicit constructions for bosonic SPTs}
\label{appendix:CGLW_bootstrap}
In Ref.~\onlinecite{CGLW}, a prescription was given to construct a ground-state wavefunction for an SPT phase protected by a finite internal symmetry group $G_{\mathrm{int}}$. As stated in the main text, we want to leverage this construction in a ``bootstrap'' procedure to construct a wavefunction for an SPT phase protected by a spatial symmetry, as outlined in Section \ref{sec_bootstrap}. For our current discussion, the important requirement is that we must be able to choose the wavefunction to be invariant under \emph{both} an internal symmetry $G_{\mathrm{int}}$ and a spatial symmetry $G_{\mathrm{spatial}}$. Ultimately, the symmetry protecting the crystalline SPT phase will be the diagonal subgroup $G_{\rm phys}$. Recall that we take orientation-reversing elements of $G_{\rm spatial}$ to also act anti-unitarily, in accordance with the CPT principle. (Thus, the orientation-reversing symmetries in $G_{\mathrm{phys}}$ are a composition of two anti-unitary operators, and so end up being unitary.)

Let us briefly review the construction of Ref.~\onlinecite{CGLW}. This construction starts from an element of the group cohomology group $\mathcal{H}^{d+1}(G_{\mathrm{int}}, \mathrm{U}(1))$. This cohomology class is represented by a $(d+1)$-cocycle in homogeneous form, which is a function $\nu : G_{\mathrm{int}}^{\times d+1} \to \mathrm{U}(1)$ satisfying
\begin{align}
g\cdot \nu(g_1, \cdots, g_{d+1}) = \nu(gg_1, \cdots, gg_{d+1}) \quad \forall g \in G_{\mathrm{int}} \\
\prod_{i=0}^{d+2} \nu^{(-1)^{i}}(g_0, \cdots, g_{i-1}, g_{i+1}, \cdots, g_{d+2}) = 1,
\end{align}
where $g \cdot \nu$ denotes the action of $G_{\mathrm{int}}$ on $U(1)$, i.e.\ anti-unitary elements of $G_{\mathrm{int}}$ act by inversion.

To construct the wavefunction on some $d$-dimensional spatial manifold, one first chooses a triangulation of the manifold. The spins will live on the vertices of this triangulation, and they will each carry a Hilbert space with basis $\{ |g\rangle : g \in G_{\mathrm{int}} \}$, on which $G_{\mathrm{int}}$ acts by left-multiplication: $\ket{h} \xrightarrow{g} \ket{gh}$.
Then one chooses a branching structure, which is a choice of direction on the edges of the triangulation, such that there are no directed cycles on any $d$ simplex. A branching structure allows us to define an ordering of the vertices on any $d$-simplex.
The wavefunction of Ref.~\onlinecite{CGLW} is then defined as a superposition
\begin{equation}
\ket{\Psi} = \sum_{\{ g_i \}} \left( \prod_{\Delta} \alpha_\Delta\left(g_{\Delta}\right) \right) \ket{ \{ g_i \} },
\end{equation}
where the sum is over all configurations $\{ g_i \}$ of group elements $g \in G_{\mathrm{int}}$ for every vertex, and the product is over all $d$-simplices. The phase factor $\alpha_\Delta$ associated to a $d$-simplex $\Delta$ is defined by
\begin{equation}
\alpha_{\Delta}(g_\Delta) = \nu^{s(\Delta)}(g_*, g_1, \cdots, g_d),
\end{equation}
where $g_1, \cdots, g_d$ are the group elements living on the vertices of the simplex (ordered according to the branching structure),  $g_* \in G_{\mathrm{int}}$ is some fixed group element which is chosen to be the same for every $d$-simplex (the resulting wavefunction turns out not to depend on $g_*$ on any closed manifold); and $s(\Delta) = \pm 1$ is the orientation of the $d$-simplex (see Ref.~\cite{CGLW} for further details). It can be verified that the wavefunction $\ket{\Psi}$ so defined is indeed invariant under the action of $G_{\mathrm{int}}$.

\begin{figure}
\input{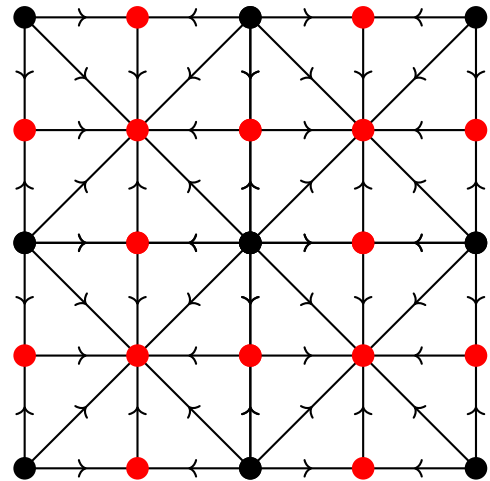}
\caption{\label{triangulation} A $p4m$-invariant triangulation and branching structure. The block dots are the vertices of the original $p4m$-invariant cellulation (the simple square lattice) and the red dots are the vertices that had to be added (through the barycentric subdivision) to get a $p4m$-invariant triangulation and branching structure.}
\end{figure}

Now it remains to show that $\ket{\Psi}$ can also be taken to be invariant under the action of a spatial symmetry $G_{\mathrm{spatial}}$. We take the 
action of $G_{\mathrm{spatial}}$ on the Hilbert space of the spins to be inherited from its action on the space manifold; that is, it simply permutes the spins. (For orientation-reversing elements of $G_{\mathrm{spatial}}$, this is followed by complex conjugation, in accordance with our stipulation that orientation-reversing elements of $G_{\mathrm{spatial}}$ should act anti-unitarily).
This will evidently be the case provided that the locations of the vertices, the triangulation, and the branching structure are all invariant under the action of $G_\mathrm{spatial}$. (For orientation-revering elements, note that the effect of the complex conjugation is cancelled by the reversal of the orientation of the simplices).
To achieve this, we can start from a $G_{\mathrm{spatial}}$-invariant cellulation of the spatial manifold (which can be obtained, for example, via the Wigner-Seitz construction), then take its barycentric subdivision, which gives a $G_{\mathrm{spatial}}$-invariant triangulation. Moreover, one can show that there is always a $G_{\mathrm{spatial}}$-invariant branching structure on this triangulation. The resulting triangulation and branching structure is illustrated in Figure \ref{triangulation} for the case $d=2$ and $G_{\mathrm{spatial}} = p4m$ (the symmetry group of the simple square lattice).

\section{Cocycles}\label{cocycles}

In this appendix, we delve a little deeper into what the equivariant cocycles mean. We will find they nicely formulate what it means for the cocycle (and hence the topological response) to be spatially-dependent. One can see this as another derivation of the equivariant cohomology classification. Importantly, we will learn how to integrate these cocycles.

\subsection{Derivatives}

We consider spacetime $X$ with its lattice it as a CW space. We want to define a ``cocycle which depends on position":
\[\omega_0(x,g_1,...,g_d) \in C^D(BG,C^0(X,U(1)^{or})).\]
vHere $D$ is the dimension of $X$. The local coefficients $U(1)^{or}$ are there because we want our cocycle to be a pseudoscalar function, so that orientation reversing elements of $G$ conjugate the phase. This is important for getting the correct action of $G$ on $C^0$ and the correct classification in the case of time reversal symmetry and other orientation-reversing spacetime symmetries. In the rest of the section, we will take this coefficient system to be understood, just writing $C^j(X)$. Note that we cannot exchange the order of $X$ and $BG$ in the above and get the right answers. In a sense, what we have is a group cochain valued in pseudoscalar functions on $X$, ie. volume forms on $X$.

We can define two differentials for such an object. The first measures a change in $\omega_0$ under a gauge transformation:
\begin{multline*}
d_G \omega_0(x,g_1,...,g_{D+1}) \\= \omega_0(x\cdot g_1,g_2,...,g_{D+1})
- \omega_0(x,g_1g_2,g_3,...,g_{D+1}) \\+ \omega_0(x,g_1,g_2g_3,...,g_{D+1}) + (-1)^{d+2} \omega_0(x,g_1,...,g_d),
\end{multline*}
where in the first term we have used the action of $G$ on $X$ which we write as a right action. This reflects that when we write $C^D(BG,C^0(X))$ we are considering $C^0(X)$ as a local coefficient system on $BG$ induced by the action of $G$ on $X$. To have a gauge invariant cocycle, we need
\[d_G \omega_0 = 0.\]
When this is true, we have a group cocycle valued in volume forms on $X$.

There is another differential that measures whether the integral of this volume form defines a topological invariant of $X$:
\[d_X \omega_0(x,y,g_1,...,g_d) = \omega_0(y,g_1,...,g_d) - \omega_0(x,g_1,...,g_d).\]
For $\omega_0$, $d_X\omega_0 = 0$ means that our cocycle is constant. However, it makes more sense to say that it is constant up to a gauge transformation. Thus, we should instead require
\[d_X \omega_0 = - d_G \omega_1\]
for some
\[\omega_1 \in C^{D-1}(BG,C^1(X)).\]
It might not be possible to solve this equation. Indeed, we are asking that
\[d_1\omega_0 := [d_X \omega_0] \in H^{d}(BG,H^1(X,U(1)^{or}))\]
be zero. If it is, then we can find $\omega_1$ and make a redefinition of our cocycle
\[\omega = \omega_0 + \omega_1.\]

To say that a homotopy of $X$ with its $G$ action can be compensated by a gauge transformation (and vice versa) we need to say $(d_G + d_X)\omega = 0$. Inspecting what we have so far, we find
\[(d_G + d_X)\omega = d_X \omega_1,\]
so we would like to find an $\omega_2 \in C^{d-2}(BG,C^2(X))$ with
\[d_X \omega_1 = - d_G \omega_2.\]
The obstruction is
\[d_2 \omega_0 := [d_X \omega_1] \in H^{d-1}(BG,H^2(X,U(1)^{or})).\]
The pattern continues until the last obstruction
\[d_d \in H^0(BG,H^{D+1}(X,U(1)^{or})).\]
If we can solve all these descent equations, then we obtain a cocycle for both $d_X$ and $d_G$, ie. both topological and gauge invariant:
\[\omega = \omega_0 + ... + \omega_D.\]
Such a thing is called an equivariant cocycle and defines a class in ordinary equivariant cohomology $H^D_G(X)$.

Note we didn't have to start with $\omega_0 \in C^D(BG,C^0(X))$ but could've instead started with a class in any of these groups:
\[\omega_0 \in C^{p}(BG,C^{q}(X))\]
with $p+q = D$. Indeed, given any class $[\omega_0] \in H^p(BG,H^q(X,U(1)^{or}))$, we can compute these differentials. If they all vanish, $\omega_0$ defines a class in equivariant cohomology $H_G^D(X,U(1)^{or})$. All classes in equivariant cohomology arise this way, but there may be new relations between classes and the group structure of $H_G^D(X)$ may be some extension, with classes in $H^p(BG,H^q(X,U(1)^{or}))$ combining into a class that comes from $H^{p+1}(BG,H^{q-1}(X,U(1)^{or}))$. The extension always goes in that direction and is always abelian.

The groups
\[E^{p,q}_2 := H^p(BG,H^q(X,U(1)^{or}))\]
have a very nice interpretation by decorating domain walls. It describes how codimension $p$ symmetry defects are decorated with $q$ forms.

\subsection{Integrals}

In this section we discuss how one computes the topological response Eq. \ref{topresp} of Section \ref{sec_topological_actions}
\[\int_M A^* \omega\]
for $A:M \to X//G$ and $\omega \in H^D_G(X)$ a cocycle described using the descent sequence of the previous section.

The integral is a sum over integrals for each $D$-facet of $M$. In order to evaluate the pullback $A^*\omega$ on a $D$-facet in $M$, we decompose $\omega = \omega_0 + \omega_1 + ... + \omega_D,$ where $\omega_j \in C^{D-j}(BG,C^j(X))$ and evaluate each $\omega_j$ and sum. The evaluation of $\omega_j$ works just like the cup product \cite{Hatcher}. This is simplest to describe in a simplical refinement of the cell structure on $M$. All simplicies in $M$ are assumed to have ordered vertices. The first $D-j+1$ vertices with the $G$ labels on edges between them form a $D-j$-simplex in $BG$ on which we evaluate $\omega_j$ to obtain an element of $C^j(X)$. The last $j+1$ vertices with their edges and faces as above are
\[x_{D-j} \leadsto y_{D-j} \xrightarrow{g_{D-j}} x_{D-j+1} \leadsto \cdots \xrightarrow{g_D} y_D\]
to which we can push all the $g$'s to the right to obtain a $j$-chain in $X$ on which we evaluate that cochain, obtaining a number. The sum of these numbers is the value of $A^*\omega$ on this $D$-simplex of $M$.

\end{appendices}

\input{main.bbl}

\end{document}

%% file: classif_pg1.txt
\begin{tabular}{|l|l|l|}
\hline
Number & Name & Classification \\ \hline
1 & $\mathrm{P1}$ & 0\\
2 & $\mathrm{P\overline{1}}$ & $\mathbb{Z}_{2}^{\times 8}$\\
3 & $\mathrm{P2}$ & $\mathbb{Z}_{2}^{\times 4}$\\
4 & $\mathrm{P2_1}$ & 0\\
5 & $\mathrm{C2}$ & $\mathbb{Z}_{2}^{\times 2}$\\
6 & $\mathrm{Pm}$ & $\mathbb{Z}_{2}^{\times 4}$\\
7 & $\mathrm{Pc}$ & 0\\
8 & $\mathrm{Cm}$ & $\mathbb{Z}_{2}^{\times 2}$\\
9 & $\mathrm{Cc}$ & 0\\
10 & $\mathrm{P2/m}$ & $\mathbb{Z}_{2}^{\times 18}$\\
11 & $\mathrm{P2_1/m}$ & $\mathbb{Z}_{2}^{\times 6}$\\
12 & $\mathrm{C2/m}$ & $\mathbb{Z}_{2}^{\times 11}$\\
13 & $\mathrm{P2/c}$ & $\mathbb{Z}_{2}^{\times 6}$\\
14 & $\mathrm{P2_1/c}$ & $\mathbb{Z}_{2}^{\times 4}$\\
15 & $\mathrm{C2/c}$ & $\mathbb{Z}_{2}^{\times 5}$\\
16 & $\mathrm{P222}$ & $\mathbb{Z}_{2}^{\times 16}$\\
17 & $\mathrm{P222_1}$ & $\mathbb{Z}_{2}^{\times 4}$\\
18 & $\mathrm{P2_12_12}$ & $\mathbb{Z}_{2}^{\times 2}$\\
19 & $\mathrm{P2_12_12_1}$ & 0\\
20 & $\mathrm{C222_1}$ & $\mathbb{Z}_{2}^{\times 2}$\\
21 & $\mathrm{C222}$ & $\mathbb{Z}_{2}^{\times 9}$\\
22 & $\mathrm{F222}$ & $\mathbb{Z}_{2}^{\times 8}$\\
23 & $\mathrm{I222}$ & $\mathbb{Z}_{2}^{\times 8}$\\
24 & $\mathrm{I2_12_12_1}$ & $\mathbb{Z}_{2}^{\times 3}$\\
25 & $\mathrm{Pmm2}$ & $\mathbb{Z}_{2}^{\times 16}$\\
26 & $\mathrm{Pmc2_1}$ & $\mathbb{Z}_{2}^{\times 4}$\\
27 & $\mathrm{Pcc2}$ & $\mathbb{Z}_{2}^{\times 4}$\\
28 & $\mathrm{Pma2}$ & $\mathbb{Z}_{2}^{\times 4}$\\
29 & $\mathrm{Pca2_1}$ & 0\\
30 & $\mathrm{Pnc2}$ & $\mathbb{Z}_{2}^{\times 2}$\\
31 & $\mathrm{Pmn2_1}$ & $\mathbb{Z}_{2}^{\times 2}$\\
32 & $\mathrm{Pba2}$ & $\mathbb{Z}_{2}^{\times 2}$\\
33 & $\mathrm{Pna2_1}$ & 0\\
34 & $\mathrm{Pnn2}$ & $\mathbb{Z}_{2}^{\times 2}$\\
35 & $\mathrm{Cmm2}$ & $\mathbb{Z}_{2}^{\times 9}$\\
36 & $\mathrm{Cmc2_1}$ & $\mathbb{Z}_{2}^{\times 2}$\\
37 & $\mathrm{Ccc2}$ & $\mathbb{Z}_{2}^{\times 3}$\\
38 & $\mathrm{Amm2}$ & $\mathbb{Z}_{2}^{\times 9}$\\
39 & $\mathrm{Aem2}$ & $\mathbb{Z}_{2}^{\times 4}$\\
\hline
\end{tabular}
\quad
\begin{tabular}{|l|l|l|}
\hline
Number & Name & Classification \\ \hline
40 & $\mathrm{Ama2}$ & $\mathbb{Z}_{2}^{\times 3}$\\
41 & $\mathrm{Aea2}$ & $\mathbb{Z}_{2}$\\
42 & $\mathrm{Fmm2}$ & $\mathbb{Z}_{2}^{\times 6}$\\
43 & $\mathrm{Fdd2}$ & $\mathbb{Z}_{2}$\\
44 & $\mathrm{Imm2}$ & $\mathbb{Z}_{2}^{\times 8}$\\
45 & $\mathrm{Iba2}$ & $\mathbb{Z}_{2}^{\times 2}$\\
46 & $\mathrm{Ima2}$ & $\mathbb{Z}_{2}^{\times 3}$\\
47 & $\mathrm{Pmmm}$ & $\mathbb{Z}_{2}^{\times 42}$\\
48 & $\mathrm{Pnnn}$ & $\mathbb{Z}_{2}^{\times 10}$\\
49 & $\mathrm{Pccm}$ & $\mathbb{Z}_{2}^{\times 17}$\\
50 & $\mathrm{Pban}$ & $\mathbb{Z}_{2}^{\times 10}$\\
51 & $\mathrm{Pmma}$ & $\mathbb{Z}_{2}^{\times 17}$\\
52 & $\mathrm{Pnna}$ & $\mathbb{Z}_{2}^{\times 4}$\\
53 & $\mathrm{Pmna}$ & $\mathbb{Z}_{2}^{\times 10}$\\
54 & $\mathrm{Pcca}$ & $\mathbb{Z}_{2}^{\times 5}$\\
55 & $\mathrm{Pbam}$ & $\mathbb{Z}_{2}^{\times 10}$\\
56 & $\mathrm{Pccn}$ & $\mathbb{Z}_{2}^{\times 4}$\\
57 & $\mathrm{Pbcm}$ & $\mathbb{Z}_{2}^{\times 5}$\\
58 & $\mathrm{Pnnm}$ & $\mathbb{Z}_{2}^{\times 9}$\\
59 & $\mathrm{Pmmn}$ & $\mathbb{Z}_{2}^{\times 10}$\\
60 & $\mathrm{Pbcn}$ & $\mathbb{Z}_{2}^{\times 3}$\\
61 & $\mathrm{Pbca}$ & $\mathbb{Z}_{2}^{\times 2}$\\
62 & $\mathrm{Pnma}$ & $\mathbb{Z}_{2}^{\times 4}$\\
63 & $\mathrm{Cmcm}$ & $\mathbb{Z}_{2}^{\times 10}$\\
64 & $\mathrm{Cmce}$ & $\mathbb{Z}_{2}^{\times 7}$\\
65 & $\mathrm{Cmmm}$ & $\mathbb{Z}_{2}^{\times 26}$\\
66 & $\mathrm{Cccm}$ & $\mathbb{Z}_{2}^{\times 13}$\\
67 & $\mathrm{Cmme}$ & $\mathbb{Z}_{2}^{\times 17}$\\
68 & $\mathrm{Ccce}$ & $\mathbb{Z}_{2}^{\times 7}$\\
69 & $\mathrm{Fmmm}$ & $\mathbb{Z}_{2}^{\times 20}$\\
70 & $\mathrm{Fddd}$ & $\mathbb{Z}_{2}^{\times 6}$\\
71 & $\mathrm{Immm}$ & $\mathbb{Z}_{2}^{\times 22}$\\
72 & $\mathrm{Ibam}$ & $\mathbb{Z}_{2}^{\times 10}$\\
73 & $\mathrm{Ibca}$ & $\mathbb{Z}_{2}^{\times 5}$\\
74 & $\mathrm{Imma}$ & $\mathbb{Z}_{2}^{\times 13}$\\
75 & $\mathrm{P4}$ & $\mathbb{Z}_{2}\times\mathbb{Z}_{4}^{\times 2}$\\
76 & $\mathrm{P4_1}$ & 0\\
77 & $\mathrm{P4_2}$ & $\mathbb{Z}_{2}^{\times 3}$\\
78 & $\mathrm{P4_3}$ & 0\\
\hline
\end{tabular}
\quad
\begin{tabular}{|l|l|l|}
\hline
Number & Name & Classification \\ \hline
79 & $\mathrm{I4}$ & $\mathbb{Z}_{2}\times\mathbb{Z}_{4}$\\
80 & $\mathrm{I4_1}$ & $\mathbb{Z}_{2}$\\
81 & $\mathrm{P\overline{4}}$ & $\mathbb{Z}_{2}^{\times 3}\times\mathbb{Z}_{4}^{\times 2}$\\
82 & $\mathrm{I\overline{4}}$ & $\mathbb{Z}_{2}^{\times 2}\times\mathbb{Z}_{4}^{\times 2}$\\
83 & $\mathrm{P4/m}$ & $\mathbb{Z}_{2}^{\times 12}\times\mathbb{Z}_{4}^{\times 2}$\\
84 & $\mathrm{P4_2/m}$ & $\mathbb{Z}_{2}^{\times 11}$\\
85 & $\mathrm{P4/n}$ & $\mathbb{Z}_{2}^{\times 3}\times\mathbb{Z}_{4}^{\times 2}$\\
86 & $\mathrm{P4_2/n}$ & $\mathbb{Z}_{2}^{\times 4}\times\mathbb{Z}_{4}$\\
87 & $\mathrm{I4/m}$ & $\mathbb{Z}_{2}^{\times 8}\times\mathbb{Z}_{4}$\\
88 & $\mathrm{I4_1/a}$ & $\mathbb{Z}_{2}^{\times 3}\times\mathbb{Z}_{4}$\\
89 & $\mathrm{P422}$ & $\mathbb{Z}_{2}^{\times 12}$\\
90 & $\mathrm{P42_12}$ & $\mathbb{Z}_{2}^{\times 4}\times\mathbb{Z}_{4}$\\
91 & $\mathrm{P4_122}$ & $\mathbb{Z}_{2}^{\times 3}$\\
92 & $\mathrm{P4_12_12}$ & $\mathbb{Z}_{2}$\\
93 & $\mathrm{P4_222}$ & $\mathbb{Z}_{2}^{\times 12}$\\
94 & $\mathrm{P4_22_12}$ & $\mathbb{Z}_{2}^{\times 5}$\\
95 & $\mathrm{P4_322}$ & $\mathbb{Z}_{2}^{\times 3}$\\
96 & $\mathrm{P4_32_12}$ & $\mathbb{Z}_{2}$\\
97 & $\mathrm{I422}$ & $\mathbb{Z}_{2}^{\times 8}$\\
98 & $\mathrm{I4_122}$ & $\mathbb{Z}_{2}^{\times 5}$\\
99 & $\mathrm{P4mm}$ & $\mathbb{Z}_{2}^{\times 12}$\\
100 & $\mathrm{P4bm}$ & $\mathbb{Z}_{2}^{\times 4}\times\mathbb{Z}_{4}$\\
101 & $\mathrm{P4_2cm}$ & $\mathbb{Z}_{2}^{\times 6}$\\
102 & $\mathrm{P4_2nm}$ & $\mathbb{Z}_{2}^{\times 5}$\\
103 & $\mathrm{P4cc}$ & $\mathbb{Z}_{2}^{\times 3}$\\
104 & $\mathrm{P4nc}$ & $\mathbb{Z}_{2}\times\mathbb{Z}_{4}$\\
105 & $\mathrm{P4_2mc}$ & $\mathbb{Z}_{2}^{\times 9}$\\
106 & $\mathrm{P4_2bc}$ & $\mathbb{Z}_{2}^{\times 2}$\\
107 & $\mathrm{I4mm}$ & $\mathbb{Z}_{2}^{\times 7}$\\
108 & $\mathrm{I4cm}$ & $\mathbb{Z}_{2}^{\times 4}$\\
109 & $\mathrm{I4_1md}$ & $\mathbb{Z}_{2}^{\times 4}$\\
110 & $\mathrm{I4_1cd}$ & $\mathbb{Z}_{2}$\\
111 & $\mathrm{P\overline{4}2m}$ & $\mathbb{Z}_{2}^{\times 13}$\\
112 & $\mathrm{P\overline{4}2c}$ & $\mathbb{Z}_{2}^{\times 10}$\\
113 & $\mathrm{P\overline{4}2_1m}$ & $\mathbb{Z}_{2}^{\times 5}\times\mathbb{Z}_{4}$\\
114 & $\mathrm{P\overline{4}2_1c}$ & $\mathbb{Z}_{2}^{\times 2}\times\mathbb{Z}_{4}$\\
115 & $\mathrm{P\overline{4}m2}$ & $\mathbb{Z}_{2}^{\times 13}$\\
116 & $\mathrm{P\overline{4}c2}$ & $\mathbb{Z}_{2}^{\times 7}$\\
117 & $\mathrm{P\overline{4}b2}$ & $\mathbb{Z}_{2}^{\times 5}\times\mathbb{Z}_{4}$\\
\hline
\end{tabular}
\quad

%% file: classif_pg2.txt
\begin{tabular}{|l|l|l|}
\hline
Number & Name & Classification \\ \hline
118 & $\mathrm{P\overline{4}n2}$ & $\mathbb{Z}_{2}^{\times 5}\times\mathbb{Z}_{4}$\\
119 & $\mathrm{I\overline{4}m2}$ & $\mathbb{Z}_{2}^{\times 9}$\\
120 & $\mathrm{I\overline{4}c2}$ & $\mathbb{Z}_{2}^{\times 6}$\\
121 & $\mathrm{I\overline{4}2m}$ & $\mathbb{Z}_{2}^{\times 8}$\\
122 & $\mathrm{I\overline{4}2d}$ & $\mathbb{Z}_{2}^{\times 2}\times\mathbb{Z}_{4}$\\
123 & $\mathrm{P4/mmm}$ & $\mathbb{Z}_{2}^{\times 32}$\\
124 & $\mathrm{P4/mcc}$ & $\mathbb{Z}_{2}^{\times 13}$\\
125 & $\mathrm{P4/nbm}$ & $\mathbb{Z}_{2}^{\times 13}$\\
126 & $\mathrm{P4/nnc}$ & $\mathbb{Z}_{2}^{\times 8}$\\
127 & $\mathrm{P4/mbm}$ & $\mathbb{Z}_{2}^{\times 15}\times\mathbb{Z}_{4}$\\
128 & $\mathrm{P4/mnc}$ & $\mathbb{Z}_{2}^{\times 8}\times\mathbb{Z}_{4}$\\
129 & $\mathrm{P4/nmm}$ & $\mathbb{Z}_{2}^{\times 13}$\\
130 & $\mathrm{P4/ncc}$ & $\mathbb{Z}_{2}^{\times 5}$\\
131 & $\mathrm{P4_2/mmc}$ & $\mathbb{Z}_{2}^{\times 24}$\\
132 & $\mathrm{P4_2/mcm}$ & $\mathbb{Z}_{2}^{\times 18}$\\
133 & $\mathrm{P4_2/nbc}$ & $\mathbb{Z}_{2}^{\times 8}$\\
134 & $\mathrm{P4_2/nnm}$ & $\mathbb{Z}_{2}^{\times 13}$\\
135 & $\mathrm{P4_2/mbc}$ & $\mathbb{Z}_{2}^{\times 8}$\\
136 & $\mathrm{P4_2/mnm}$ & $\mathbb{Z}_{2}^{\times 14}$\\
137 & $\mathrm{P4_2/nmc}$ & $\mathbb{Z}_{2}^{\times 8}$\\
138 & $\mathrm{P4_2/ncm}$ & $\mathbb{Z}_{2}^{\times 10}$\\
139 & $\mathrm{I4/mmm}$ & $\mathbb{Z}_{2}^{\times 20}$\\
140 & $\mathrm{I4/mcm}$ & $\mathbb{Z}_{2}^{\times 14}$\\
141 & $\mathrm{I4_1/amd}$ & $\mathbb{Z}_{2}^{\times 9}$\\
142 & $\mathrm{I4_1/acd}$ & $\mathbb{Z}_{2}^{\times 5}$\\
143 & $\mathrm{P3}$ & $\mathbb{Z}_{3}^{\times 3}$\\
144 & $\mathrm{P3_1}$ & 0\\
145 & $\mathrm{P3_2}$ & 0\\
146 & $\mathrm{R3}$ & $\mathbb{Z}_{3}$\\
147 & $\mathrm{P\overline{3}}$ & $\mathbb{Z}_{2}^{\times 4}\times\mathbb{Z}_{3}^{\times 2}$\\
148 & $\mathrm{R\overline{3}}$ & $\mathbb{Z}_{2}^{\times 4}\times\mathbb{Z}_{3}$\\
149 & $\mathrm{P312}$ & $\mathbb{Z}_{2}^{\times 2}$\\
150 & $\mathrm{P321}$ & $\mathbb{Z}_{2}^{\times 2}\times\mathbb{Z}_{3}$\\
151 & $\mathrm{P3_112}$ & $\mathbb{Z}_{2}^{\times 2}$\\
152 & $\mathrm{P3_121}$ & $\mathbb{Z}_{2}^{\times 2}$\\
153 & $\mathrm{P3_212}$ & $\mathbb{Z}_{2}^{\times 2}$\\
154 & $\mathrm{P3_221}$ & $\mathbb{Z}_{2}^{\times 2}$\\
155 & $\mathrm{R32}$ & $\mathbb{Z}_{2}^{\times 2}$\\
156 & $\mathrm{P3m1}$ & $\mathbb{Z}_{2}^{\times 2}$\\
\hline
\end{tabular}
\quad
\begin{tabular}{|l|l|l|}
\hline
Number & Name & Classification \\ \hline
157 & $\mathrm{P31m}$ & $\mathbb{Z}_{2}^{\times 2}\times\mathbb{Z}_{3}$\\
158 & $\mathrm{P3c1}$ & 0\\
159 & $\mathrm{P31c}$ & $\mathbb{Z}_{3}$\\
160 & $\mathrm{R3m}$ & $\mathbb{Z}_{2}^{\times 2}$\\
161 & $\mathrm{R3c}$ & 0\\
162 & $\mathrm{P\overline{3}1m}$ & $\mathbb{Z}_{2}^{\times 9}$\\
163 & $\mathrm{P\overline{3}1c}$ & $\mathbb{Z}_{2}^{\times 3}$\\
164 & $\mathrm{P\overline{3}m1}$ & $\mathbb{Z}_{2}^{\times 9}$\\
165 & $\mathrm{P\overline{3}c1}$ & $\mathbb{Z}_{2}^{\times 3}$\\
166 & $\mathrm{R\overline{3}m}$ & $\mathbb{Z}_{2}^{\times 9}$\\
167 & $\mathrm{R\overline{3}c}$ & $\mathbb{Z}_{2}^{\times 3}$\\
168 & $\mathrm{P6}$ & $\mathbb{Z}_{2}^{\times 2}\times\mathbb{Z}_{3}^{\times 2}$\\
169 & $\mathrm{P6_1}$ & 0\\
170 & $\mathrm{P6_5}$ & 0\\
171 & $\mathrm{P6_2}$ & $\mathbb{Z}_{2}^{\times 2}$\\
172 & $\mathrm{P6_4}$ & $\mathbb{Z}_{2}^{\times 2}$\\
173 & $\mathrm{P6_3}$ & $\mathbb{Z}_{3}^{\times 2}$\\
174 & $\mathrm{P\overline{6}}$ & $\mathbb{Z}_{2}^{\times 4}\times\mathbb{Z}_{3}^{\times 3}$\\
175 & $\mathrm{P6/m}$ & $\mathbb{Z}_{2}^{\times 10}\times\mathbb{Z}_{3}^{\times 2}$\\
176 & $\mathrm{P6_3/m}$ & $\mathbb{Z}_{2}^{\times 4}\times\mathbb{Z}_{3}^{\times 2}$\\
177 & $\mathrm{P622}$ & $\mathbb{Z}_{2}^{\times 8}$\\
178 & $\mathrm{P6_122}$ & $\mathbb{Z}_{2}^{\times 2}$\\
179 & $\mathrm{P6_522}$ & $\mathbb{Z}_{2}^{\times 2}$\\
180 & $\mathrm{P6_222}$ & $\mathbb{Z}_{2}^{\times 8}$\\
181 & $\mathrm{P6_422}$ & $\mathbb{Z}_{2}^{\times 8}$\\
182 & $\mathrm{P6_322}$ & $\mathbb{Z}_{2}^{\times 2}$\\
183 & $\mathrm{P6mm}$ & $\mathbb{Z}_{2}^{\times 8}$\\
184 & $\mathrm{P6cc}$ & $\mathbb{Z}_{2}^{\times 2}$\\
185 & $\mathrm{P6_3cm}$ & $\mathbb{Z}_{2}^{\times 2}$\\
186 & $\mathrm{P6_3mc}$ & $\mathbb{Z}_{2}^{\times 2}$\\
187 & $\mathrm{P\overline{6}m2}$ & $\mathbb{Z}_{2}^{\times 9}$\\
188 & $\mathrm{P\overline{6}c2}$ & $\mathbb{Z}_{2}^{\times 3}$\\
189 & $\mathrm{P\overline{6}2m}$ & $\mathbb{Z}_{2}^{\times 9}\times\mathbb{Z}_{3}$\\
190 & $\mathrm{P\overline{6}2c}$ & $\mathbb{Z}_{2}^{\times 3}\times\mathbb{Z}_{3}$\\
191 & $\mathrm{P6/mmm}$ & $\mathbb{Z}_{2}^{\times 22}$\\
192 & $\mathrm{P6/mcc}$ & $\mathbb{Z}_{2}^{\times 9}$\\
193 & $\mathrm{P6_3/mcm}$ & $\mathbb{Z}_{2}^{\times 9}$\\
194 & $\mathrm{P6_3/mmc}$ & $\mathbb{Z}_{2}^{\times 9}$\\
195 & $\mathrm{P23}$ & $\mathbb{Z}_{2}^{\times 4}\times\mathbb{Z}_{3}$\\
\hline
\end{tabular}
\quad
\begin{tabular}{|l|l|l|}
\hline
Number & Name & Classification \\ \hline
196 & $\mathrm{F23}$ & $\mathbb{Z}_{3}$\\
197 & $\mathrm{I23}$ & $\mathbb{Z}_{2}^{\times 2}\times\mathbb{Z}_{3}$\\
198 & $\mathrm{P2_13}$ & $\mathbb{Z}_{3}$\\
199 & $\mathrm{I2_13}$ & $\mathbb{Z}_{2}\times\mathbb{Z}_{3}$\\
200 & $\mathrm{Pm\overline{3}}$ & $\mathbb{Z}_{2}^{\times 14}\times\mathbb{Z}_{3}$\\
201 & $\mathrm{Pn\overline{3}}$ & $\mathbb{Z}_{2}^{\times 4}\times\mathbb{Z}_{3}$\\
202 & $\mathrm{Fm\overline{3}}$ & $\mathbb{Z}_{2}^{\times 6}\times\mathbb{Z}_{3}$\\
203 & $\mathrm{Fd\overline{3}}$ & $\mathbb{Z}_{2}^{\times 2}\times\mathbb{Z}_{3}$\\
204 & $\mathrm{Im\overline{3}}$ & $\mathbb{Z}_{2}^{\times 8}\times\mathbb{Z}_{3}$\\
205 & $\mathrm{Pa\overline{3}}$ & $\mathbb{Z}_{2}^{\times 2}\times\mathbb{Z}_{3}$\\
206 & $\mathrm{Ia\overline{3}}$ & $\mathbb{Z}_{2}^{\times 3}\times\mathbb{Z}_{3}$\\
207 & $\mathrm{P432}$ & $\mathbb{Z}_{2}^{\times 6}$\\
208 & $\mathrm{P4_232}$ & $\mathbb{Z}_{2}^{\times 6}$\\
209 & $\mathrm{F432}$ & $\mathbb{Z}_{2}^{\times 4}$\\
210 & $\mathrm{F4_132}$ & $\mathbb{Z}_{2}$\\
211 & $\mathrm{I432}$ & $\mathbb{Z}_{2}^{\times 5}$\\
212 & $\mathrm{P4_332}$ & $\mathbb{Z}_{2}$\\
213 & $\mathrm{P4_132}$ & $\mathbb{Z}_{2}$\\
214 & $\mathrm{I4_132}$ & $\mathbb{Z}_{2}^{\times 4}$\\
215 & $\mathrm{P\overline{4}3m}$ & $\mathbb{Z}_{2}^{\times 7}$\\
216 & $\mathrm{F\overline{4}3m}$ & $\mathbb{Z}_{2}^{\times 5}$\\
217 & $\mathrm{I\overline{4}3m}$ & $\mathbb{Z}_{2}^{\times 5}$\\
218 & $\mathrm{P\overline{4}3n}$ & $\mathbb{Z}_{2}^{\times 4}$\\
219 & $\mathrm{F\overline{4}3c}$ & $\mathbb{Z}_{2}^{\times 2}$\\
220 & $\mathrm{I\overline{4}3d}$ & $\mathbb{Z}_{2}\times\mathbb{Z}_{4}$\\
221 & $\mathrm{Pm\overline{3}m}$ & $\mathbb{Z}_{2}^{\times 18}$\\
222 & $\mathrm{Pn\overline{3}n}$ & $\mathbb{Z}_{2}^{\times 5}$\\
223 & $\mathrm{Pm\overline{3}n}$ & $\mathbb{Z}_{2}^{\times 10}$\\
224 & $\mathrm{Pn\overline{3}m}$ & $\mathbb{Z}_{2}^{\times 10}$\\
225 & $\mathrm{Fm\overline{3}m}$ & $\mathbb{Z}_{2}^{\times 13}$\\
226 & $\mathrm{Fm\overline{3}c}$ & $\mathbb{Z}_{2}^{\times 7}$\\
227 & $\mathrm{Fd\overline{3}m}$ & ???\\
228 & $\mathrm{Fd\overline{3}c}$ & ???\\
229 & $\mathrm{Im\overline{3}m}$ & $\mathbb{Z}_{2}^{\times 13}$\\
230 & $\mathrm{Ia\overline{3}d}$ & ???\\
\hline
\end{tabular}
\quad

%% file: main.bbl
%